\documentclass[12pt]{article}
\usepackage{amsmath,amssymb,amsfonts,amsthm,bbm}
\usepackage{epic,eepic,epsfig,longtable}
\usepackage{multirow,verbatim}
\usepackage{array}
\usepackage{graphicx}
\usepackage{floatrow}
\usepackage{subcaption}
\usepackage{paralist}
\usepackage{latexsym}
\usepackage{comment}
\usepackage{epsfig}
\usepackage{setspace}
\usepackage{CJK}
\usepackage{color}
\usepackage{geometry}
\usepackage{algorithm}
\usepackage{algpseudocode}

\usepackage{rotating}

\usepackage[authoryear, round]{natbib}
\def\spacingset#1{\renewcommand{\baselinestretch}%
{#1}\small\normalsize} \spacingset{1}

\makeatletter
\def\singlespace{\def\baselinestretch{1}\@normalsize}

\numberwithin{equation}{section}

\renewcommand{\hat}{\widehat}

\renewcommand{\hat}{\widehat}

\newcommand{\bfm}[1]{\ensuremath{\mathbf{#1}}}

\def\ba{\bfm a}   \def\bA{\bfm A}  
\def\bb{\bfm b}   \def\bB{\bfm B}  
     
   \def\bD{\bfm D}  
\def\be{\bfm e}     
\def\bff{\bfm f}    \def\FF{\mathbb{F}}

   \def\bI{\bfm I}  
   \def\bJ{\bfm J}

   \def\bM{\bfm M}

\def\bq{\bfm q}   \def\bQ{\bfm Q}  
   \def\bR{\bfm R}

\def\bu{\bfm u}   \def\bU{\bfm U}  
     
\def\bw{\bfm w}   \def\bW{\bfm W}  
\def\bx{\bfm x}   \def\bX{\bfm X}  
\def\by{\bfm y}   \def\bY{\bfm Y}

\newcommand{\bfsym}[1]{\ensuremath{\boldsymbol{#1}}}

 \def\bbeta{\bfsym \beta}
 \def\bgamma{\bfsym \gamma}             \def\bGamma{\bfsym \Gamma}

 \def\bmu{\bfsym {\mu}}                 
 \def\bnu{\bfsym {\nu}}
 \def\btheta{\bfsym {\theta}}           
 
 \def\bepsilon{\bfsym \epsilon}        		
              \def\bSigma{\bfsym \Sigma}
 \def\blambda {\bfsym {\lambda}}        \def\bLambda {\bfsym {\Lambda}}

 				\def\bpi {\bfsym {\pi}}
 \def\bxi{\bfsym {\xi}}			
 \def\bzeta{\bfsym {\zeta}}				\def\bPhi{\bfsym {\Phi}}

\def\bvartheta{\bfsym{\vartheta}}	
\DeclareMathOperator{\corr}{corr}
\DeclareMathOperator{\cov}{cov}
\DeclareMathOperator{\Cov}{Cov}

\DeclareMathOperator{\Diag}{Diag}
\DeclareMathOperator{\vech}{vech}
\DeclareMathOperator{\E}{E}

\DeclareMathOperator{\Var}{Var}
\DeclareMathOperator{\var}{var}

\DeclareMathOperator{\tr}{tr}

\def\newpage{\vfill\eject}

\def\var{\mbox{var}}
\def\corr{\mbox{corr}}

\def\today{\ifcase\month\or
  January\or February\or March\or April\or May\or June\or
  July\or August\or September\or October\or November\or December\fi
  \space\number\day, \number\year}
\def\Cov{\mbox{Cov}}

\newdimen\biblioindent    \biblioindent=30pt

 at 8truept

\newcommand{\beq}{\begin{equation}}
  \newcommand{\eeq}{\end{equation}}
\newcommand{\beqn}{\begin{eqnarray}}
  \newcommand{\eeqn}{\end{eqnarray}}
\newcommand{\beqnn}{\begin{eqnarray*}}
  \newcommand{\eeqnn}{\end{eqnarray*}}

\allowdisplaybreaks
\usepackage{verbatim}
\renewcommand{\baselinestretch}{1.66}
\def\tilde{\widetilde}

\def\FF{\mathcal{F}}
\def\[{\left [}  \def\]{\right ]} \def\({\left (}  \def\){\right )}
 \def\endpf{$\blacksquare$}
\def\hat{\widehat}

 \def \E {\mathrm{E}}    
 \def \1 {\mathbf{1}}

\newtheorem{thm}{Theorem}

\newtheorem{lemma}{Lemma}

\newtheorem{remark}{Remark}
\newtheorem{proposition}{Proposition}
\newtheorem{mydef}{Definition}
\newtheorem{assumption}{Assumption}

\usepackage{lipsum}
\title{Factor and Idiosyncratic VAR Volatility Matrix Models for  Heavy-Tailed High-Frequency Financial Observations\footnote{
Minseok Shin is Assistant Professor, Department of Industrial and Management Engineering, POSTECH, Pohang 37673, South Korea.
Donggyu Kim is Professor,  Department of Economics, University of California, Riverside, CA 92521, USA.
His research was supported by the National Research Foundation of Korea (NRF) grant funded by the Korea government (MSIT) [RS-2024-00343129].
Yazhen Wang is Chair and Professor, Department of Statistics, University of Wisconsin-Madison, 1300 University Avenue, Madison, WI 53706, USA.
His research was supported by NSF [DMS-1707605, DMS-1913149].
Jianqing Fan is Frederick L. Moore'18 Professor of Finance, Department of Operations Research and Financial Engineering, Princeton University, Princeton, NJ 08544,  USA.  
His research was supported by NSFC [71991471, 71991470].}
}

\author{ Minseok Shin$^a$,   Donggyu Kim$^b$\footnote{corresponding authors. E-mail addresses: minseokshin@postech.ac.kr (M. Shin),  donggyu.kim@ucr.edu (D. Kim), yzwang@stat.wisc.edu (Y. Wang), jqfan@princeton.edu (J. Fan). }, Yazhen Wang$^c$, and  Jianqing Fan$^{d\dag}$ \\
$^a$Pohang University of Science and Technology (POSTECH) \\
$^b$University of California, Riverside\\
$^c$University of Wisconsin-Madison \\
$^d$Princeton University
}

\begin{document}

\maketitle
\begin{spacing}{1.65}

\begin{abstract}
	This paper introduces a novel process for both factor and idiosyncratic volatility matrices whose eigenvalues follow the vector auto-regressive (VAR) model.
	We call it the factor and idiosyncratic VAR (FIVAR) model.
	The FIVAR model accounts for the dynamics of the factor and idiosyncratic volatilities and includes many parameters.   
	In addition, many empirical studies have shown that high-frequency stock returns and volatilities often exhibit heavy tails.
	To handle these two problems simultaneously, we propose a penalized optimization procedure with a truncation scheme for parameter estimation.   
	We apply the proposed parameter estimation procedure to predicting large volatility matrices and establish its asymptotic properties.  
	%Using high-frequency trading data,  the proposed method is applied to large volatility matrix prediction and minimum variance portfolio allocation.
\end{abstract}

\noindent \textbf{Keywords:}  diffusion process,  robust estimation, high-dimensionality, POET, Huber loss, LASSO. \\
\noindent \textbf{JEL classification codes:} C14, C22, C55, C58.

\section{Introduction}

Volatility analysis for high-frequency financial data is a vibrant research area in financial econometrics and statistics.
With the wide availability of high-frequency financial data, several well-performing non-parametric estimation methods have been developed to estimate integrated volatilities \citep{ait2010high,  barndorff2008designing, barndorff2011multivariate, bibinger2014estimating, christensen2010pre,fan2018robust, fan2007multi, jacod2009microstructure,  shin2023adaptive, xiu2010quasi, zhang2005tale, zhang2006efficient, zhang2011estimating}.
%Examples include two-scales estimator \citep{zhang2005tale}, multi-scale realized volatility (MSRV) \citep{zhang2006efficient, zhang2011estimating}, wavelet estimator \citep{fan2007multi}, pre-averaging realized volatility (PRV) \citep{christensen2010pre, jacod2009microstructure}, kernel realized volatility (KRV) \citep{barndorff2008designing, barndorff2011multivariate}, quasi-maximum likelihood estimator (QMLE) \citep{ait2010high, xiu2010quasi},  local method of moments \citep{bibinger2014estimating},  and robust pre-averaging realized volatility \citep{fan2018robust, shin2023adaptive}.
With these non-parametric (daily) realized volatility estimators, parametric models have been developed to account for volatility dynamics over time.
Examples include the realized volatility-based modeling approaches \citep{andersen2003modeling}, the heterogeneous auto-regressive (HAR) models \citep{corsi2009simple}, the realized GARCH models \citep{hansen2012realized},  the high-frequency-based volatility (HEAVY) models \citep{shephard2010realising},  and the unified GARCH-It\^o models \citep{kim2016unified, song2021volatility}.
Their empirical studies showed that incorporating high-frequency information, such as realized volatility, helps capture the volatility dynamics for a finite number of assets.
However, in financial practice, we often need to handle a large number of assets, which leads to an excessive number of parameters for typical sample sizes.
To overcome this problem, the approximate factor model structure is often imposed on volatility matrices \citep{fan2013large}.
For example, high-dimensional factor-based It\^{o} processes are widely utilized with the sparsity assumption on the idiosyncratic volatility \citep{ait2017using, fan2016incorporating, kim2018Large, kong2018systematic}. 
Recently, \citet{kim2019factor}  developed the factor GARCH-It\^o model, based on the high-dimensional factor-based It\^o processes.
The factor GARCH-It\^o model assumes that the eigenvalue sequence of the latent factor volatility matrices admits some unified GARCH-It\^o model structure  \citep{kim2016unified} so that the dynamics of the volatility can be explained by the factors.
See also \citet{hetland2023dynamic} for the low-dimensional low-frequency setting and \citet{kim2022unified, kim2023factor} for the high-dimensional high-frequency setting.
	We note that when employing the approximate factor model structure, the existing literature does not model the idiosyncratic volatility and assumes that the idiosyncratic volatility process is martingale.

\begin{figure}[!ht]
	\centering
	\includegraphics[width = 0.98\textwidth]{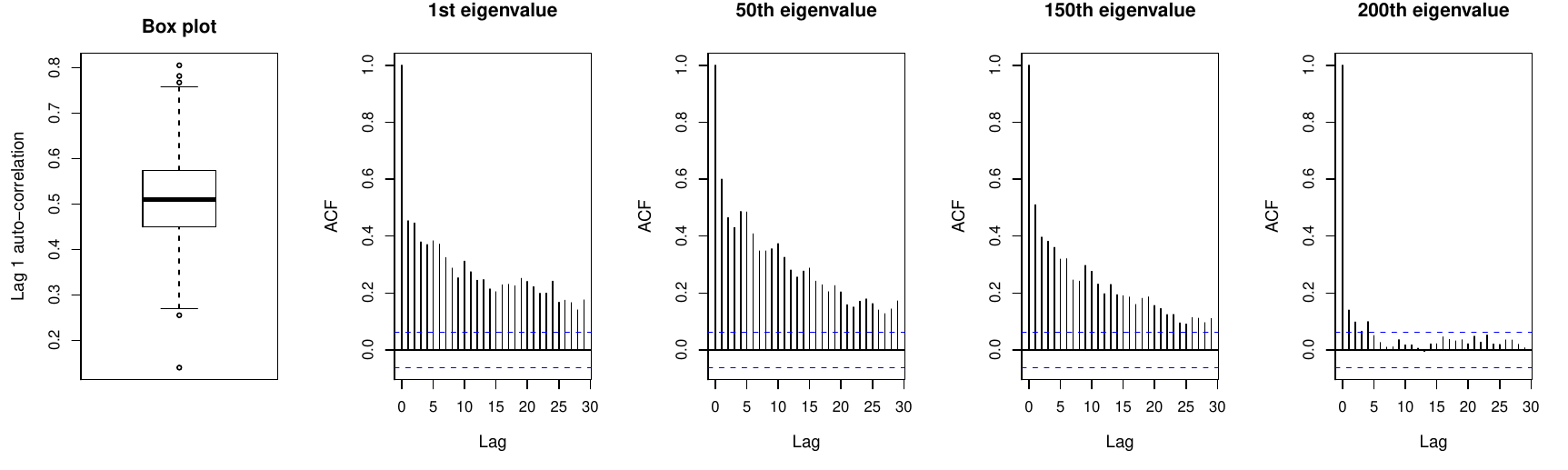}
	\caption{The box plot of the first-order auto-correlations for the time series of 200 daily estimated eigenvalues of the idiosyncratic volatility matrix and the ACF plots for the time series of the 1st, 50th, 150th, and 200th eigenvalues.
	}\label{Fig-1}
\end{figure}

\begin{figure}[!ht]
	\centering
	\includegraphics[width = 0.8\textwidth]{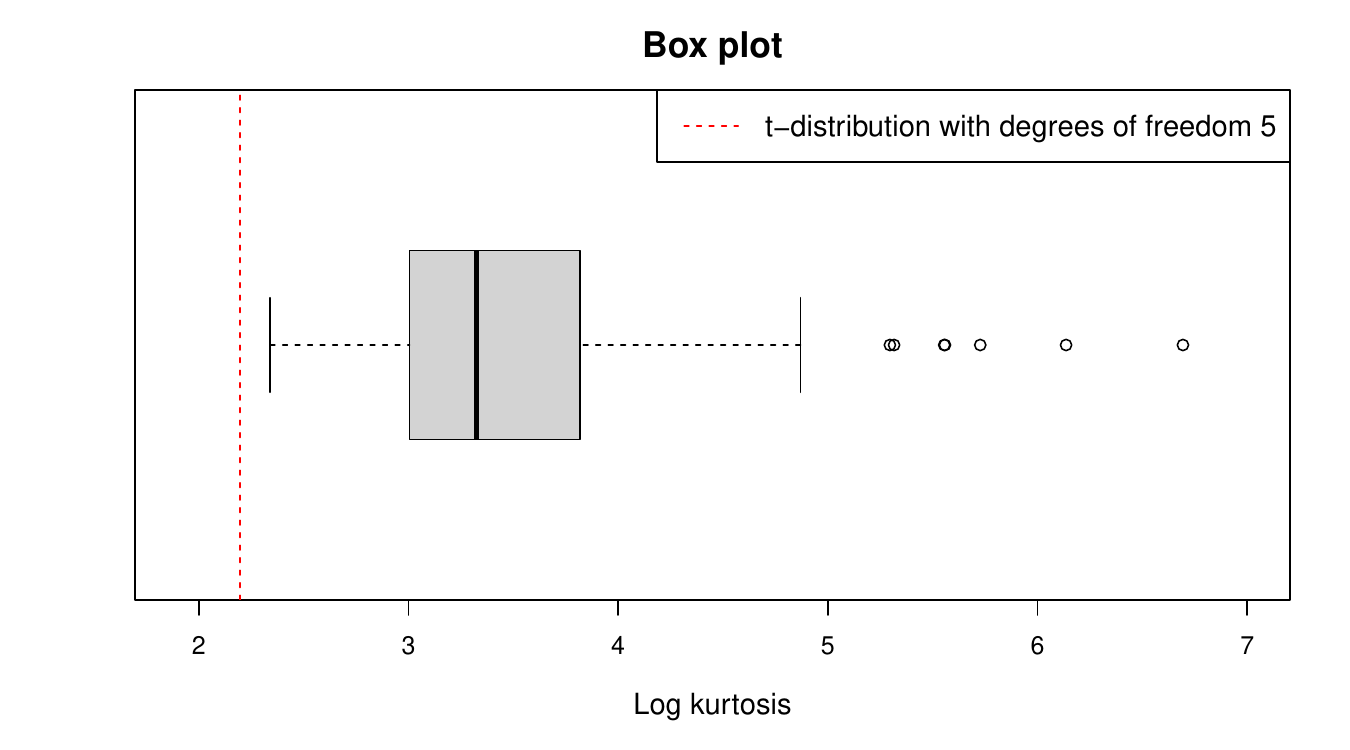}
	\caption{The boxplot of the 200 log kurtoses obtained from the daily jump adjusted pre-averaging realized volatility estimators for 997 trading days in the period 2016--2019.
	 The daily jump adjusted pre-averaging realized volatility estimators are estimated using 1-min log-returns of the most liquid 200 assets in the S\&P 500 index. 
	 The red dash represents the kurtosis of the $t_5$-distribution.}\label{Fig-2}
\end{figure}

However, several empirical studies indicate that idiosyncratic volatility also has a dynamic structure, and it comprises a large proportion of the total volatility  \citep{barigozzi2016generalized, connor2006common, herskovic2016common}.
To provide evidence of the existence of the dynamics in the idiosyncratic process for the high-frequency financial returns, we estimated the 200 daily eigenvalues of the idiosyncratic volatility matrix based on the top 200 large trading volume stocks in the S\&P 500 index.
The estimation procedure will be described in Section \ref{SEC-4.2} and Section \ref{SEC-5.2}.
Figure \ref{Fig-1} depicts the distribution of the first-order auto-correlations of the 200 time series of 200 daily estimated eigenvalues as well as the ACF plots for the time series of daily eigenvalue estimates of the 1st, 50th, 150th, and 200th eigenvalues.
We note that other eigenvalues also have similar time series structures.
Figure \ref{Fig-1} shows that the lag-1 autocorrelations are quite strong, which supports a dynamic structure in the eigenvalue processes of the idiosyncratic volatility.  
In addition, these estimated eigenvalues exhibit fairly long memories, with significant autocorrelation of lags of about 1 to 4 weeks.
Thus, simultaneously modeling the idiosyncratic volatility as well as the factor volatility is important to capture volatility dynamics.
On the other hand, since the dimension of the idiosyncratic volatility is large, modeling the factor and idiosyncratic volatilities simultaneously results in the problem of over-parameterization.
To address this issue, the sparsity of model parameters is often imposed,  and high-dimensional estimation procedures, such as LASSO \citep{tibshirani1996regression} and SCAD \citep{fan2001variable}, which are usually developed under a sub-Gaussian tail condition, are employed.  
However, this sub-Gaussian assumption is at odds with the empirical observations that the financial market exhibits heavy tails.
For example, Figure \ref{Fig-2} shows the boxplot of the 200 log kurtoses for the daily jump adjusted pre-averaging realized volatility estimators \citep{ait2016increased,christensen2010pre, jacod2009microstructure} for 997 trading days in the period 2016--2019.
	 The daily jump adjusted pre-averaging realized volatility estimators are estimated using 1-min log-returns of the most liquid 200 assets in the S\&P 500 index. 
The detailed estimation procedure is presented in \eqref{eq-A.1} in the Appendix.
From Figure \ref{Fig-2}, we can see that the volatility processes have heavy-tailed distributions.
See also \citet{cont2001empirical, fan2018robust, mao2018stochastic, shin2023adaptive}.
Thus, the high-dimensional estimation procedure developed under the sub-Gaussian tail condition is inappropriate.
These stylized features lead to the demands for developing a diffusion process for both factor and idiosyncratic volatilities with heavy-tailed observations.

In this paper, we introduce a novel process to account for the dynamics in the factor and idiosyncratic volatilities based on the VAR model with heavy-tailed innovations.
Specifically, it is assumed that the eigenvectors of the latent factor and idiosyncratic volatility matrices do not vary over a time period.
In contrast, we allow the eigenvalues to evolve with time and impose a parametric dynamic structure.
%In particular, the instantaneous eigenvalue processes of the latent factor and idiosyncratic instantaneous volatility matrices are continuous with respect to time and have a VAR structure at integer time points so that it is a form of some interpolation of the VAR structure.
%This structure makes the integrated eigenvalues have the VAR structure, and thus, the dynamics of the volatility can be explained by both the factor and idiosyncratic components.
In particular, the daily integrated eigenvalues of the factor and idiosyncratic volatility matrices have the VAR structure, and thus, the dynamics of the volatility can be explained by both the factor and idiosyncratic components.
We call it the factor and idiosyncratic VAR (FIVAR) model.
When it comes to estimating model parameters,  the high dimensionality of the idiosyncratic volatility matrix causes over-parameterization.
Furthermore, we allow the heavy-tailedness based on the bounded $c_\epsilon$-th moment condition for $c_{\epsilon} >4$.
It is assumed that the model parameters are sparse so that an $\ell_{1}$-penalty, such as LASSO, can be employed.
The usual $\ell_1$-penalty method does not work under the heavy-tailedness \citep{sun2020adaptive}, and a Huber loss is employed to address this issue \citep{huber1964robust}.
We show that the proposed estimation procedure has robustness with the desirable convergence rate.
We also propose a procedure for large volatility matrix prediction and investigate its asymptotic properties.

The rest of the paper is organized as follows.
Section \ref{SEC-2}  introduces the FIVAR model, based on the high-dimensional factor-based It\^{o} diffusion process, and investigates its properties.
Section \ref{SEC-3} proposes the robust parameter estimation method for a high-dimensional VAR model with heavy-tailedness and establishes its concentration properties.
In Section \ref{SEC-4}, we apply the proposed estimator to large volatility matrix prediction.
In Section \ref{SEC-5}, we conduct a simulation study to check the finite sample performance of the proposed estimator and apply the estimation method to high-frequency trading data.
The conclusion is presented in Section \ref{SEC-6}, and the technical proofs and miscellaneous materials are presented in the  Appendix.

Before closing this section, let us introduce some notations.
For a given  $p_1 \times p_2$ matrix $\bM = \left(M_{ij}\right)$, let
\begin{equation*}
	\|\bM\|_1 = \max\limits_{1 \leq j \leq p_2}\sum\limits_{i = 1}^{p_1}|M_{ij}|,\hspace{0.5cm} \|\bM\|_\infty = \max\limits_{1 \leq i \leq p_1}\sum\limits_{j = 1}^{p_2}|M_{ij}| , \hspace{0.5cm}
	\| \bM \| _{\max} = \max_{i,j} | M_{ij}|.
\end{equation*}
Note that $\|\cdot\|_{\max}$ is not a matrix norm in general, but can be interpreted as a vector norm.
The matrix spectral norm $\|\bM\|_2$ is the square root of the largest eigenvalue of $\bM\bM^\top$  and the Frobenius norm of $\bM$ is denoted by $\|\bM\|_F = \sqrt{ \mathrm{tr}(\bM^{\top} \bM) }$.
When $\bM$ is a square matrix, the spectral radius $\rho(\bM)$ is the largest value of the absolute eigenvalues of $\bM$.
For any vector $\bx=(x_1, \ldots, x_{p})^\top\in \mathbb{R}^{p}$ and $q \geq 1$, the $\ell_q$ norm $\left\| \bx\right\| _{q}=\left( \sum ^{p}_{i=1}\left| x_{i}\right| ^{q}\right) ^{1/q}$.
For any vectors $\bx, \by \in \mathbb{R}^{p}$, we set $\langle \bx, \by \rangle = \bx^{\top}\by$.
For a function $f:\mathbb{R}^{p} \rightarrow \mathbb{R}$, its gradient vector is denoted by $\nabla f \in \mathbb{R}^{p}$ as long as it exists.
We denote $\left\| Z \right\|_{L_q}=\left\{ \E \left(\left| Z\right|^{q}\right)\right\} ^{1/q}$ for a random variable $Z \in \mathbb{R}$ and $q \geq 1$.
The half-vectorization, $\vech \(\bM\)$, of the matrix $\bM$ is the column vector obtained by vectorizing only the lower triangular part of $\bM$. 
Also, $\tr\(\bM\)$ is the trace of $\bM$ and $\det\(\bM\)$ is the determinant of  $\bM$.
$\Diag \(\bM\)$ denotes the square diagonal matrix with the elements of the main diagonal of $\bM$.
$C$'s denote generic positive constants whose values are free of other parameters and may change from appearance to appearance.

\section{FIVAR model} \label{SEC-2}

Let $\bX(t)= (X_1 (t), \ldots, X_p(t))^\top$ be the vector of true log-prices of $p$ assets at time $t$.
To account for the cross-sectional dependence in financial asset prices, we employ the following factor-based jump-diffusion model:
\begin{equation}\label{eq-2.1}
	d\bX(t) = \bmu(t)dt+\bB(t)d\bff(t)+d\bu(t)+\bJ(t)d\bLambda(t),
\end{equation}
where $\bmu(t)\in \mathbb{R}^p$ is a drift vector, $\bB(t)\in \mathbb{R}^{p \times r}$ is an unknown factor loading matrix, $\bff(t)\in \mathbb{R}^r$ is a latent factor process, and $\bu(t) \in \mathbb{R}^{p}$ is an idiosyncratic process.
For the jump part, $\bJ(t) = (J_1(t), \ldots, J_p(t))^\top$ is a jump size vector, and $\bLambda(t) = (\Lambda_1(t), \ldots, \Lambda_p(t))^\top$ is a $p$-dimensional Poisson process with an intensity $\bI(t) = (I_1(t), \ldots, I_p(t))^\top$.
It is  assumed that the factor and idiosyncratic processes $\bff(t)$ and $\bu(t)$ follow the continuous-time diffusion models:
\begin{equation*}
	d\bff(t) = \bvartheta ^\top(t)d\bW(t) \quad \text{and} \quad d\bu(t) = \bPhi ^\top(t)d \bW^{\ast}(t),
\end{equation*}
where  $\bvartheta(t)$ and  $\bPhi(t)$ are  $r$ by $r$ and $p$ by $p$ instantaneous volatility matrices, respectively, and $\bW(t)$ and $\bW^{\ast}(t)$ are $r$-dimensional and $p$-dimensional independent Brownian motions, respectively.
Stochastic processes $\bX(t)$, $\bmu(t)$, $\bB(t)$, $\bff(t)$, $\bu(t)$, $\bvartheta(t)$, and $\bPhi(t)$ are defined on a filtered probability space $(\Omega, \FF, \{\FF_t, t \in [0, \infty)\},  P)$ with filtration $\FF_t$ satisfying the usual conditions, that is, the filtered probability space $(\Omega, \FF, \{\FF_t, t \in [0, \infty)\},  P)$ is complete and the filtration  $\FF_t$ is right-continuous. 
The instantaneous volatility matrix of the log-price $\bX(t)$ is
\begin{equation}\label{eq-2.2}
	\bgamma(t) = \left(\gamma_{ij}(t)\right)_{1\leq i,j\leq p} = \bB(t)\bvartheta^\top(t)\bvartheta(t)\bB^\top(t) + \bPhi^\top(t)\bPhi(t).
\end{equation}
We assume that $\bgamma(t)$ is continuous.
The integrated volatility for the $d$th day is
\begin{equation*}
	\bGamma_d = \left( \Gamma_{d,ij}\right)_{i,j=1,\ldots ,p}= \int_{d-1}^{d}\bgamma(t) dt =  \mathbf{\Psi}_{d} + \mathbf{\Sigma}_{d},
\end{equation*}
where $\mathbf{\Psi}_{d} = \int_{d-1}^{d} \bB(t)\bvartheta^\top(t)\bvartheta(t)\bB^\top(t)dt$ and $\mathbf{\Sigma}_{d} = \int_{d-1}^{d} \bPhi^\top(t)\bPhi(t)dt$ for $d\in \mathbb{N}$.

Let $\mathbf{q}_{t,1}^F,\ldots ,\mathbf{q}_{t,r}^F$ and  $\lambda _{t,1}\left( \btheta _{1}\right),\ldots ,\lambda _{t,r}\left( \btheta _{r}\right)$ be the eigenvectors and eigenvalues of the instantaneous factor volatility matrix $\bB(t)\bvartheta^\top(t)\bvartheta(t)\bB^\top(t)/p$, respectively, and $\mathbf{q}_{t,1}^I,\ldots ,\mathbf{q}_{t,p}^I$ and $\lambda _{t,r+1}\left( \btheta _{r+1}\right),$ \\ $\ldots ,\lambda _{t,p+r}\left( \btheta _{p+r}\right)$ be the eigenvectors and eigenvalues of the instantaneous idiosyncratic volatility matrix $\bPhi^\top(t)\bPhi(t)$, respectively.
We note that in the high-dimensional factor model, the eigenvalues of the factor volatility matrix are usually assumed to diverge at the order of the dimension $p$.
Therefore, to match the sizes of the factor and idiosyncratic parts, we divide the factor part by $p$.
In this paper, to distinguish notations for the factor and idiosyncratic parts, we use subscript and superscript of  $F$ and $I$ to their associated quantities, respectively.
In the latent factor model, to identify the latent factor loading matrix and factors, it is often assumed that the latent factor loading matrix is orthonormal and the latent factors have a diagonal covariance matrix, which implies that the eigenvectors and eigenvalues are related to the factor loading matrix and factors, respectively \citep{ait2017using,  fan2013large, kim2019factor}.
In this paper, we also consider the eigenvalues as the latent factor and idiosyncratic associated variables.
{It is assumed that the eigenvectors, $\mathbf{q}_{t,i}^F$, $i=1,\ldots , r$, are constant over time, that is,  $\mathbf{q}_{t,i}^F=\mathbf{q}_{i}^F$ for $i=1,\ldots , r$ and  $t \in [0, \infty)$.
	Also, it is assumed that $\mathbf{q}_{t,i}^I=\mathbf{q}_{i}^I$ for $i=1,\ldots , p$ and  $t \in [0, \infty)$.
We note that the constant assumption can be relaxed to the constant eigenvectors for each day.
However, \citet{kim2019factor} shows that the estimation procedures with time-invariant eigenvectors perform better.}
In light of this, it is assumed that the eigenvectors are constant over time, and hence, the volatility dynamics are driven by those of the eigenvalues.
Thus, to capture the daily volatility dynamics, we model the daily integrated eigenvalues of the factor and idiosyncratic volatilities by the following factor and idiosyncratic VAR (FIVAR) model.

\begin{mydef} \label{def-1}
	We call a log-price vector $\bX(t)$, $t\in [0,\infty)$, to follow a FIVAR($h$) model if its associated values satisfy the following iterative relations:
	\begin{equation}\label{eq-2.3}
		\bxi_{d} = \bnu + \sum ^{h}_{k=1}\bA_{k}\bxi_{d-k}     +\bepsilon_{d} \text{ a.s.},
	\end{equation}
	where $\bxi_{d} = \left(\xi_{d,1}, \ldots , \xi_{d,p+r}\right)^{\top}=\int ^{d}_{d-1}\blambda _{t}\left( \btheta\right)dt$,  $\blambda _{t}\left( \btheta\right)=\left(\lambda _{t,1}\left( \btheta_{1}\right), \ldots,  \blambda _{t,p+r}\left( \btheta_{p+r}\right)\right)^{\top}$,  $\bnu=(\nu_1, \ldots,$ $\nu_{p+r})^{\top}$, $\bA_{k} = (A_{k,i,j})_{1\leq i,j \leq p+r}$ for all $1\leq k \leq h$, and $\bepsilon_{d}=\left(\epsilon_{d,1},\ldots ,\epsilon_{d,p+r}\right) ^{\top}$ is i.i.d. innovation at time $d$ with $\E(\bepsilon_{d})=\mathbf{0}_{p+r}$, which is independent of $\bxi_{d-l}$ for all $l \in \mathbb{N}$.
\end{mydef}

Definition \ref{def-1} indicates that the daily integrated eigenvalues $\bxi_{d}$ follow the VAR model under the FIVAR model.
In Appendix \ref{existence}, we show the existence of the continuous eigenvalue process that satisfies the VAR model structure of the integrated eigenvalue process.
In the simulation study, we use the continuous eigenvalue diffusion process to generate simulated data.
%We note that  $\bnu$ is the the vector of the intercepts, $\left(a_{1}, \ldots , a_{p+r}\right)^{\top}$ of instantaneous eigenvalues in \eqref{eq-2.3} and the expected value, $\left(\E\left[z^2_{1,t}\right], \ldots , \E\left[z^2_{p+r,t}\right]\right)^{\top}$, of the random fluctuations, and  the coefficient, $\bzeta_{1}$, of the first AR order in \eqref{eq-2.3} coming from handling the iterative relationship of the integrated eigenvalue. 
%The coefficients $\bA_{k}$'s are determined by the coefficients,  $\bzeta_{k}$'s, of the instantaneous eigenvalue in \eqref{eq-2.3}.
%It is worth noting that we cannot specify all interpolation parameters in the instantaneous process because some quantities, such as   $\left(a_{1}, \ldots , a_{p+r}\right)^{\top}$  and the expected value of the random fluctuations, are not uniquely defined.
Unlike the factor GARCH-It\^o model \citep{kim2019factor}, the FIVAR model considers not only the factor component but also the idiosyncratic component.
In the empirical study, we find that the idiosyncratic eigenvalues have a time series structure, and incorporating the idiosyncratic dynamics helps capture the volatility dynamics.
Details can be found in Section \ref{SEC-5.2}.
We note that the proposed model is not the unique way to explain the observed auto-correlation structure in the empirical study (see \citet{bollerslev2016exploiting, cipollini2021realized, hansen2014estimating}). 
That is, the FIVAR model is one of the possible solutions, and we find its empirical benefits. 
However, incorporating the idiosyncratic component causes high-dimensionality.
Furthermore, to account for the heavy-tailedness, we allow that the martingale noise $\bepsilon_{d}$ has heavy tails.
That is, when it comes to statistical inferences for the proposed FIVAR model, we face two problems: the heavy-tailedness and over-parameterization.
In the following section, we propose an estimation procedure that can handle the heavy-tailedness and high dimensionality.

\begin{remark}
In this paper, we assume that the rank $r$ is constant over time.
However, it may be more realistic to allow the rank $r$ to vary over time.
To handle the time-varying rank $r$, we can consider a state heterogeneous structure of the volatility process as in \citet{chun2022state}.
For example, we can assume that the number of common factors is the same under the same state.   
Then, we need to extend the one-dimensional case in \citet{chun2022state} to the high-dimensional case. 
However, the extension to the high-dimensional case is not straightforward.
Thus, we leave this for a future study.
\end{remark}

\section{Estimation procedure for the heavy-tailed VAR model} \label{SEC-3}

In this section, we propose a robust parameter estimation method for the high-dimensional VAR model in \eqref{eq-2.3}.
%\begin{equation*}
%    \bxi_{d} = \bnu + \sum ^{h}_{k=1}\bA_{k}\bxi_{d-k} +\bepsilon_{d} \text{ a.s.},
%\end{equation*}
Our idea is basically to robustly fit this model for each component.
Let  $\bbeta =\begin{pmatrix}\bnu & \bA_{1} & \cdots & \bA_{h} \end{pmatrix}$ and  we denote $\bbeta_{i}$ by the $i$th row of $\bbeta$.
To overcome the curse of dimensionality, the sparsity of $\bbeta_{i}$ is assumed:  the number of nonzero elements in each $\bbeta_{i}$ is bounded by a small number $s_{\beta} \geq 1$.
In contrast, for the factor-related parameter, the factor model usually assumes that the idiosyncratic variables do not affect the factor variable.
To reflect this prior,  $A_{k,i,j}=0$ for $k=1, \ldots, h$, $i=1, \ldots, r$, and $j=r+1, \ldots, p+r$ is assumed.
That is, the factor-related coefficients $\bbeta_{i}$'s, $i=1,\ldots,r$, have the specific sparse structure.
We denote the true model parameter by $\bbeta_{0}$ and its $i$th row by $\bbeta_{i0}$.
It is worth mentioning that the sparsity implies the Granger non-causality between the related variables.
In practice, we do not know the number, $r$, of latent factors and AR lag $h$. 
In this section, it is assumed that $r$ and $h$ are given, and we will discuss how to choose them in Section \ref{SEC-5.1}.

To accommodate the sparsity structure, we often employ the penalized regression model, such as  LASSO \citep{tibshirani1996regression} and SCAD \citep{fan2001variable}.
When analyzing data with the LASSO procedure, we need some sub-Gaussian tail conditions.
However, as shown in Figure \ref{Fig-2}, the volatilities often exhibit heavy tails in financial applications.
To tackle this heavy-tailedness, we often employ a robustification method \citep{catoni2012challenging, fan2017estimation, minsker2018sub, sun2020adaptive}.
In this paper, we employ the Huber loss $l_{\tau}$ \citep{huber1964robust}
\begin{equation*}
	l_{\tau}\left( x\right) = x^{2}/2 I(\left| x\right| \leq \tau) + (\tau\left| x\right| -\tau^{2}/2) I(\left| x\right| > \tau), 
	%\begin{cases}x^{2}/2 & \text{ if } \left| x\right| \leq \tau \\ \tau\left| x\right| -\tau^{2}/2 & \text{ if } \left| x\right| > \tau, \end{cases}
\end{equation*}
where $\tau>0$ is the robustification parameter, and  the truncation (Winsorization) method
\begin{equation*}
	\psi_{\varpi}\left( x\right) =  x I( \left| x\right| \leq \varpi ) +  \mbox{sign}(x) \varpi I( \left| x\right| > \varpi),
	%\begin{cases}x & \text{ if } \left| x\right| \leq \varpi \\  \mbox{sign}(x)\varpi & \text{ if } \left| x\right| > \varpi, \end{cases}
\end{equation*}
where $\varpi > 0$ is a truncation parameter.
We denote $\psi_{\varpi}\left(\bx \right) =\left(\psi_{\varpi}\left( x_1\right), \ldots, \psi_{\varpi}\left( x_{p_1}\right)\right)^{\top}$ for any vector $\bx=\left(x_1, \ldots, x_{p_1}\right)^{\top}\in \mathbb{R}^{p_{1}}$.

By combining the truncation and $\ell_{1}$-regularization methods, we can simultaneously deal with robustness and the curse of dimensionality.
Specifically, we estimate the true sparse coefficient $\bbeta_{i0}$ as follows:
\begin{equation}\label{eq-3.1}
	\hat{\bbeta}_{i} = \arg \min_{\bbeta_i \in \mathbb{R}^{h(p+r)+1}}{\mathcal{L}}^{I,i}_{\tau,\varpi}(\bbeta_i) + \eta_I \left\| \bbeta_i \right\| _{1} \quad \text{ for } i=r+1, \ldots, p+r,
\end{equation}
where $\eta_I>0$ is the regularization parameter, the empirical loss function is
\begin{equation}\label{eq-3.2}
	{\mathcal{L}}^{I,i}_{\tau,\varpi}(\bbeta_i)=(n-h)^{-1}\sum ^{n}_{d=h+1}l_{\tau_I}\left( \hat{\xi}_{d,i}-\langle \psi_{\varpi_I}(\hat{\bxi}_{d-1}^{I}), \bbeta_i\rangle \right),
\end{equation}
{$n$ is the number of days in the sample}, $\hat{\bxi}_{d}^{I} = \left(1, \hat{\bxi}^{\top}_{d}, \ldots, \hat{\bxi}^{\top}_{d-h+1}\right)^{\top}$, and $\hat{\bxi}_{d}=\left(\hat{\xi}_{d,1}, \ldots, \hat{\xi}_{d,p+r}\right)^{\top}$ is a non-parametric estimator for ${\bxi}_{d}$.  
Note that in \eqref{eq-3.2}, the Huber loss $l_{\tau_I}$ is used to handle the heavy-tailedness of  $\bepsilon_{d}$ and the truncation function $\psi_{\varpi}$ is used to guard against the tail of $\bxi_{d}$.
In contrast, since the sparsity structure of the coefficients for the factor part is known, it is a low-dimensional problem.  
We do not need the $\ell_1$ penalty term.
However, we still need the truncation parts to handle the heavy-tailedness as follows:
\begin{equation}\label{eq-3.3}
	\hat{\bbeta}_{i} = \arg \min_{\bbeta_i \in \mathbb{R}^{hr+1}}{\mathcal{L}}^{F,i}_{\tau,\varpi}(\bbeta_i) \quad  \text{ for } i=1, \ldots, r,
\end{equation}
where
\begin{equation}\label{eq-3.4}
	{\mathcal{L}}^{F,i}_{\tau,\varpi}(\bbeta_i)=(n-h)^{-1}\sum ^{n}_{d=h+1}l_{\tau_{F}}\left( \hat{\xi}_{d,i}-\langle  \psi_{\varpi_{F}}(\hat{\bxi}_{d-1}^{F}), \bbeta_i \rangle \right),
\end{equation}
$\hat{\bxi}_{d}^{F}$ is an $(hr+1)$ by $1$ vector obtained by stacking $1$ and the first $r$ elements of each $\hat{\bxi}_{d-k}$, $k=0, \ldots, h-1$.
We note that, in financial practice, we cannot observe the true price or volatility process, so we employ the non-parametric estimator $\hat{\bxi}_{d}$ of $\bxi_{d}$.
We discuss the non-parametric estimators in Section \ref{SEC-4}.

%\begin{remark}
%We can alternatively use the following estimator for the whole true parameter matrix $\bbeta_{0}$:
%\begin{equation}\label{eq-3.5}
%	 \(\hat{\bbeta}_{1}, \ldots, \hat{\bbeta}_{p+r}\) = \arg \min_{\bbeta_{1}, \ldots, \bbeta_{p+r}}\sum^{r}_{i=1} {\mathcal{L}}^{F,i}_{\tau,\varpi}(\bbeta_i)+\sum^{p+r}_{i=r+1} \left({\mathcal{L}}^{I,i}_{\tau,\varpi}(\bbeta_i) + \eta_I \left\| \bbeta_i \right\| _{1}\right).
%\end{equation}
%Since the optimization for each $\hat{\bbeta}_{i}$ is separable,
%the solution of \eqref{eq-3.5} is the same as that of \eqref{eq-3.1} and \eqref{eq-3.3}.
%However, the proposed estimators facilitate parallel computation, which saves computation time. Thus, we use lower-dimensional subproblems \eqref{eq-3.1} and \eqref{eq-3.3}.
%\end{remark}

We investigate the theoretical properties of $\hat{\bbeta}$  under the following assumptions.

\begin{assumption} \label{Assumption1}
	~
	\begin{itemize}
		\item[(a)]
		The process $(\tilde{\bxi}_{d})_{d=1,2,\ldots}$ is strictly stationary and the spectral radius of $\tilde{\bA}$, $\rho(\tilde{\bA})$, is less than 1, where $\tilde{\bxi}_{d}$ and $\tilde{\bA}$ are the vectorization of $\bxi_d$ and its corresponding coefficient matrix defined in \eqref{eq-A.3} in the Appendix, respectively.
		
		\item[(b)] The number of nonzero elements in each $\bbeta_{i0}$ is bounded by a  number $s_{\beta}$.
		
		\item[(c)] $\epsilon_{d,i}$ and $\xi_{d,i}$ satisfy  $  \max_{1\leq i \leq p+r}   \E (|\epsilon_{d,i}|^{c_{\epsilon}})<\infty$ and $\max_{1\leq i \leq p+r}   \E (|\xi_{d,i}|^{c_{\epsilon}})<\infty$ for some constant $c_{\epsilon}>4$.
		
		\item[(d)] The process $(\tilde{\bxi}_{d})_{d=1,2,\ldots}$ is $\alpha$-mixing  and the $\alpha$-mixing coefficients satisfy  $\alpha(k) = O\left(\varphi^k \right)$  for some $\varphi \in (0,1).$
		
		\item[(e)] The non-parametric estimator $\hat{\bxi}_{d}$ satisfies
		\begin{equation*}
			\max_{1\leq d \leq n}\max_{1\leq i \leq r}\left| \hat{\xi}_{d,i}-\xi_{d,i} \right|\leq b^{F}_{m,n,p} \quad \text{and} \quad
			\max_{1\leq d \leq n}\max_{r+1\leq i \leq p+r}\left| \hat{\xi}_{d,i}-\xi_{d,i} \right|\leq b^{I}_{m,n,p} \text{ a.s.},
		\end{equation*}
		where $m$ represents the number of observations for estimating $\xi_{d,i}$,  and  $b_{m,n,p}^{F}$ and $b_{m,n,p}^{I}$ converge to zero as $m$, $n$, and $p$ increase.
		
		\item[(f)] There exists a constant $\kappa>0$ such that the following inequality holds for some $D_F\geq \(hr+1\)\eta_F / \kappa$ and   $1 \leq i \leq r$, where the bound of  $\eta_F$ is given in Theorem \ref{Theorem1}:
		\begin{equation*}
			\inf\{ \bw^{\top}\nabla ^{2}{\mathcal{L}}^{F,i}_{\tau,\varpi}(\bbeta_i) \bw: \text{ } \left\| \bw\right\| _{2}= 1,\text{ } \left\| \bbeta_i -\bbeta_{i0}\right\| _{1}\leq D_F
			\} \geq \kappa.
		\end{equation*}
		
		\item[(g)] Define the $\ell_{1}$-cone $\mathcal W_{i} = \left\{ \bw \in \mathbb{R}^{hp+1}:\text{ }\left\| \bw_{S_{i}^{c}} \right\|_{1} \leq 3\left\| \bw_{S_{i}}\right\|_{1}\right\}$, where $\bw_{S_{i}^{c}}$ is the subvector obtained by stacking $\left\{\bw_{j}: \text{ } j \in S_{i}^{c} \right\}$, $\bw_{S_{i}}$ is the subvector obtained by stacking $\left\{\bw_{j}: \text{ } j \in S_{i} \right\}$, and $S_{i} = \left\{j: \text{ j-th element of } \bbeta_{i0} \neq 0 \right\}$.
		Then, there exists a constant $\kappa>0$ such that the following inequality holds for some $D_I \geq 48s_{\beta}\eta_I / \kappa$ and $1 \leq i \leq p$, where the bound of $\eta_I$ is given in Theorem \ref{Theorem1}:
		\begin{equation*}
			\inf\{ \bw^{\top}\nabla ^{2}{\mathcal{L}}^{I,i}_{\tau,\varpi}(\bbeta_i) \bw: \text{ } \bw \in \mathcal W_{i}, \text{ } \left\| \bw\right\| _{2}= 1,\text{ } \left\| \bbeta_i -\bbeta_{i0}\right\| _{1}\leq D_I
			\} \geq \kappa.
		\end{equation*}
	\end{itemize}
\end{assumption}

\begin{remark}
	Assumption \ref{Assumption1}(a) is the strictly stationary and stable conditions for the VAR(1) representation of the model \eqref{eq-2.3}.
	Assumption \ref{Assumption1}(c) allows the heavy-tailedness in the VAR model.
	Since we consider the high-dimensional VAR model, we need the moment condition for $\xi_{d,i}$, such as  $\max_{1\leq i \leq p+r}  \E (|\xi_{d,i}|^{c_{\epsilon}})<\infty$.
	However, under Assumption \ref{Assumption1}(a)--(b), the condition $\max_{1\leq i \leq p+r}  \E (|\epsilon_{d,i}|^{c_{\epsilon}})$ $<\infty$ implies the condition $\max_{1\leq i \leq p+r} \E (|\xi_{d,i}|^{c_{\epsilon}}) <\infty$ when $s_{\beta}$ is bounded by some positive constant (see Lemma \ref{Lemma1} in the Appendix).
	We note that we do not impose the bounded $s_{\beta}$ throughout the paper; thus, we need the moment condition for $\xi_{d,i}$. 
	Assumption \ref{Assumption1}(d) is required to handle the dependency in the VAR model.
	Under Assumption \ref{Assumption1}(a), Assumption \ref{Assumption1}(d) holds if the process $(\tilde{\bxi}_{d})_{d=1,2,\ldots}$ is geometric ergodic (see Proposition 2 in \citet{liebscher2005towards} and Fact 5 in the online Appendix of \citet{wong2020lasso}). 
	We note that the geometric ergodicity can be obtained under the mild condition on $\bepsilon_{d}$ (see Example 3 in \citet{wong2020lasso}).
	Assumption \ref{Assumption1}(e) represents the concentration property of the non-parametric estimator $\hat{\bxi}_{d}$.  
	In Section \ref{SEC-4}, we propose a method for constructing $\hat{\bxi}_{d}$ and show its associated inequality holds with high probability.
	Assumptions \ref{Assumption1}(f)--(g) are the eigenvalue conditions for the Hessian matrices $\nabla ^{2}{\mathcal{L}}^{F,i}_{\tau,\varpi}(\bbeta_i)$ and $\nabla ^{2}{\mathcal{L}}^{I,i}_{\tau,\varpi}(\bbeta_i)$, respectively.
	This is called the localized restricted eigenvalue ($LRE$) condition \citep{fan2018lamm, sun2020adaptive}, which implies strictly positive restricted eigenvalues over a local neighborhood.
\end{remark}

The following theorem provides the convergence rate of $\hat{\bbeta}_i$ defined in \eqref{eq-3.1} and \eqref{eq-3.3}.

\begin{thm}\label{Theorem1}
	Under the model \eqref{eq-2.3}, Assumption \ref{Assumption1}, $n \geq 3$, $\delta \geq 1$, $\sqrt{n \delta} +\left(\tau_F+\varpi_F\right)\left(\log n \right)^{2}\delta = O \left(n\right)$, and $\eta_F \geq C [b^{F}_{m,n,p} + \tau_{F}^{-2}+\varpi_{F}^{-2} + \frac{\tau_F\varpi_F\left(\log n \right)^2 \delta + \sqrt{n \delta}}{n} ]$,  we have,  for $i= 1, \ldots, r$,  with probability at least $1-4hre^{-\delta}$,
	\begin{equation}\label{eq-3.5}
		\left\| \hat{\bbeta}_{i} -\bbeta_{i0}\right\|_{2}  \leq \frac{\(hr+1\)^{1/2}\eta_F}{\kappa}.
	\end{equation}
	Furthermore, we assume that $s_{\beta}\sqrt{n \delta} + \left(\tau_I+\varpi_I\right)\left(\log n \right)^{2}\delta = O \left(n\right)$ and $\eta_I \geq C [s_{\beta}\(b_{m,n,p}^{F}+ b_{m,n,p}^{I}\) + s_{\beta}^{3}\tau_I^{-2}+s_{\beta}\varpi_I^{-2} + \frac{\tau_I\varpi_I\left(\log n \right)^2 \delta + s_{\beta}\sqrt{n \delta}}{n}  ].$
	Then, we have, for $i= r+1, \ldots, p+r$, with probability at least $1-4h\(p+r\)e^{-\delta}$,
	\begin{equation}\label{eq-3.6}
		\left\| \hat{\bbeta}_{i} -\bbeta_{i0}\right\|_{2}  \leq \frac{12s_{\beta}^{1/2}\eta_I}{\kappa}.
	\end{equation}
\end{thm}

\begin{remark}
	Theorem \ref{Theorem1} shows the convergence rates for the general setting of the low-dimensional and high-dimensional VAR models, where the covariates are not observable and observations are  heavy-tailed.
	Specifically,  $b_{m,n,p}^{F}$ and $b_{m,n,p}^{I}$ in $\eta_F $ and $\eta_I $ are the costs to estimate the true covariates.
	When $\bxi_{d}$ is directly observable, $b_{m,n,p}^{F}$ and $b_{m,n,p}^{I}$ become zero.
	Take $\delta= 2\log p$, $\tau_F = \varpi_F= C\left(n / \log p\right)^{1/4}$, $\eta_F = C \left(\log n\right)^2\sqrt{ \log p / n}$, $\tau_I = Cs_{\beta}\left(n / \log p\right)^{1/4}$,  $\varpi_I = C \left(n / \log p\right)^{1/4}$, and $\eta_I = Cs_{\beta} \left(\log n\right)^2 \sqrt{\log p/ n}$.
	Then, $\hat{\bbeta}_{i}$ for the factor and idiosyncratic parts have a near-optimal convergence rate of $\left(\log n\right)^2\sqrt{\log p/ n}$ and $s_{\beta}^{3/2}\left(\log n\right)^2\sqrt{\log p/ n}$, respectively \citep{sun2020adaptive}.
	The additional  $\left(\log n\right)^2$  term  comes from handling the dependency in the process $(\bxi_{d})_{d=1,2,\ldots}$.
	When comparing to the optimal rate for the high-dimensional case, established in \citet{sun2020adaptive}, we have the additional $s_{\beta} \left(\log n\right)^2$.
	Usually, the sparsity level is small; thus,  the proposed method does not lose significant efficiency, even for the dependent and heavy-tailed case.
	%When the covariate vector does not have a dependency,
	%\citet{sun2020adaptive} showed that the optimal convergence rates are  $\sqrt{ \log p / n}$ and  $\sqrt{s_{\beta} \log p / n}$
	%for the low-dimensional and high-dimensional cases, respectively.
	%The difference term   $\left(\log n\right)^2$  comes from handling the dependency in the process $(\bxi_{d})_{d=1,2,\ldots}$, which is relatively negligible.
\end{remark}

\begin{remark}
In addition to obtaining concentration inequalities for the model parameter estimation method, obtaining confidence intervals for the model parameters is also important.
To do this, we need to adjust the bias of the proposed estimator and obtain a debiased estimator.
This bias is coming from Huber loss, truncation, high-dimensional observation error, and $\ell_1$ regularization.
It is a demanding task to simultaneously handle them.
We leave this issue for a future study.
\end{remark}

\section{Large volatility matrix prediction}\label{SEC-4}
\subsection{A model set-up}\label{SEC-4.1}

In this section, using the estimation procedure in Section \ref{SEC-3}, we discuss how to predict the large volatility matrix, based on the FIVAR model.
Given the observations of $n$ days, the parameter of interest is the conditional expected volatility matrix  $\E \left(\bGamma_{n+1} | \FF_{n} \right)$.
Recall that the integrated volatility matrix $\bGamma_d$ has the following low-rank plus sparse structure:
\begin{equation*}
	\bGamma_d= \mathbf{\Psi}_{d} + \mathbf{\Sigma}_{d} = p\sum^{r}_{i=1}\xi_{d,i}\bq_{i}^F \(\bq_{i}^{F}\)^{\top}+\sum^{p}_{i=1}\xi_{d,i+r}\bq_{i}^I \(\bq_{i}^I\)^{\top} \,\text{ a.s.},
\end{equation*}
{where  $p\xi_{d,i}$'s are the $i$-th largest eigenvalues of $\mathbf{\Psi}_{d}$ for $i=1, \ldots, r$ and $\xi_{d,i+r}$'s are the $i$-th largest eigenvalues of $\mathbf{\Sigma}_{d}$  for $i=1, \ldots, p$.}
It is assumed that the rank, $r$, of $\mathbf{\Psi}_{d}$ is bounded and the idiosyncratic volatility matrix $\mathbf{\Sigma}_{d}=\left( \Sigma_{d,ij}\right)_{i,j=1,\ldots ,p}$ satisfies the following sparse condition:
\begin{equation}\label{eq-4.1}
	\max_{ 1 \leq d \leq n}\max_{ 1 \leq i \leq p} \sum_{j=1}^p | \Sigma_{d,ij}| ^{\Upsilon}   (\Sigma_{d,ii} \Sigma_{d,jj} ) ^{(1-\Upsilon)/2}  \leq M_{I} s_{I} \, \text{ a.s.},
\end{equation}
where $M_I$ is a bounded positive random variable, $\Upsilon \in [0,1)$, and $s_{I}$  is a deterministic function of $p$, which grows slowly in $p$.
This low-rank plus sparse structure is widely employed when analyzing the large matrices \citep{ait2017using,  bai2002determining, fan2018robust, fan2013large, kim2018Large,  stock2002forecasting, shin2023adaptive}.
We note that when we directly use the total volatility without decomposition, we cannot explain the sparse structure of the eigenvectors of the idiosyncratic volatility matrices.
This may introduce numerous parameters and lead to higher complexity.

Unfortunately, the true log-price $\bX(t)$ cannot be directly observed since the high-frequency data are contaminated by microstructure noise.
To account for this, it is assumed that the observed log-price $Y_i(t_k)$ has the following additive noise structure:
\begin{equation}  \label{eq-4.2}
	Y_i (t_{d,k}) = X_i(t_{d,k}) + e_{i}(t_{d,k}) \quad  \text{ for } i = 1, \ldots, p, d = 1, \ldots, n , k = 0, \ldots, m ,
\end{equation}
where $d-1=t_{d,0}<\cdots<t_{d,m}=d$, and the microstructure noise $e_{i}(t_{d,k})$  is a stationary random variable with mean zero.
Empirical studies have shown that microstructure noise is serially dependent and endogenous \citep{ait2011ultra,hansen2006realized,jacod2017statistical,li2022remedi,ubukata2009estimation}.
Fortunately, as long as non-parametric integrated volatility matrix estimators satisfy \eqref{eq-4.3} {presented below}, the dependent structure of the microstructure noise does not affect the main results of this paper. 
There are several estimation procedures that are robust to dependent structures of the microstructure noise \citep{barndorff2011subsampling, jacod2017statistical, kim2016asymptotic, li2020robust}.
Similarly, the assumptions on the jumps do not affect the main results of this paper as long as \eqref{eq-4.3} holds.
There are also several estimation methods that can handle jumps when estimating integrated volatilities \citep{ait2016increased, shin2023adaptive}.
Thus, we only require condition  \eqref{eq-4.3}.
On the other hand, for simplicity, the observation time points are assumed to be synchronized and equally spaced: $t_{d,k}- t_{d,k-1} = m^{-1}$ for $d=1,\ldots, n$ and $k=1,\ldots, m$.

\begin{remark}
	{In this paper, we mainly focus on the parametric structure of the volatility process, so it is assumed that the observation time points are synchronized and equally spaced for simplicity.
		The conditions for the observation time points can be relaxed to the non-synchronized and unequally spaced conditions by using generalized sampling time \citep{ait2010high}, refresh time \citep{barndorff2011multivariate}, and previous tick \citep{andersen2003modeling, barndorff2011multivariate, zhang2011estimating} schemes.
		See also \citet{bibinger2014estimating, fan2019structured, park2016estimating}.}
\end{remark}

\subsection{Large volatility matrix prediction}\label{SEC-4.2}

To predict the large volatility matrix, we first employ a non-parametric integrated volatility matrix estimator $\hat{\bGamma}_d$, which is robust to jumps and dependent structures of the microstructure noise  \citep{ait2016increased, barndorff2011subsampling, bibinger2015econometrics, jacod2009microstructure, kim2016asymptotic, koike2016quadratic,  li2020robust, shin2023adaptive}.
Based on the non-parametric estimator $\hat{\bGamma}_d$, we estimate the eigenvectors and eigenvalues of factor and idiosyncratic volatility matrices as follows.
For estimating the `daily' integrated eigenvalues $\bxi_d$ on the factor volatility matrix $\mathbf{\Psi}_{d}$, based on the assumption of time-invariance of eigenvectors,  we calculate $r$  eigenvectors $\hat{\bq}^F_1, \ldots, \hat{\bq}^F_{r}$ of the average of the recent $\ell$ days' non-parametric integrated volatility matrix estimators, $\frac{1}{\ell}\sum_{d=n-\ell+1}^{n}\hat{\bGamma}_d$, where  $\ell$ is the window length for the eigenvector estimation.
Then, we obtain the estimators of time-dependent eigenvalues $\hat{\xi}_{d,i}=\(\hat{\bq}^F_{i}\)^{\top}\hat{\bGamma}_d \hat{\bq}^F_i/p$ for $d=1, \ldots, n$ and $i=1, \ldots, r$.   
This provides a part of inputs for \eqref{eq-3.2} and \eqref{eq-3.4}. 
We note that the eigenvectors are computed once based on the most recent $\ell$ days among the total $n$ observations, which provides better empirical performance than computing eigenvectors using a rolling-window scheme.
This may be because the above fixed-window approach provides a stable basis for constructing the daily eigenvalue sequence, which is more effective for predicting the target future volatility matrix.

To provide the rest of the inputs, namely $\hat{\xi}_{d,i}$ for $i > r$ for the idiosyncratic volatility matrix $\bSigma_d$, we apply the principal orthogonal complement thresholding (POET) method \citep{fan2013large}  as follows.
First, we decompose the input volatility matrix
$$\hat{\bGamma}_d= \sum_{k=1}^p \bar{\xi}_{d,k} \bar{ \bq}  _{d,k}  \bar{\bq}_{d,k} ^\top, $$
where $\bar{\xi}_{d,k}$ is the $k$-th largest eigenvalue of $\hat{\bGamma}_d$ and $\bar{\bq}_{d,k}$ is its corresponding eigenvector.
We then obtain the input idiosyncratic volatility matrix estimator  $\bar{\bSigma}_d =(\bar{\Sigma}_{d,ij})_{1\leq i,j\leq p} = \hat{\bGamma}_d - \sum_{k=1}^r \bar{\xi}_{d,k} \bar{\bq} _{d,k} \bar{\bq}_{d,k} ^\top$ and apply the adaptive thresholding method to $\bar{\bSigma}_d$  by computing
\begin{equation*}
	\hat{\Sigma}_{d,ij} =
	\begin{cases}
		\bar{\Sigma}_{d,ij} \vee 0   & \text{ if } i= j\\
		g_{ij} ( \bar{\Sigma}_{d,ij}) \1 ( |\bar{\Sigma}_{d,ij}| \geq \upsilon_{ij} )   & \text{ if } i \neq j
	\end{cases}
	\quad \text{ and } \quad \hat{\bSigma}_d =(\hat{\Sigma}_{d,ij})_{1\leq i,j\leq p},
\end{equation*}
where the thresholding function $g_{ij} (\cdot) $ satisfies $|g_{ij} (x) -x |\leq \upsilon_{ij}$, and the adaptive thresholding level $\upsilon_{ij} = \upsilon_{m}\, \sqrt{ ( \bar{\Sigma}_{d,ii} \vee 0 ) ( \bar{\Sigma}_{d,jj} \vee 0 )}$.
For example, we often utilize the soft thresholding function $g_{ij}(x)=x-\mbox{sign}(x) \upsilon_{ij}$ and the hard thresholding function $g_{ij}(x)=x$.
The thresholding parameter $\upsilon_{m}$ will be specified in Proposition \ref{Proposition1}.
With the idiosyncratic  volatility matrix estimator  $\hat{\bSigma}_d$, we calculate $p$  eigenvectors, $\hat{\bq}^I_{1}, \ldots, \hat{\bq}^I_{p}$, of $\frac{1}{\ell}\sum_{d=n-\ell+1}^{n}\hat{\bSigma}_d$ and obtain $\hat{\xi}_{d,i+r}=\(\hat{\bq}_{i}^{I}\)^{\top}\hat{\bSigma}_d \hat{\bq}_i^{I}$ for $d=1, \ldots, n$ and $i=1, \ldots, p$.
Again, we here use the time-invariant assumption of the eigenvectors.

With these inputs, we can estimate the true model parameter $\bbeta_{0}$ using the VAR model parameter estimation procedure in Section \ref{SEC-3} and calculate the predicted eigenvalue estimator by $\hat{\bxi}_{n+1}=\left(\hat{\xi}_{n+1,1},\ldots, \hat{\xi}_{n+1,p+r}\right)^{\top}=\hat{\bnu}+\sum_{k=1}^{h}\hat{\bA}_{k}\hat{\bxi}_{n+1-k}$ using \eqref{eq-2.3}.
Finally, we estimate the conditional expected volatility matrix by
\begin{equation*}
	\tilde{\bGamma}_{n+1}=\hat{\mathbf{\Psi}}_{n+1} + \hat{\mathbf{\Sigma}}_{n+1} = p\sum^{r}_{i=1}\hat{\xi}_{n+1,i}\hat{\bq}_{i}^{F} \(\hat{\bq}_{i}^{F}\)^{\top}+\sum^{p}_{i=1}\hat{\xi}_{n+1,i+r}\hat{\bq}_{i}^I \(\hat{\bq}_{i}^I\)^{\top}.
\end{equation*}
We describe the estimation procedure in Algorithm \ref{algorithm1} in the Appendix.

\begin{remark}
	To estimate the eigenvectors, the constant eigenvector over time is assumed, and the window length, $\ell$, for the eigenvector estimation can be from 1 to $n$. 
	%Theoretically, to predict the future volatility matrix, we only require the constant condition during the intraday, and we can allow the  time-varying eigenvector on a daily basis, that is, the eigenvector process has the form of a step function, as long as the eigenvectors are martingale. 
	%Specifically, for the time-varying eigenvector, we estimate the eigenvectors and eigenvalues using the daily realized volatility estimator and employ the previous day's eigenvector to predict the future volatility matrix.  
	In the empirical study, we investigate the effect of the eigenvector estimation methods, and we find that the volatility matrix estimator with the previous 22-day observations (one month) shows the best performance.
	This shows that the averaging step helps mitigate  volatile fluctuations in the volatility process, and by using the recent 22-day instead of the longer period, such as the whole period, we can explain the effect of  the eigenvector dynamics.  
	On the other hand, in high-frequency finance literature, the intraday time-varying patterns are often observed \citep{andersen2019time, andersen2021recalcitrant, kong2021discrepancy}.
	We also conducted hypothesis tests for the constant eigenvector, and the constant eigenvector hypothesis is often rejected (see Section \ref{empirical-study1}).
	Thus, it is more natural to assume intraday time-varying eigenvectors. 
	However, under this condition, we need to calculate a lot of local eigenvalues and eigenvectors and accumulate the local estimators.
	Furthermore, the intraday dynamic structure of the volatility process becomes more complex since we need to consider two different dynamic structures.
	This complexity may cause large estimation errors and the possibility of over-parameterization. 
	Thus, it is a demanding task to develop a parametric model that can explain the intraday and interday dynamics simultaneously and obtain robust prediction performance. 
	We leave this for a future study. 
\end{remark}

We investigate the theoretical properties of the POET estimator under the following assumptions.
These conditions are often used when analyzing the asymptotic behaviors of the POET estimator \citep{fan2018robust, shin2023adaptive}.
\begin{assumption}\label{Assumption2}
	~
	\begin{itemize}
		\item[(a)] For some fixed constant $C_1$,  we have
		$
		\frac{p}{r} \max_{ 1\leq  i \leq p} \sum_{j=1}^r {q}_{ij} ^2  \leq C_1 \text{ a.s.},
		$
		where ${\bq}_j = ({q}_{1j}, \ldots, {q}_{pj})^\top$ is the $j$th eigenvector of $\mathbf{\Psi}_{d}$.
		
		\item [(b)] For $d=1, \ldots, n$, let $D_{d,\xi}=\min \{\xi_{d,i} -  \xi_{d,i+1}, i=1, \ldots, r-1\}$, $D_{d,\xi}$ and $\xi_{d,r} \geq C_2$ a.s., and $\xi_{d,1} \leq C_3$ a.s. for some generic positive constants $C_2$ and $C_3$.
		
		\item[(c)] For $d=1, \ldots, n$, $\xi_{d,r+1}$ is bounded by some positive constant and $\xi_{d,p+r}$ stays away from zero almost surely.
		
		\item[(d)] $ s_{I}/ \sqrt{p}  +\sqrt{ \log p/m^{1/2}} =o(1)$.
	\end{itemize}
\end{assumption}

The following proposition derives the concentration properties of the POET estimator.

\begin{proposition}\label{Proposition1}
	Under the FIVAR($h$) model, suppose that the concentration inequality,
	\begin{equation}\label{eq-4.3}
		\Pr \left \{\max_{1 \leq d \leq n} \max_{1\leq i, j \leq p} \left| \hat{\Gamma}_{d,ij} - \Gamma_{d,ij} \right| \geq  C  \sqrt{\frac{\log \left( pn \vee m\right) }{m^{1/2}}} \right \} \leq  p^{-1},
	\end{equation}
	Assumption \ref{Assumption2}, and the sparsity condition \eqref{eq-4.1} are met.
	Take $ \upsilon_{m}=C_{\varpi} H_m$ for some large fixed constant $C_{\varpi}$, where $H_m =  s_{I}/ p +  \sqrt{\log \left( pn \vee m\right)/ m^{1/2}}$.
	Then, we have, for a sufficiently large m, with probability at least $1-p^{-1}$,
	\begin{eqnarray}
		&& \max_{1\leq d \leq n}\max_{1\leq i \leq r}\left| \hat{\xi}_{d,i}-\xi_{d,i} \right|\leq C H_m, \label{eq-4.4} \\
		&& \max_{1\leq d \leq n}\max_{r+1\leq i \leq p+r}\left| \hat{\xi}_{d,i}-\xi_{d,i} \right|\leq  C   s_{I}  H_m^{1-\Upsilon}, \label{eq-4.5} \\
		&& \max_{1\leq d \leq n}\| \hat{\bSigma}_d  - \bSigma_d  \|_{2} \leq C  s_{I}   H_m^{1-\Upsilon}, \quad \text{and} \label{eq-4.6}\\
		&& \max_{1\leq d \leq n}\| \hat{\bSigma}_d  - \bSigma_d  \|_{\max} \leq C   H_m . \label{eq-4.7}
	\end{eqnarray}
\end{proposition}

\begin{remark}
	To investigate the asymptotic behavior of the POET estimator, we need the concentration inequality \eqref{eq-4.3}. 
This concentration inequality can be obtained under the local boundedness condition of the instantaneous volatility process even with heavy-tailed observations and serially dependent noises. 
	For example, using the truncation method, \citet{shin2023adaptive} established the concentration inequality \eqref{eq-4.3} with heavy-tailed microstructure noises and jumps. 
See also \citet{fan2018robust}.
On the other hand, by employing the local averaging method, \citet{jacod2019estimating} developed an integrated volatility estimator that can handle serially dependent noise with polynomially $\rho$-mixing property. 
By combining these truncation and local averaging methods, we can obtain \eqref{eq-4.3}  when the noise is serially dependent and both the noise and jumps are heavy-tailed.
	We note that continuous adapted processes are locally bounded, and more generally, left-continuous adapted processes are almost surely locally bounded on every finite time interval.
	The proposed  FIVAR model is continuous; thus, the locally bounded condition is satisfied.
Thus, the concentration tail condition \eqref{eq-4.3} is not restrictive.
\end{remark}

The concentration inequalities \eqref{eq-4.4}--\eqref{eq-4.5} show that Assumption \ref{Assumption1}(e) is satisfied with high probability.
For example, we have $b_{m,n,p}^{F}=CH_m$ and $b_{m,n,p}^{I}=C s_{I}  H_m^{1-\Upsilon}$.
Using Theorem \ref{Theorem1}, we can derive the following result.

\begin{thm}\label{Theorem2}\
	Under the assumptions in Proposition \ref{Proposition1} and Assumption \ref{Assumption1} (except for Assumption \ref{Assumption1}(e)), 
	let $n \geq 3$, $s_{\beta}\sqrt{n \log p} + s_{\beta}n^{1/4}\left(\log n\right)^2 \left(\log p\right)^{3/4} = O \left(n\right)$, $\tau_F = \varpi_F= C\left(n / \log p\right)^{1/4}$, $\tau_I = Cs_{\beta}\left(n / \log p\right)^{1/4}$,  and $\varpi_I = C \left(n / \log p\right)^{1/4}$.
	Suppose that Assumption \ref{Assumption1}(f)--(g) hold with
	$
	\eta_F = C \left\{H_m + \left(\log n\right)^2\sqrt{ \log p / n}\right\}   \quad \text{and}  \quad \eta_I = C\left\{s_{\beta}  s_{I}H_m^{1-\Upsilon} + s_{\beta}\left(\log n\right)^2 \sqrt{\log p/ n}\right\}.
	$
	Then, we have, for a sufficiently large m, with probability at least $1-2p^{-1}$,
	\begin{eqnarray}
		&& \max_{1\leq i \leq r}\left\| \hat{\bbeta}_{i} -\bbeta_{i0}\right\|_{2}  \leq C \left\{H_m + \left(\log n\right)^2\sqrt{ \log p / n}\right\},  \quad \text{and} \label{eq-4.8} \\
		&& \max_{r+1\leq i \leq p+r}\left\| \hat{\bbeta}_{i} -\bbeta_{i0}\right\|_{2}  \leq C\left\{s_{\beta}^{3/2}  s_{I}H_m^{1-\Upsilon} + s_{\beta}^{3/2}\left(\log n\right)^2 \sqrt{\log p/ n}\right\}. \label{eq-4.9}
	\end{eqnarray}
\end{thm}

\begin{remark}
	Theorem \ref{Theorem2} shows that the low-dimensional factor VAR has the convergence rate  $s_{I}/ p +  \sqrt{\log \left( pn \vee m\right)/ m^{1/2}} +\left(\log n\right)^2\sqrt{ \log p / n}$.
	The $s_{I}/ p $ term is the cost to identify the latent factor, and the $m^{-1/4}$ term comes from estimating the integrated volatility matrix.
	Finally, the $n^{-1/2}$ term is the usual convergence rate of estimating model parameters in low-frequency time series.
	In contrast, the high-dimensional factor VAR has the convergence rate $s_{\beta}^{3/2}  s_{I}H_m^{1-\Upsilon} + s_{\beta}^{3/2}\left(\log n\right)^2 \sqrt{\log p/ n}$.
	The first term, $s_{\beta}^{3/2}  s_{I}H_m^{1-\Upsilon}$, is the cost to estimate the latent idiosyncratic volatility matrix with the noisy high-frequency data.
	The second term, $s_{\beta}^{3/2}\left(\log n\right)^2 \sqrt{\log p/ n}$, is the convergence rate of the sparse high-dimensional regression.
	We note that the $\left(\log n\right)^2$ term is the cost of handling the dependency in the eigenvalue process.
\end{remark}

With the results in Theorem \ref{Theorem2}, we investigate the theoretical properties of the future volatility matrix estimator $\tilde{\bGamma}_{n+1}$.
To study the future idiosyncratic volatility matrix estimator $ \hat{\mathbf{\Sigma}}_{n+1}$, we need the additional condition for the eigen-gap as follows.

\begin{assumption}\label{Assumption3}
	For  some $\chi \in (0,1)$ and $i= r+1, \ldots,  p+r-1$, we have $C_4 \chi^{i} \leq \xi_{d,i} -  \xi_{d,i+1} \leq C_5 \chi^{i}$ for some positive constants $C_4$ and $C_5$.
\end{assumption}

\begin{remark}
	To have the bounded eigenvalues for the idiosyncratic volatility matrices such as Assumption \ref{Assumption2}(c), we cannot have that all eigen-gaps are some positive constants.
	Specifically, several eigen-gaps can be constant, but most of them may need to converge to zero.
	To check the behavior of the eigen-gaps, we draw the plot of the eigen-gaps of the idiosyncratic volatility matrix using high-frequency trading data (see Figure \ref{Fig-7} in the Appendix).
	We find that the eigen-gaps have an exponentially decaying pattern.
	Thus, to account for this, we impose Assumption \ref{Assumption3}.
	We note that even if the finite number of $ \xi_{d,i}$'s do not satisfy this condition, we can obtain the same results in Theorem \ref{Theorem3}.
\end{remark}

The following theorem establishes the convergence rates of the future volatility matrix estimator.

\begin{thm}\label{Theorem3}
	Under the assumptions in Theorem \ref{Theorem2} and Assumption \ref{Assumption3}, we have with probability at least $1-2p^{-1}$,
	\begin{eqnarray}
		\max_{1\leq i \leq r}\left| \hat{\xi}_{n+1,i}-\E \left(\xi_{n+1,i} | \FF_{n} \right) \right|&\leq& C \Big[H_m + \left(\log n\right)^2\sqrt{ \log p / n}\Big], \label{eq-4.10} \\
		\max_{r+1\leq i \leq p+r}\left| \hat{\xi}_{n+1,i}-\E \left(\xi_{n+1,i} | \FF_{n} \right)  \right|&\leq& C\Big[s_{\beta}^{2}  s_{I}H_m^{1-\Upsilon} + s_{\beta}^{2}\left(\log n\right)^2 \sqrt{\log p/ n}\Big], \label{eq-4.11} \\
		\|\tilde{\bGamma}_{n+1} - \E \left(\bGamma_{n+1} | \FF_{n} \right) \|_{\bGamma^{\ast}} &\leq& C\Big[p^{1/2}H_m^{2}+ p^{1/2} \log p \left(\log n\right)^4/n \cr
		&& \quad + s_{\beta}^{2}  s_{I}H_m^{1-\Upsilon} + s_{\beta}^{2}\left(\log n\right)^2\sqrt{ \log p / n} \Big] \label{eq-4.12},
	\end{eqnarray}
	where the relative Frobenius norm $\| \bM \| _{\bGamma^{\ast}} ^2 = p^{-1} \|\bGamma^{\ast-1/2}  \bM  \bGamma ^{\ast-1/2}  \| _F^2$ and $\bGamma^{\ast}=\E \left(\bGamma_{n+1} | \FF_{n} \right)$.
\end{thm}

\begin{remark}
	The relative Frobenius norm is used in Theorem \ref{Theorem3} since the top eigenvalues of $\bGamma^*$ are diverging  (see \cite{fan2008high}).
	Theorem \ref{Theorem3} indicates that the proposed estimator $\tilde{\bGamma}_{n+1}$ is consistent as long as $p=o(m \wedge n^2)$ in terms of the relative Frobenius norm.
	Its convergence rate is similar to that of \citet{kim2019factor} except for  the additional terms, $p^{1/2} \log p \left(\log n\right)^4/n$ and $s_{\beta}^{2}\left(\log n\right)^2 \sqrt{\log p/ n}$, which come from handling the VAR model structure in the factor and idiosyncratic volatility matrices, respectively.
\end{remark}

\subsection{Discussion on the  tuning parameter selection}\label{SEC-4.3}

To implement the proposed robust estimation method, we need to choose the tuning parameters.
In this section,  we discuss how to select the tuning parameters  in \eqref{eq-3.1}--\eqref{eq-3.4}.
For the factor part, let $\sigma_{F}=\sqrt{\sum^{r}_{i=1}\sum^{n}_{d=1}\hat{\xi}^2_{d,i}/\(nr\)}$.
We choose
\begin{equation}\label{eq-4.13}
	\varpi_{F}= c_{F,1}\sigma_{F}\(\frac{n}{\log p}\)^{1/4} \quad \text{ and } \quad \tau_F= c_{F,2}\sigma_{F}\(\frac{n}{\log p}\)^{1/4},
\end{equation}
where $c_{F,1}$ and $c_{F,2}$ are tuning parameters.
For the idiosyncratic part, we first standardize the variables, $\hat{\xi}_{d,i}$, $i=1, \ldots, p+r$, to have mean zero and variance 1. 
Then, we choose
\begin{equation}\label{eq-4.14}
	\varpi_I= c_{I,1}\(\frac{n}{\log p}\)^{1/4},  \tau_I= c_{I,2}\(\frac{n}{\log p}\)^{1/4}, \quad \text{ and } \quad \eta_I=c_{\eta}\(\frac{\log p}{n}\)^{1/2},
\end{equation}
where  $c_{I,1}$, $c_{I,2}$, and $c_{\eta}$ are tuning parameters. 
We select $c_{\eta} \in [0.1, 10]$  by minimizing the corresponding Bayesian information criterion (BIC).
In the simulation and empirical studies, we choose $c_{F,1}=4$, $c_{F,2}=1/4$, $c_{I,1}=4$, and $c_{I,2}=4$.
These choices are  based on the empirical study (Section \ref{SEC-5.2}).
Specifically, we choose $c_{F,1}$, $c_{F,2}$, $c_{I,1}$, and $c_{I,2}$ which minimize the corresponding mean squared prediction error (MSPE).
Details can be found in Section \ref{SEC-5.2}.

\section{Numerical study}\label{SEC-5}
\subsection{A simulation study}\label{SEC-5.1}

In this section, we conducted simulations to validate the finite sample performance of the proposed estimation methods.
We generated the data for $n$ days with frequency $1/m^{all}$ on each day and let $t_{d,j}=d-1+j/m^{all}$ for  $d=1, \ldots, n$ and $j=0, \ldots, m^{all}$.
We considered the jump-diffusion process with the FIVAR($h$) model in Definition \ref{def-1} and generated heavy-tailed and sub-Gaussian processes. 
The specific simulation setup is described in Appendix \ref{simulation-setup}.
The noise-contaminated high-frequency data were generated from model \eqref{eq-4.2}, where the noise $e_{i}(t_{i,k})$ was obtained from the independent Gaussian distribution with mean zero and standard deviation $0.01 \sqrt{ \int _{0}^1  \gamma _{ii}(t)  dt }.$ 
This choice is inspired by  \citet{wang2010vast}  who found that the relative noise level is typically around $1\%$ for the stock index for high-frequency trading data.
We first generated the data for $500$ days, and we varied $n$ from $100$ to $500$.
For each $n$, we obtained the data from the last $n$ days among the $500$ days.

To estimate the integrated volatility matrix $\bGamma_d=\left(\Gamma_{d,ij}\right)_{i,j=1, \ldots, p}$, we utilized the jump adjusted pre-averaging realized volatility matrix (PRVM) estimator \citep{ait2016increased,christensen2010pre, jacod2009microstructure} defined in \eqref{eq-A.1} in the Appendix.
Then, we estimated the conditional expected volatility matrix $\E \left(\bGamma_{n+1} | \FF_{n} \right)$, based on the estimation procedure in Section \ref{SEC-4.2}.
Specifically, we first projected  $\hat{\bGamma}_d$ onto the positive semi-definite cone in the spectral norm to make it positive semi-definite.
Since the eigenvectors are constant over time, we estimated them using the $n$ period observations.
To determine the rank $r$, we employed the procedure in \citet{ait2017using} as follows:
\begin{equation}\label{eq-5.1}
	\hat{r}=\arg \min_{1 \leq j \leq r_{\text{max}}}\sum_{d=1}^{n}\left[p^{-1}\bar{\xi}_{d,j}+ j \times c_1 \left\{\sqrt{\log p / m^{1/2}} +p^{-1}\log p \right\}^{c_2}\right]-1,
\end{equation}
where $\bar{\xi}_{d,j}$ is the $j$-th largest eigenvalue of PRVM, $r_{\text{max}}=30$, {$c_1=0.02 \times \bar{\xi}_{d,30}$}, and $c_2=0.5$.
For the POET estimation procedure, we employed the soft thresholding scheme  and selected the thresholding level that minimizes the corresponding Frobenius norm.
When estimating $\bbeta_{i0}$'s, we used the tuning parameter selection method discussed in Section \ref{SEC-4.3}.
To select the lag $h$, we utilized the Bayesian information criterion (BIC). 
We calculated the future volatility matrix estimator with $\hat{\bbeta}_{i}$ and call it the Huber-LASSO (H-LASSO) estimator.

For comparison, we employ the ordinary least squares (OLS) and LASSO estimators as follows.
The OLS estimator only considers the dynamics of the factor volatility matrix and obtains $\hat{\bbeta}_{i}$, $i=1, \ldots, \hat{r},$ using the OLS method.
The OLS estimator  predicts the future idiosyncratic volatility matrix by the average of the previous 22-day's idiosyncratic volatility matrices to smooth random fluctuations.
On the other hand, the LASSO estimator considers the dynamics in both factor and idiosyncratic volatility matrices.
The LASSO estimator uses the same estimation procedure as the H-LASSO estimator, except for the truncation method.
That is, the OLS estimator can explain only the dynamics from the factor component, while the LASSO estimator can account for the dynamics from both the factor and idiosyncratic components.
However, they cannot account for the heavy-tailedness.
We also investigated the previous day's PRVM estimator from the POET procedure as the non-parametric benchmark.
We call it the POET-PRVM.
We calculated the average estimation errors under the Frobenius norm, the max norm, the relative Frobenius norm (see Theorem \ref{Theorem3} for the definition), and the spectral norm by 500 iterations.
{Note that we conducted one-day-ahead forecasts for each of 500 iterations.}

We first checked the performance of the methods for model parameter estimation.
The parameter of interest is the true parameter matrix $\bbeta_0$.
Table \ref{Table-1} reports the  Frobenius, max, and spectral norm errors of the LASSO and H-LASSO estimators, with $n=100, 200, 500$ and $m=250, 500, 2000$.
We note that for both heavy-tailed and sub-Gaussian processes,  the number of factors $r$ and lag $h$ are estimated without errors for all $n$ and $m$.  The reason is that the data generation process has a large eigen-gap between the factor and idiosyncratic volatility matrices and a strong time series structure.
From Table \ref{Table-1}, we find that the estimation errors of the proposed H-LASSO estimator are usually decreasing as the number of low-frequency or high-frequency observations increases.
%In contrast, the estimation errors of the LASSO estimator usually decrease as the number of low-frequency observations increases, but not for the high-frequency observations.
The exception is the max norm error for $n=500$ and $m=2000$, while the overall error performances, such as Frobenius and spectral norm errors, always decrease as $n$ or $m$ increases. 
An explanation is that bigger outliers for heavy tails are more frequently observed as the high-frequency observation increases.
For example, when $m$ is small, the relative frequency of outliers may be low due to the smoothing effect from the subsampling.
Furthermore, the max norm measure is highly affected by the outlier.
When comparing two estimation methods, the H-LASSO estimator performs better than the LASSO estimator for both heavy-tailed and sub-Gaussian processes.
One possible explanation for this is that, even if the process is generated by the sub-Gaussian variables, the log-prices process can  still have some heavy tails.
The truncation method can reduce the variance of the estimator, which is larger than that of the increase in estimation bias, even for the sub-Gaussian case.
From this result, we find the benefit of handling the heavy-tailedness.
These results support the theoretical findings in Section \ref{SEC-3}.

\begin{table}[!ht]
	\caption{The Frobenius, max, and spectral norm errors of the LASSO and H-LASSO estimators with $n=100, 200, 500$ and $m=250, 500, 2000$.}\label{Table-1}
	\centering
	\scalebox{0.85}{
		\begin{tabular}{l l c c c c c c c c c c c c c  c c c c  c c c}
			\hline
			\multicolumn{6} {c}{} & \multicolumn{2} {c}{Frobenius}  && \multicolumn{2} {c}{Max} && \multicolumn{2} {c}{Spectral} \\ \cline{7-8} \cline{10-11} \cline{13-14}
			Tail &&                 $n$   &&  $m$   &&   LASSO   &  H-LASSO   &&  LASSO   &  H-LASSO   &&   LASSO   &  H-LASSO  \\ \hline
			
			Heavy                && 100  && 250    &&   0.802    &  0.666    && 0.446 &  0.312 && 0.579 &  0.420      \\
			&&                           && 500    &&   0.724    &  0.594    && 0.422 &  0.293 && 0.535 &  0.380      \\
			&&                           && 2000   &&   0.691    &  0.556    && 0.406 &  0.270 && 0.511 &  0.339      \\
			&&                           &&        &&            &         \\
			&&                     200   && 250    &&   0.684    &  0.594    && 0.355 &  0.244 && 0.453 &  0.335      \\
			&&                           && 500    &&   0.625    &  0.527    && 0.349 &  0.229 && 0.426 &  0.290      \\
			&&                           && 2000   &&   0.609    &  0.512    && 0.335 &  0.225 && 0.413 &  0.280      \\
			&&                           &&        &&            &         \\
			&&                     500   && 250    &&   0.576    &  0.556    && 0.216 &  0.179 && 0.321 &  0.294      \\
			&&                           && 500    &&   0.516    &  0.490    && 0.216 &  0.173 && 0.281 &  0.229      \\
			&&                           && 2000   &&   0.512    &  0.476    && 0.227 &  0.179 && 0.284 &  0.224      \\
			&&                           &&        &&            &         \\
			Sub-Gaussian        && 100   && 250    &&   0.878    &  0.713    && 0.525 &  0.360 && 0.683 &  0.474      \\
			&&                           && 500    &&   0.838    &  0.651    && 0.554 &  0.355 && 0.681 &  0.450      \\
			&&                           && 2000   &&   0.830    &  0.616    && 0.579 &  0.334 && 0.674 &  0.419      \\
			&&                           &&        &&            &         \\
			&&                     200   && 250    &&   0.728    &  0.636    && 0.376 &  0.270 && 0.499 &  0.374     \\
			&&                           && 500    &&   0.666    &  0.562    && 0.373 &  0.253 && 0.479 &  0.338      \\
			&&                           && 2000   &&   0.637    &  0.529    && 0.365 &  0.238 && 0.454 &  0.309      \\
			&&                           &&        &&            &         \\
			&&                     500   && 250    &&   0.615    &  0.583    && 0.252 &  0.202 && 0.356 &  0.305      \\
			&&                           && 500    &&   0.548    &  0.513    && 0.246 &  0.196 && 0.322 &  0.264      \\
			&&                           && 2000   &&   0.526    &  0.491    && 0.248 &  0.198 && 0.305 &  0.249      \\   \hline
		\end{tabular}
	}
\end{table}

\begin{figure}[!ht]
	\centering
	\includegraphics[width = 0.82\textwidth]{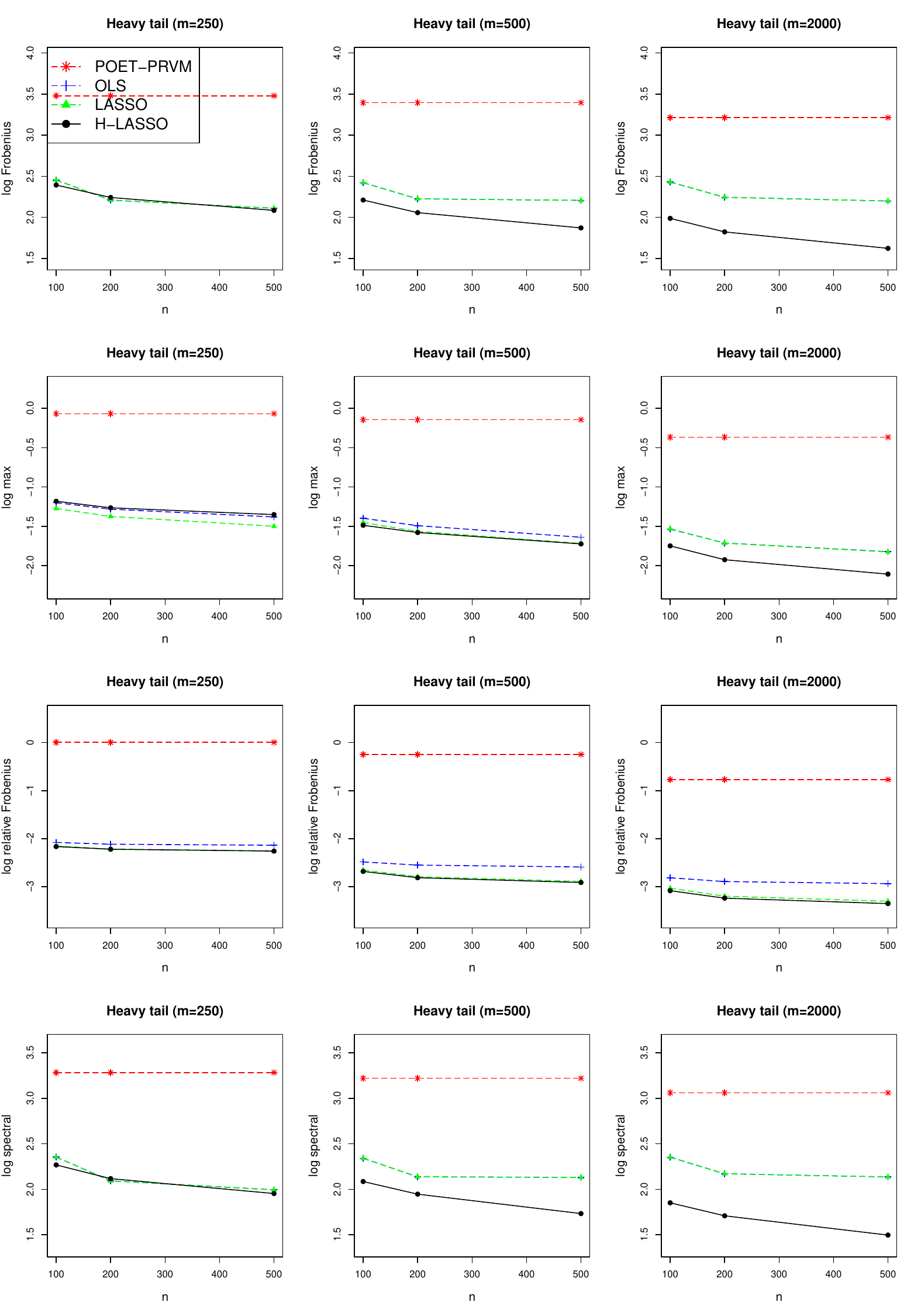}
	\caption{The log Frobenius, max, relative Frobenius, and spectral norm error plots of the POET-PRVM, OLS, LASSO, and H-LASSO estimators for the conditional expected integrated volatility matrix estimation with the heavy-tailed process, given $n=100, 200, 500$ and $m=250, 500, 2000$.}\label{Fig-3}
\end{figure}

\begin{figure}[!ht]
	\centering
	\includegraphics[width = 0.82\textwidth]{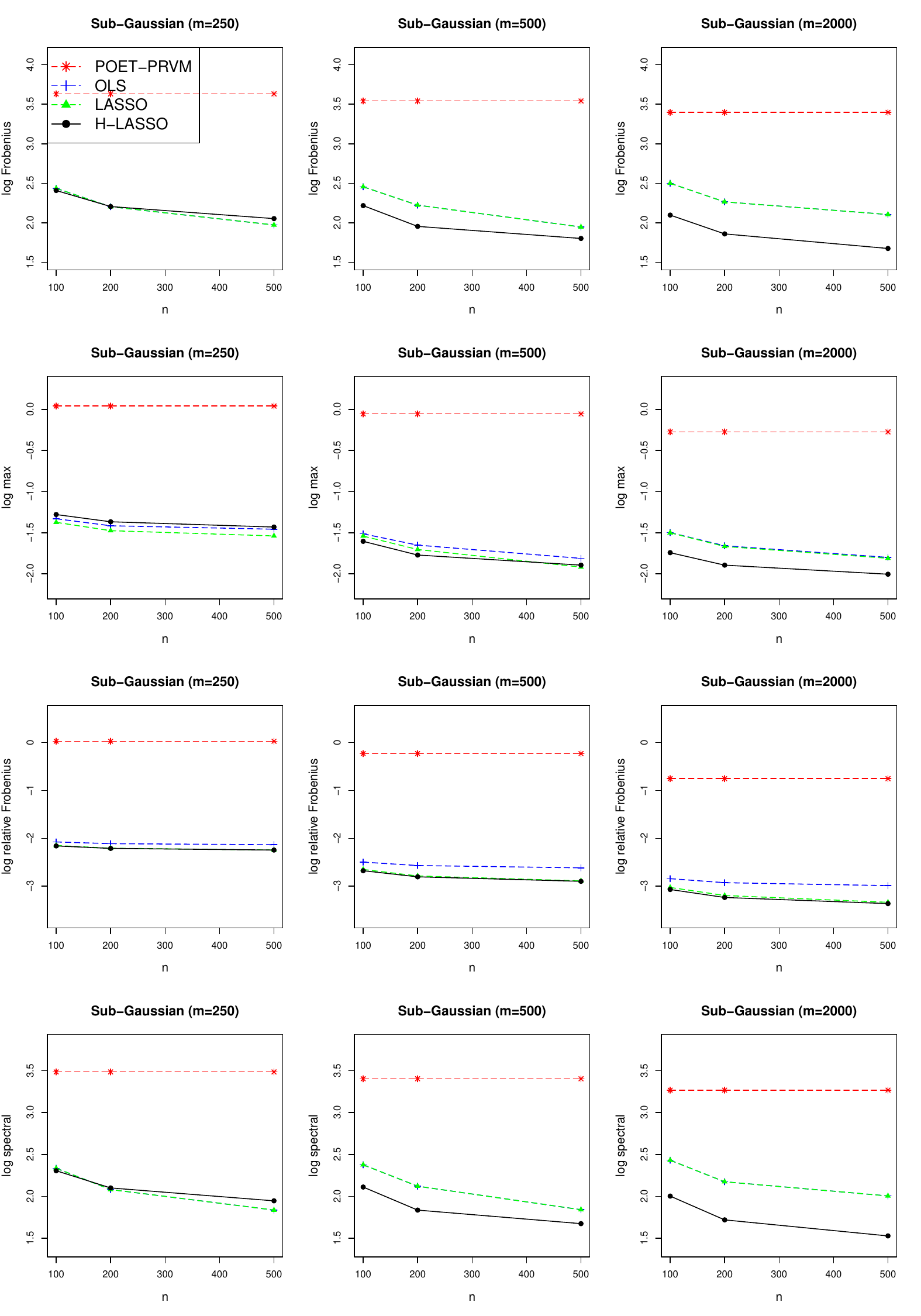}
	\caption{The log Frobenius, max, relative Frobenius, and spectral norm error plots of the POET-PRVM, OLS, LASSO, and H-LASSO estimators for the conditional expected integrated volatility matrix estimation with the sub-Gaussian process, given $n=100, 200, 500$ and $m=250, 500, 2000$.}\label{Fig-4}
\end{figure}

One of the main objectives of this paper is to predict  future volatility.
Therefore, we checked the performance of predicting future volatility.
Figures \ref{Fig-3} and \ref{Fig-4} plot the log Frobenius, max, relative Frobenius, and spectral norm errors of the future volatility matrix estimators with $n=100, 200, 500$ and $m=250, 500, 2000$ for the heavy-tailed and sub-Gaussian processes.
From Figures \ref{Fig-3} and \ref{Fig-4}, we find that the parametric estimation methods show better performance than the non-parametric POET-PRVM estimator.
{When comparing the OLS and LASSO estimators, the LASSO estimator performs better than the OLS estimator in terms of the relative Frobenius norm.
	One possible explanation for this is that the OLS estimator can partially  explain the volatility dynamics via the factor component, but fails to explain the whole dynamics.
	On the other hand, the Frobenius and spectral norm errors are similar for the OLS and LASSO estimators.
	This may be because they are highly affected by the errors in estimating large eigenvalues, such as the eigenvalues of the factor volatility matrix.
	Furthermore, the max norm error is also similar for the OLS and LASSO estimators.
	This may be because the OLS estimator does not have as many outliers as the LASSO estimator since the OLS estimator uses the average of the previous 22-day's idiosyncratic volatility matrices.
	Finally, the H-LASSO estimator shows the best performance for the heavy-tailed and sub-Gaussian processes.}
These results are consistent with our notion that the H-LASSO estimator is robust to the heavy-tailedness, and it can explain the dynamics from the factor and idiosyncratic components.

\subsection{An empirical study}\label{SEC-5.2}
We applied the proposed FIVAR($h$) model to real high-frequency trading data for 200 assets from January 2016 to December 2019 (997 trading days).
The top 200 large trading volume stocks among the S\&P 500  were selected from the Wharton Data Service (WRDS) system.
{The trading volumes were calculated using the data from  years 2015 to 2020.}
To synchronize the high-frequency data, we used the previous tick scheme \citep{andersen2003modeling, barndorff2011multivariate, zhang2011estimating} with equal distance intervals, {which helps mitigate the errors from the irregularity of the observation times.}
When applying the refresh time scheme for the 200 assets, we find that the average number of daily synchronized samples is 593.49, which corresponds to 39.42-sec sampling frequency. 
Hence, we chose 1-min sampling frequency to enjoy the benefit of large samples, {which corresponds to $m=390$.}
We excluded days with half trading hours.
The data of year $2020$ was excluded to avoid the effects of extreme market conditions.
	Specifically, we performed the structural break test for the eigenvalue process in Section \ref{empirical-study2} and found the non-stationarity when including 2020 data.  
	We note that approximately 10 CPU cores with 2 GHz and 200 GB of RAM are required to conduct the whole empirical study within 3 days.

To apply the proposed estimation procedures, we first calculated 997 daily integrated volatility matrices using the jump adjusted pre-averaging realized volatility matrix (PRVM) estimator in \eqref{eq-A.1}.
	We chose $K$ as 19 and $c_{i,u}$ as 4 times the sample standard deviation of the pre-averaged variables $m^{1/4}\bar{Y}_{i}\left(t_{d,k}\right)$.
	We projected the daily PRVM estimators onto the positive semi-definite cone in the spectral norm to make them positive semi-definite.
	In the empirical study, to predict the future volatility matrix, we employed the rolloing window scheme with the in-sample period $n=251$.
	For each in-sample period, we estimated the rank $r$ based on the rank estimation procedure in \eqref{eq-5.1} with $n=251$.
   We note that $r$ was always estimated to be 3.

%The result is $\hat{r}=3$.
%Also, Figure \ref{Fig-4} shows the scree plot drawn using the first 100 eigenvalues of the sum of 997 PRVM estimates.
%From Figure \ref{Fig-4}, we can see that the possible values of the rank $r$ are $1,2,3,4,5$.
%From these results, we conducted the empirical study for $r=3$.
%\begin{figure}[!ht]
%\centering
%\includegraphics[width = 0.77\textwidth]{Scree_plot.pdf}
%\vspace{-0.65cm}
%\caption{The scree plot of the first 100  eigenvalues of the sum of 997 PRVM estimates.}\label{Fig-4}
%\end{figure}

To estimate the idiosyncratic volatility matrix $\bSigma_d$, we utilized the hard thresholding scheme based on the  Global Industry Classification Standard (GICS) proposed by \citet{fan2016incorporating}.
Specifically, the idiosyncratic components for the different sectors were set to zero, and we maintained those for the same sector.
This corresponds to the hard-thresholding scheme with the sector information.

To choose the tuning parameters $\ell$, $c_{F,1}$, $c_{F,2}$, $c_{I,1}$, and $c_{I,2}$, we defined 
	\begin{eqnarray*}
		&& \Lambda^{F}(\ell, c_{F,1}, c_{F,2}) = \frac{1}{T}\sum_{d=1}^{T}\| \hat{\mathbf{\Psi}}_d^{\text{H-LASSO}}(\ell, c_{F,1}, c_{F,2}) - \hat{\mathbf{\Psi}}_{d+1}^{\text{POET}}\|^2_F, \cr
		&& \Lambda^{I}(\ell, c_{I,1}, c_{I,2}) = \frac{1}{T}\sum_{d=1}^{T}\| \hat{\mathbf{\Sigma}}_d^{\text{H-LASSO}}(\ell, c_{I,1}, c_{I,2}) - \hat{\mathbf{\Sigma}}_{d+1}^{\text{POET}}\|^2_F, 
	\end{eqnarray*}
	where $\hat{\mathbf{\Psi}}_d^{\text{H-LASSO}}(\ell, c_{F,1}, c_{F,2})$ is the factor volatility matrix forecast from the H-LASSO estimator with the tuning parameters $\ell$, $c_{F,1}$, and $c_{F,2}$ for the $d$-th day, and $\hat{\mathbf{\Sigma}}_d^{\text{H-LASSO}}(\ell, c_{I,1}, c_{I,2})$ is the  idiosyncratic volatility matrix forecast from the H-LASSO estimator with the tuning parameters $\ell$, $c_{I,1}$, and $c_{I,2}$ for the $d$-th day. 
	Also, $\hat{\mathbf{\Psi}}_{d}^{\text{POET}}$ and $\hat{\mathbf{\Sigma}}_{d}^{\text{POET}}$ are the factor and idiosyncratic volatility matrix estimators from the POET-PRVM estimator for the $d$-th day, respectively.
	Then, we selected $\ell$, $c_{F,1}$, and $c_{F,2}$ by minimizing $\Lambda^{F}(\ell, c_{F,1}, c_{F,2})$ over $\ell \in \left\{ 1, 5, 22, 251 \right\} $ and $c_{F,1}, c_{F,2} \in \left\{ 1/4, 1/2, 1, 2, 4\right\}.$
	Similarly, we chose  $\ell$, $c_{I,1}$, and $c_{I,2}$ by minimizing $\Lambda^{I}(\ell, c_{I,1}, c_{I,2})$ over $\ell \in \left\{ 1, 5, 22, 251 \right\}$ and $c_{I,1}, c_{I,2} \in \left\{ 1/4, 1/2, 1, 2, 4\right\}.$
	For the choice of tuning parameters, we chose the in-sample period as $n=251$ and out-of-sample period as day 252 to day 500 (year 2017).
	The selected parameters are $\ell=22$, $c_{F,1}=4$, $c_{F,2}=1/4$, $c_{I,1}=4$, and $c_{I,2}=4$.
	We note that $\ell=22$ was chosen for both $\Lambda^{F}(\ell, c_{F,1}, c_{F,2})$ and $\Lambda^{I}(\ell, c_{I,1}, c_{I,2})$.
	We also note that the stationarity of the volatility process is a reasonable assumption, which justifies the above tuning parameter choice procedure.
	To determine the lag $h$, we applied the Bayesian information criterion (BIC) to the VAR model. 
	It leads to $h=1$ for all in-sample period.
	Then, we estimated the conditional expected volatility matrix $\E \left(\bGamma_{n+1} | \FF_{n} \right)$ with the POET-PRVM, OLS, LASSO, and H-LASSO estimators.

%Specifically, we used the rolling window scheme with the past $n$ period observations to estimate the model parameters.
%To mitigate the potential effect of dynamics and fluctuations in the eigenvector process, we employed the previous 22 days' observations to estimate the eigenvectors.
%In  Appendix \ref{empirical-study2}, we compared the eigenvector estimators with different averaging days.

For a comparison, we employed the DCC-NL estimator \citep{ de2021factor, engle2019large, ledoit2015spectrum, ledoit2022power}, which employs the nonlinear shrinkage estimator and the dynamic conditional correlation (DCC) model \citep{engle2002dynamic}.
Specifically, let  $\Delta \bY_{d}=\(\Delta Y_{1,d}, \ldots, \Delta Y_{p,d}\)^{\top}$ and  $\Delta Y_{i,d}$ be the daily return for the $i$-th asset and $d$-th day.
To obtain the DCC-NL estimator,  we first employed the following GARCH$(1,1)$ model:
\begin{equation*}
	U_{i,d}^2=a_i+b_{1,i}\Delta Y_{i,d-1}^2 +b_{2,i} U_{i,d-1}^2,
\end{equation*}
where the conditional variance $U_{i,d}^2=\var\(\Delta Y_{i,d} | \FF_{d-1}\)$.
Based on the GARCH model, we calculated the conditional variance for the next day, $U_{i,n+1}^2$, and obtained the devolatilized returns
\begin{equation*}
	\Delta \bY^s_{d}= \(\Delta Y_{1,d} /U_{1,d}, \ldots, \Delta Y_{p,d} /U_{p,d}\)^{\top}.
\end{equation*}
With this devolatilized return series $\left\{\Delta \bY^s_{d}\right\}$, we obtained $\text{cov}\(\Delta \bY^s_{d}\)$ based on the nonlinear shrinkage.
Then, we applied the DCC model with $\text{cov}\(\Delta \bY^s_{i,d}\)$ being used for correlation targeting, and calculated the conditional correlation matrix for the next day, $\bR_{n+1} = \corr\(\Delta \bY_{n+1} | \FF_{n}\).$
Finally, we estimated the conditional covariance matrix for the next day, $\cov\(\Delta \bY_{n+1} | \FF_{n}\)$, as $\bU_{n+1} \bR_{n+1} \bU_{n+1}$, where $\bU_{n+1}=\Diag\(U_{1,n+1}, \ldots, U_{p,n+1}\)$. 
Detailed estimation procedure can be found in \citet{engle2019large}.
%We note that the DCC-NL estimator can handle the curse of dimensionality.
We also employed the HAR-DRD model \citep{oh2016high} based on the POET-PRVM estimator. Specifically, we first decomposed the POET-PRVM estimator for the $d$-th day, $\hat{\bGamma}_{d}^{POET}=\(\hat{\Gamma}_{d,ij}^{POET}\)_{1 \leq i,j \leq p}$, into
\begin{equation*}
	\hat{\bGamma}_{d}^{POET} =  \sqrt{\bD_d}  \bR_d \sqrt{\bD_d},
\end{equation*}
where $\bD_d=\(D_{d,ij}\)_{1 \leq i,j \leq p} =\Diag \(\hat{\Gamma}_{d,11}^{POET}, \ldots, \hat{\Gamma}_{d,pp}^{POET} \)$ is the diagonal matrix of integrated volatility estimators and $\bR_d$ is the correlation matrix estimator. 
Then, we applied the following HAR model to each integrated volatility estimator:
\begin{equation*}
	D_{d+1,ii}=a_i+b_{i}^{(day)} D_{d,ii} + b_{i}^{(week)} \frac{1}{5}\sum^{4}_{j=0}D_{d-j,ii} + b_{i}^{(month)}\frac{1}{22}\sum^{21}_{j=0}D_{d-j,ii}+e_{d+1,i}, \quad i=1, \ldots, p.
\end{equation*}
We then obtained the conditional integrated volatility estimators for the next day.
To ensure that the  volatility forecasts are positive, we set their lower bound as $10^{-7}$.
For the correlation matrix estimator, we applied the following HAR-type model:
\begin{eqnarray*}
	\vech\(\bR_{d+1}\) &=& \vech\(\bar{\bR}_d\)\(1-a-b-c\) +a  \cdot \vech\(\bR_{d}\) \cr
	&&+ b \cdot \frac{1}{5}\sum^{4}_{j=0} \vech\(\bR_{d-j}\)+  c \cdot \frac{1}{22}\sum^{21}_{j=0}\vech\(\bR_{d-j}\) + \be_{d+1},
\end{eqnarray*}
where $\bar{\bR}_d = \frac{1}{d}\sum^{d}_{j=1} \bR_{j}$ and $\(a,b,c\) \in \mathbb{R}^3$.
Then, we forecast the  next day's integrated volatility matrix  based on the conditional integrated volatility and correlation estimators.
We call it the HAR-DRD estimator. 
To predict the future volatility matrix, we also employed the rolling window scheme with the in-sample period of $251$ days for the DCC-NL and HAR-DRD models.
We note that for all estimators, including the HAR-DRD estimator, the logarithm of the integrated volatility estimator is not used to check the effect of modeling idiosyncratic volatilities in linear modeling approaches.
In fact, there are some cases in which the logarithm improves the performance of the volatility estimators.
However, it is difficult to model the log-volatility in the high-dimensional high-frequency set-up.
For example, \citet{kim2022exponential} introduced the exponential GARCH-It\^{o} volatility model for the one-dimensional case, but the extension to the high-dimensional case is not straightforward.
We leave this issue for a future study.
We note that all estimators except the DCC-NL estimator use the POET-PRVM estimator as an input.
Since the DCC-NL estimator uses the daily total returns, only the DCC-NL estimator forecasts the total volatility.
Thus, we adjusted the jump component in the following applications.
To do this, we obtained the jump covariation matrix estimator by subtracting the jump adjusted PRVM estimator from the PRVM estimator.
The PRVM estimator can be obtained by setting the truncation parameters as infinity in \eqref{eq-A.1}.
We note that \citet{ait2016increased} showed that the PRVM estimator converges to the total volatility.

To investigate the performance of the future volatility matrix estimators, we employed the high-frequency data from 2017 to 2019. 
We chose the in-sample period as $n=251$ (one year), and we used three different out-of-sample periods: 2018 and 2019 (period 1), 2018 only (period 2), and 2019 only (period 3).
For the period 1, we calculated the average number of non-zero elements in $\hat{\bbeta}_{i}$ excluding the intercept term over $i=4, \ldots, 203$.
The results are 2.860 and 2.910 for the H-LASSO and LASSO estimators, respectively.

To check the performance of the proposed estimation procedures, we first investigated the following mean squared prediction error (MSPE) and QLIKE \citep{bollerslev2018modeling, laurent2013loss}:
\begin{eqnarray}\label{eq-5.2}
	&&	 \text{MSPE}(\tilde{\bGamma}) = \frac{1}{T}\sum_{d=1}^{T}\| \tilde{\bGamma}_d -\hat{\bGamma}_{d}^{POET}\|^2_F, \cr
	&&	 \text{QLIKE}(\tilde{\bGamma}) = \frac{1}{T}\sum_{d=1}^{T}\log\( \det\(\tilde{\bGamma}_d\)\) + \tr\(\tilde{\bGamma}^{-1}_d \hat{\bGamma}_{d}^{POET}\),
\end{eqnarray}
where $T$ is the number of days in the out-of-sample period, $\tilde{\bGamma}_{d}$ is one of the future volatility matrix forecasts from the POET-PRVM, OLS, LASSO, H-LASSO, DCC-NL, and HAR-DRD estimators for the $d$-th day of the out-of-sample period, and $\hat{\bGamma}_{d}^{POET}$ is the POET-PRVM estimator for the $d$-th day, which is a proxy of the ground truth.
For the DCC-NL estimator, we subtracted $\tilde{\textbf{JV}}_d$ from $\tilde{\bGamma}_d$, where $\tilde{\textbf{JV}}_d$ is the future jump covariation matrix forecast for the $d$-th day obtained by previous day's jump covariation matrix estimator.
This adjustment helps improve the performance of the DCC-NL estimator.
We note that MSPE is a form of the mean squared error that is one of the robust loss functions for volatility comparisons \citep{hansen2006consistent, patton2011volatility, patton2009evaluating}.
Also, QLIKE is robust to the presence of noise in the volatility proxy \citep{hansen2006consistent, laurent2013loss, patton2011volatility, patton2009evaluating}. 
Table \ref{Table-2} reports the MSPE and QLIKE results of the POET-PRVM, OLS, LASSO, H-LASSO, DCC-NL, and HAR-DRD estimators for three out-of-sample periods.
We find that the H-LASSO estimator shows good performance in terms of both MSPE and QLIKE. 
We note that the LASSO and H-LASSO estimators have similar performance in terms of the QLIKE. 
This may be because the QLIKE is highly affected by the small eigenvalue estimation, and the small eigenvalues are less affected by the heavy-tailedness.
On the other hand, the HAR-DRD estimator shows the best performance in terms of MSPE, but it did not perform well for the QLIKE loss.
These results show the proposed H-LASSO estimator can help explain the dynamics of the idiosyncratic volatility matrix under the sparsity condition and the heavy-tailedness of the financial data.

\begin{table}[!ht]
	\caption{The MSPE and QLIKE  of   the  POET-PRVM, OLS, LASSO, H-LASSO, DCC-NL, and HAR-DRD estimators (period 1, from 2018 to 2019; period 2, 2018; period 3, 2019).}\label{Table-2}
	\centering
	\scalebox{0.82}{
		\begin{tabular}{l l l l c c c c c c c c c c c  c c c c c c}
			\hline
			&&		&&	POET-PRVM 	&&	OLS	&&	LASSO    &&  H-LASSO && DCC-NL  &&  HAR-DRD \\ \hline
			Period 1 &&	  MSPE $\times 10^4$    &&   	2.525    &&  2.390 	&&	2.393	    &&	2.052    &&	2.788	   &&	2.043	    \\
			&& QLIKE $\times 10^{-3}$            &&	-1.247     && -1.681 	&&	-1.688    &&	-1.688	  &&	-1.672   &&	-1.616	 \\ 
			&&                         &&            &&         &&           &&           &&          &&     \\
			Period 2	&&	  MSPE $\times 10^4$    &&  3.941  	     &&  3.972 	&&	3.977     &&	3.354	   &&	4.094    &&	3.343     \\
			&& QLIKE $\times 10^{-3}$            &&	-1.220     && -1.662 	&&	 -1.665	 &&	-1.665  	&&  -1.633  &&	-1.562	 \\ 
			&&                         &&            &&         &&           &&           &&          &&     \\
			Period 3	&&	  MSPE $\times 10^4$    &&  	1.104   	&&  0.802 	&&	0.802      && 0.746  &&	1.476    &&	0.737     \\ 
			&& QLIKE $\times 10^{-3}$            &&	-1.274    && -1.701 	&&	 -1.711	 &&	-1.711   	&&	-1.711   &&	-1.670	 \\  \hline
		\end{tabular}
	}
\end{table}

\begin{figure}[!ht]
	\centering
	\includegraphics[width = 0.95\textwidth]{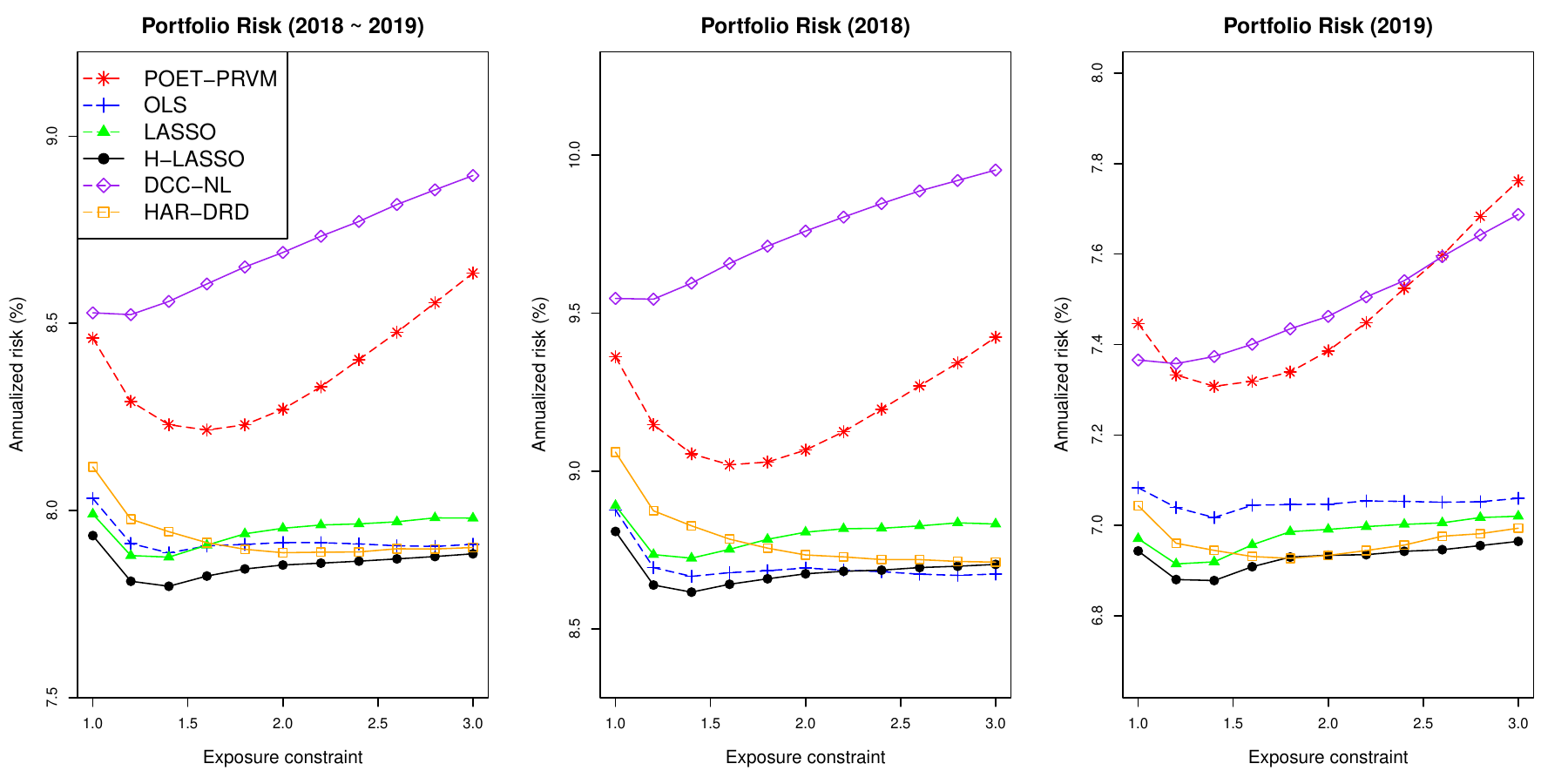}
	\caption{The out-of-sample risks of the minimum variance portfolios constructed by the POET-PRVM, OLS, LASSO, H-LASSO, DCC-NL, and HAR-DRD estimators.}\label{Fig-5}
\end{figure}

To investigate the out-of-sample portfolio allocation performance, we applied the proposed estimators to the following minimum variance portfolio allocation problem:
\begin{equation*}
	\min_{\omega} \omega^\top\(\tilde{\bGamma}_{d}+\tilde{\textbf{JV}}_d\)\omega,\quad \text{ subject to } \omega^\top\mathbf{J}=1 \text{ and } \left\| \omega\right\|_{1} \leq c_{0},
\end{equation*}
where $\mathbf{J}=\left(1, \ldots ,1\right)^\top \in \mathbb{R}^{p}$, $c_{0}$ is the gross exposure constraint that changed from 1 to 3, $\tilde{\bGamma}_{d}$ is one of the future volatility matrix estimators from POET-PRVM, OLS, LASSO, H-LASSO, DCC-NL, and HAR-DRD for the $d$-th day, and $\tilde{\textbf{JV}}_d$ is the future jump covariation matrix estimator for the $d$-th day.
Except for the DCC-NL estimator, we used the jump covariation matrix estimator for the $(d-1)$-th day as $\tilde{\textbf{JV}}_d$.
We note that in many studies \citep{andersen2007roughing, corsi2010threshold, duong2015empirical, patton2015good, wang2016forecasting}, decomposing total volatility into its continuous and jump components has been shown to improve the accuracy of volatility forecasting.
In this paper, we focus on predicting continuous volatility and use the non-parametric jump covariation matrix estimator for portfolio allocation.
It is worth noting that adding $\tilde{\textbf{JV}}_d$ slightly improves the performances of the future volatility matrix estimators.
This may be because the jumps provide additional risk information.
In this paper, since the pattern of the prediction performance is similar, we only report the results of adding the jump component.
To obtain the out-of-sample risks, we constructed the portfolios at the beginning of each trading day, based on the stock weights calculated using each future volatility matrix estimator.
The portfolios were maintained for one day, and we calculated the realized volatility using the 10-min portfolio log-returns to mitigate the microstructural noise effect.
Specifically, the realized volatility for the $d$-th day is obtained by
\begin{equation*}
\sum_{k=1}^{39} \(\hat{\omega}_d^\top  \Delta \bY^{\text{10-min}}_{d,k} \)^2,
\end{equation*}
where $\hat{\omega}_d$ is the stock weights for the $d$-th day,  $\Delta \bY^{\text{10-min}}_{d,k}=\(\Delta Y^{\text{10-min}}_{1,d,k}, \ldots, \Delta Y^{\text{10-min}}_{p,d,k}\)^{\top}$, and  $\Delta Y^{\text{10-min}}_{i,d,k}$ is the $k$-th 10-min return for the $i$-th asset and $d$-th day.
Then, we measured the out-of-sample risk using the square root of their average for each out-of-sample period.
Figure \ref{Fig-5} depicts the out-of-sample risks of the portfolios constructed using the POET-PRVM, OLS, LASSO, H-LASSO, DCC-NL, and HAR-DRD estimators.
From Figure \ref{Fig-5}, we find that the POET-PRVM and DCC-NL estimators become unstable as the gross exposure constraint increases.
This may be because the POET-PRVM estimator cannot explain the dynamics of the volatility process, and the  DCC-NL estimator only uses the low-frequency information.
On the other hand, the H-LASSO estimator has a stable result and has the smallest risk. 
These results indicate that considering both dynamic structures in idiosyncratic volatility and heavy-tailedness in financial data helps account for the dynamics of large volatility matrix processes.

\section{Conclusion}\label{SEC-6}

In this paper, we develop a novel factor and idiosyncratic VAR (FIVAR) model to account for the dynamic structure of the large volatility matrix, which has a low-rank plus sparse structure.
Under the FIVAR model,  the daily eigenvalues of the factor and idiosyncratic volatility matrices have the VAR model structure.
To further allow the heavy-tailedness in financial data, we use the bounded moment condition for the VAR model.
Then, we propose a robust estimation procedure for the VAR model parameters, which employs the truncation method and $\ell_{1}$-penalty to deal with the heavy-tailedness and explore the sparsity.
We show that it can handle the heavy-tailedness, observation error, and high dimensionality with a desirable convergence rate.
We also propose the large volatility prediction procedure and investigate its asymptotic properties.

%In the empirical study, in terms of the prediction error and portfolio allocation, the proposed estimator shows the best performance overall except for the MSPE measurement. 
%We note that for MSPE, H-LASSO shows the second-best performance.
In the empirical study, in terms of prediction error and portfolio allocation, the proposed estimator shows the best performance overall, except for the MSPE measurement. 
We note that for MSPE, H-LASSO shows the second-best performance.
It reveals that, when predicting large volatility matrices, the proposed estimation method helps handle the heavy-tailedness of financial data and explains the dynamic structure of factor and idiosyncratic volatility matrices.
On the other hand, we simply used today's jump covariation matrix estimator to forecast the next day's jump covariation matrix.
However, jumps often occur due to unexpected events; thus, they exhibit nonstationary behavior and different dynamics compared to continuous processes.
This challenge is further complicated by its high dimensionality.
Therefore, modeling a high-dimensional jump process is important but challenging. 
%We believe that a thorough high-dimensional jump dynamics analysis could improve predictive performance.
Thus, we leave this issue for a future study.

On the other hand, one of the key assumptions in the proposed model is the sparsity condition of the model parameters. 
Thus, it would be interesting to construct a test procedure for the sparsity condition. 
To do this, we may need to debias the biased H-LASSO estimator and to derive its asymptotic distribution under the sparsity hypothesis. 
This is a theoretically demanding task.
It would be interesting and important to develop a tuning parameter choice procedure that has rigorous theoretical properties and works well in practice. 
However, developing a tuning parameter selection procedure that works well from both practical and theoretical perspectives may be challenging.
We leave these topics for future studies.

%On the other hand, in this paper, we assume that the eigenvectors are constant over time, which are not flexible for the application  of our proposed method over a relatively long period.
%%may cause some dynamics in the volatility process to be missed.
%Thus, it would be interesting and important to develop the parametric model for the eigenvector process.
%We leave this issue for future study.
\begin{singlespace}
\bibliography{myReferences}
\end{singlespace}

\newpage

\appendix
\section{Appendix}\label{Appendix}
\subsection{Existence of a continuous eigenvalue process}\label{existence}
In this section, we propose a continuous eigenvalue process whose integrated eigenvalue process satisfies the VAR model structure.
\begin{proposition}\label{Proposition2}
Let $\bzeta_k = (\zeta_{k,i,j})_{1\leq i,j \leq p+r}$ for all $1\leq k \leq h$, $\det(\bzeta_{1})\neq 0$, and the spectral radius of $\bzeta_{1}$, $\rho(\bzeta_{1})<1$.
Then,  the integrated eigenvalue process $\bxi_{d}$ satisfies the VAR$(h)$ model for the following process defined for $i=1, \ldots, p+r$:
	\begin{eqnarray*}
		\lambda _{t,i}\left( \btheta _{i}\right)&=&(\lceil t \rceil -t)\lambda _{\lceil t-1 \rceil,i}\left( \btheta _{i}\right)
		+\sum ^{p+r}_{j=1}\zeta_{1,i,j}\left\{ \int ^{t}_{\lceil t-1 \rceil}\lambda _{s,j}\left( \btheta _{j}\right) ds\right\}  +(\lceil t \rceil -t )   \int ^{t}_{\lceil t-1 \rceil} J_{\lambda,i} (s) d \tilde{\Lambda}_{\lambda, i} (s)\cr
		&& +(t-\lceil t \rceil+1)\left(a_{i}+\sum ^{h}_{k=2}\sum ^{p+r}_{j=1}\zeta_{k,i,j}\left\{ \int ^{\lceil t \rceil - k +1}_{\lceil t \rceil - k}\lambda _{s,j}\left( \btheta _{j}\right) ds\right\}\right)   +(\lceil t \rceil -t)   Z^{2}_{i,t},
	\end{eqnarray*}
	where 
	$\btheta _{i}=(\zeta_{k,i,j})_{1\leq k \leq h, 1\leq j \leq p+r}$ is the model parameter, {$\lceil t \rceil$ is a ceiling function, which is the smallest integer greater than or equal to $t$}, $J_{\lambda,i} (t)$ is a jump size process, $\tilde{\Lambda}_{\lambda, i} (t)$ is a compensated Poisson process,  and $Z_{i,t}=\int ^{t}_{\lceil t-1 \rceil}z_{i,t}dW_{i,t}$, where $W_{i,t}$ is a standard Brownian motion and $z_{i,t}$ is a continuous process over each integer time interval.
\end{proposition}
%In the proposed FIVAR model, the instantaneous eigenvalues at the integer time points (time unit is usually day) satisfy the VAR($h$)-type structure as follows:
%\begin{equation*}
%	\lambda _{d,i}\left( \btheta _{i}\right)=a_{i}+\sum ^{h}_{k=1}\sum ^{p+r}_{j=1}\zeta_{k,i,j}\left\{ \int ^{d-k+1}_{d-k}\lambda _{t,j}\left( \btheta _{j}\right) dt\right\}.
%\end{equation*}
%Thus, under the FIVAR model, the instantaneous eigenvalue process is an interpolation of the VAR structure.
%By introducing the jump process for the instantaneous eigenvalue process, we can account for the co-jumps with the return process \citep{bandi2016price, bibinger2018common, jacod2017testing, jacod2010price}. 
We note that to guarantee the positiveness of the eigenvalue process, we need some lower bound condition for the jump process, such as $ J_{\lambda,i} (t) \tilde{ \Lambda}_{\lambda,i} (t) \geq -c$ a.s. for any $t$ and some positive constant $c$ that is related with the lower bound of the continuous part of the instantaneous eigenvalue process (e.g., $a_i$).
%The random fluctuation is modeled by  $Z^{2}_{i,t}$, which is included in integrated eigenvalues.
%Thus, the volatility process is not deterministic. %, and, so, we can explain the randomness of integrated volatilities.
The above continuous eigenvalue process will be used to conduct a simulation study based on high-frequency simulated data.
\begin{remark}
In this paper, we propose a prediction procedure for integrated volatility matrices based on the VAR model structure of daily integrated eigenvalues. 
On the other hand, the intraday dynamic structure of the eigenvalues is not used for the prediction procedure.
Specifically, since we use only the VAR model structure for the interday prediction procedure, the intraday dynamic structure does not affect the prediction method as long as the integrated eigenvalues follow the proposed VAR model.
%That is, the intraday dynamics in Definition \ref{def-1} are proposed to show the existence of the continuous eigenvalue process that satisfies the VAR model structure of the integrated eigenvalue process.
%This enables us to accommodate both a continuous-time It\^o process and a discrete-time VAR model and to conduct a simulation study based on high-frequency simulated data.
It is interesting and important to develop a unified model that can explain interday and intraday dynamics simultaneously. 
We leave this for a future study.
\end{remark}

%Furthermore, we model the heavy-tailedness by the heavy-tailed random fluctuation $Z^{2}_{i,t}$.
\textbf{Proof of Proposition \ref{Proposition2}.}
For non-negative integer $l \in \mathbb{N}_{0}$, let
\begin{eqnarray*}
\bR(l)=\left(R_1(l),\ldots ,R_{p+r}(l)\right)^{\top} \text{ with } R_{i}(l)=\int ^{d}_{d-1}\frac{\left( d-t\right) ^{l}}{l!}\lambda _{t,i}\left( \btheta _{i}\right) dt
\end{eqnarray*}
and $\bR(0)$ is the quantity that we would like to obtain.
We have
\begin{equation*}
	\lambda _{d,i}\left( \btheta _{i}\right)=a_{i}+\sum ^{h}_{k=1}\sum ^{p+r}_{j=1}\zeta_{k,i,j}\left\{ \int ^{d-k+1}_{d-k}\lambda _{t,j}\left( \btheta _{j}\right) dt\right\}.
\end{equation*}
Then, using the It\^{o}'s lemma, we have
\begin{eqnarray*}
R_{i}(l)&=&\frac{a_{i}}{(l+2)!} + \frac{\lambda_{d-1,i}\left( \btheta _{i}\right)}{l!(l+2)}+\sum ^{p+r}_{j=1}\zeta _{1,i,j}\int ^{d}_{d-1}\dfrac{\left( d-t\right) ^{l+1}}{\left( l+1\right) !}\lambda _{t,j}\left( \btheta _{j}\right) dt \cr
&&+\sum ^{h}_{k=2}\sum ^{p+r}_{j=1}\frac{\zeta_{k,i,j}}{(l+2)!}\int ^{d-k+1}_{d-k}\lambda _{t,j}\left( \btheta _{j}\right) dt+\int ^{d}_{d-1}\dfrac{\left( d-t\right) ^{l+2}}{l!\left( l+2\right) }z_{i,t}^{2} dt \cr
&&+\int ^{d}_{d-1} \frac{\left( d-t\right) ^{l+2}}{ l! (l+2)} J_{\lambda,i} (t) d \tilde{\Lambda}_{\lambda, i} (t) + 2\int ^{d}_{d-1}\dfrac{\left( d-t\right) ^{l+2}}{l!\left( l+2\right) }\int ^{t}_{d-1}z_{i,s}dW_{i,s} z_{i,t}dW_{i,t} \cr
&=&\frac{a_{i}}{(l+1)!} + \sum ^{h-1}_{k=1}\sum ^{p+r}_{j=1}\left( \dfrac{\zeta _{k,i,j}}{l!\left( l+2\right) }+\dfrac{\zeta _{k+1,i,j}}{\left( l+2\right) !}\right)\int ^{d-k}_{d-k-1}\lambda _{t,j}\left( \btheta _{j}\right) dt \cr
&&+\sum ^{p+r}_{j=1}\dfrac{\zeta _{h,i,j}}{l!\left( l+2\right)}\int ^{d-h}_{d-h-1}\lambda _{t,j}\left( \btheta _{j}\right) dt+\int ^{d}_{d-1}\dfrac{\left( d-t\right) ^{l+2}}{l!\left( l+2\right) }z_{i,t}^{2} dt\cr
&&+2\int ^{d}_{d-1}\dfrac{\left( d-t\right) ^{l+2}}{l!\left( l+2\right) }\int ^{t}_{d-1}z_{i,s}dW_{i,s} z_{i,t}dW_{i,t} +\int ^{d}_{d-1} \frac{\left( d-t\right) ^{l+2}}{l! (l+2)} J_{\lambda,i} (t) d \tilde{\Lambda}_{\lambda, i} (t) \cr
&&+\sum ^{p+r}_{j=1}\zeta _{1,i,j}R_{j}(l+1) \text{ a.s.}
\end{eqnarray*}
Thus, we have
\begin{eqnarray*}
\bR(l)&=&\frac{\ba}{(l+1)!}+\sum ^{h-1}_{k=1}\left( \dfrac{\bzeta _{k}}{l!\left( l+2\right) }+\dfrac{\bzeta _{k+1}}{\left( l+2\right) !}\right)\int ^{d-k}_{d-k-1}\blambda _{t}\left( \btheta\right) dt +\dfrac{\bzeta _{h}}{l!\left( l+2\right) }\int ^{d-h}_{d-h-1}\blambda _{t}\left( \btheta\right) dt\cr
&&+ \Biggl[\int ^{d}_{d-1}\dfrac{\left( d-t\right) ^{l+2}}{l!\left( l+2\right) }z_{i,t}^{2} dt +\int ^{d}_{d-1} \frac{\left( d-t\right) ^{l+2}}{l! (l+2)} J_{\lambda,i} (t) d \tilde{\Lambda}_{\lambda, i} (t) \cr
&& \qquad \qquad \qquad \qquad \qquad \qquad \qquad +2\int ^{d}_{d-1}\dfrac{\left( d-t\right) ^{l+2}}{l!\left( l+2\right) }\int ^{t}_{d-1}z_{i,s}dW_{i,s} z_{i,t}dW_{i,t} \Biggr]^{\top}_{i=1,\ldots, p+r}\cr
&&+\bzeta_{1}\bR(l+1) \text{ a.s.},
\end{eqnarray*}
where $\ba = \left(a_{1}, \ldots , a_{p+r}\right)^{\top}$.
Define
\begin{equation*}
\bpi_{1}=\sum^{\infty}_{l=0}\dfrac{\bzeta_{1}^{l}}{(l+1)!}  =\bzeta^{-1}_{1}\left( e^{\bzeta_{1}}-\bI_{p+r}\right)  \quad \text{ and } \quad \bpi_{2}=\sum^{\infty}_{l=0}\dfrac{\bzeta_{1}^{l}}{(l+2)!}= \bzeta^{-2}_{1} ( e^{\bzeta_{1}}-\bI_{p+r}-\bzeta_{1} ),
\end{equation*}
where $\bI_{p+r}$ is the $(p+r)$-dimensional identity matrix and $e^{\bzeta_{1}}=\sum ^{\infty }_{l=0}\bzeta^{l}_{1}/l!$.
Then,  iterativing the above formula, we have
\begin{eqnarray*}
\bR(0)&=&\int ^{d}_{d-1}\blambda _{t}\left( \btheta\right) dt \cr
&=&\bpi_{1}\ba+\sum^{h-1}_{k=1}\left((\bpi_{1}-\bpi_{2})\bzeta_k+\bpi_{2}\bzeta_{k+1} \right)\int ^{d-k}_{d-k-1}\blambda _{t}\left( \btheta\right) dt +(\bpi_{1}-\bpi_{2})\bzeta_h \int ^{d-h}_{d-h-1}\blambda _{t}\left( \btheta\right) dt \cr
&&+\sum^{\infty}_{l=0}\bzeta_{1}^{l}\Biggl[\int ^{d}_{d-1}\dfrac{\left( d-t\right) ^{l+2}}{l!\left( l+2\right) }z_{i,t}^{2} dt +\int ^{d}_{d-1} \frac{\left( d-t\right) ^{l+2}}{l! (l+2)} J_{\lambda,i} (t) d \tilde{\Lambda}_{\lambda, i} (t)  \cr
&&	\qquad \qquad \qquad \qquad \qquad \qquad  +2\int ^{d}_{d-1}\dfrac{\left( d-t\right) ^{l+2}}{l!\left( l+2\right) }\int ^{t}_{d-1}z_{i,s}dW_{i,s} z_{i,t}dW_{i,t} \Biggr]^{\top}_{i=1,\ldots, p+r} \cr
&=& \bnu + \sum ^{h}_{k=1}\bA_{k}\bxi_{d-k}     +\bepsilon_{d} \quad  \text{ a.s.},
\end{eqnarray*}
where
\begin{eqnarray} \label{eq-A.2}
	&&  \bA_{k} = ( (\bpi_{1}-\bpi_{2})\bzeta_k+\bpi_{2}\bzeta_{k+1}  )  \text{ for }  1\leq k \leq h-1, \cr
	&& \bA_{h}=(\bpi_{1}-\bpi_{2})\bzeta_h, \cr
    &&\bnu=\bpi_{1}\ba + \sum ^{\infty }_{l=0}\bzeta_{1}^{l}\Biggl[\int ^{d}_{d-1}\left( \frac {\left( d-t\right) ^{l+2}}{\left( l+1\right) !}-\frac {\left( d-t\right) ^{l+2}}{\left( l+2\right) !}\right) \E\left[z^2_{i,t}\right]dt \Biggr]^{\top}_{i=1,\ldots, p+r}, \quad \text{ and} \cr
    &&\bepsilon_{d}=\sum ^{\infty }_{l=0}\bzeta_{1}^{l}\Biggl[\int ^{d}_{d-1}\left( \frac {\left( d-t\right) ^{l+2}}{\left( l+1\right) !}-\frac {\left( d-t\right) ^{l+2}}{\left( l+2\right) !}\right) \left(z^2_{i,t}-\E\left[z^2_{i,t}\right]\right)dt +\int ^{d}_{d-1} \frac{\left( d-t\right) ^{l+1}}{(l+1)!} J_{\lambda,i} (t) d \tilde{\Lambda}_{\lambda, i} (t)  \cr
    &&\qquad \qquad \quad  + 2\int ^{d}_{d-1}\left( \frac {\left( d-t\right) ^{l+2}}{\left( l+1\right) !}-\frac {\left( d-t\right) ^{l+2}}{\left( l+2\right) !}\right)\int ^{t}_{d-1}z_{i,s}dW_{i,s} z_{i,t}dW_{i,t}
 \Biggr]^{\top}_{i=1,\ldots, p+r}.
\end{eqnarray}
\endpf

\subsection{A  simulation setup}\label{simulation-setup}

We considered the following jump diffusion process that satisfies the FIVAR($h$) model:
\begin{eqnarray*}
&&  d\bX(t) = \bQ_{F}\blambda_{F}^{1/2}(t)d\bW(t)+\bQ_{I}\blambda_{I}^{1/2}(t)\bW^{\ast}(t)+\bJ(t)d\bLambda(t), \cr
&&\lambda _{t,i}\left( \btheta _{i}\right)=(\lceil t\rceil -t)\lambda _{\lceil t-1\rceil,i}\left( \btheta _{i}\right)
   +\sum ^{p+r}_{j=1}\zeta_{1,i,j}\left\{ \int ^{t}_{\lceil t-1\rceil}\lambda _{s,j}\left( \btheta _{j}\right) ds\right\} +(\lceil t\rceil -t)   \int ^{t}_{\lceil t-1\rceil} J_{\lambda,i} (s) d \tilde{\Lambda}_{\lambda, i} (s)\cr
&& \qquad \qquad \quad +(t-\lceil t\rceil +1)\left(a_{i}+\sum ^{h}_{k=2}\sum ^{p+r}_{j=1}\zeta_{k,i,j}\left\{ \int ^{\lceil t\rceil - k +1}_{\lceil t\rceil - k}\lambda _{s,j}\left( \btheta _{j}\right) ds\right\}\right)+(\lceil t\rceil -t)Z^{2}_{i,t},
\end{eqnarray*}
where $\blambda_{F}(t)=\Diag(p\lambda_{t,1}(\btheta _{1}),...,p\lambda_{t,r}(\btheta _{r}))$, $\blambda_{I}(t)=\Diag(\lambda_{t,r+1}(\btheta_{r+1}),...,\lambda_{t,p+r}(\btheta _{p+r}))$, and $\bW(t)$ and $\bW^{\ast}(t)$ are $r$-dimensional and $p$-dimensional independent Brownian motions, respectively, $\bJ(t) = (J_1(t), \ldots, J_p(t))^\top$ is the jump size vector, and $\bLambda(t) = (\Lambda_1(t), \ldots, \Lambda_p(t))^\top$ is the Poisson process with intensity $\bI(t) =  (5, \ldots, 5)^\top$.
The jump size $J_i(t)$  was obtained from the independent Gaussian distribution with mean zero and standard deviation $0.05 \sqrt{ \int _{0}^1  \gamma _{ii}(t)  dt }$.
For $t \in [d-1, d)$, $i=1, \ldots, p+r$, and $d=1,\ldots,n$, we set $J_{\lambda,i} (t)=J_{\lambda,i} (d-1)$, and $J_{\lambda,i} (d)$'s were generated from independent unif($-a_i/100, a_i/100$).
Also, the compensated Poisson process $\tilde{\Lambda}_{\lambda, i} (t)$ has the intensity  $I_{\lambda,i}(t) =  (10, \ldots, 10)^\top$.
To obtain the eigenvector matrix for the factor part, $\bQ_{F}$, we first generated the symmetric $p$ by $p$ matrix whose elements were obtained from  i.i.d. unif(0, 1).
Then, we chose its first $r$ eigenvectors as $\bQ_{F}$.
We chose the eigenvector matrix for the idiosyncratic part, $\bQ_{I}$, as the $p$-dimensional identity matrix.
For $t \in [d-1, d)$, $i=1, \ldots, p+r$, and $d=1,\ldots,n$, we set $z_{i,t}=z_{i,d-1}$.
Let $v_1=0.6$, $v_i=0.3$ for $i=2, \ldots, r$, and $v_i=0.1$ for $i=r+1, \ldots, p+r$.
For the heavy-tailed process, $z_{i,d}$'s  were obtained from $v_i$ times independent t-distribution with degrees of freedom $9$,  while for the sub-Gaussian process, $z_{i,d}$'s  were generated from  $v_i$ times independent unif($-2, 2$).
We chose  $p=200$, $r=3$, $h=1$, $m^{all} = 2000$, and we varied $m$ from $250$ to $2000$.
The model parameters are chosen as follows.
We set $a_1=0.8$, $a_2=a_3=0.6$, $a_i=(14-i)/10$ for $4\leq i \leq 13$, $a_i=0.1$ for $14\leq i\leq 203$,
\begin{eqnarray*}
&& (\zeta_{1,i,j})_{1\leq i,j \leq 3} = \begin{pmatrix} 0.5 & 0.15 & 0 \\ 0 & 0.45 & 0.1 \\ 0 & 0 & 0.4 \end{pmatrix},  \quad (\zeta_{1,i,j})_{2k\leq i,j \leq 2k+1} = \begin{pmatrix} 0.19-0.02k & 0.02 \\ 0.02 & 0.18-0.02k \end{pmatrix}
\end{eqnarray*}
for $2 \leq k \leq 6$, and $(\zeta_{1,i,j})_{14\leq i,j \leq 203}$ as 0.05 times $190$-dimensional identity matrix.
Other elements of $(\zeta_{1,i,j})$ were set as zero.
We took $\bX(0)=(0, \ldots, 0)^\top$ and $\lambda_{0,i}(\btheta _{i})=\E(\lambda_{1,i}(\btheta_{i}))$.

We calculated the jump adjusted pre-averaging realized volatility matrix (PRVM) estimator \citep{ait2016increased, christensen2010pre, jacod2009microstructure} as follows:
\begin{equation}\label{eq-A.1}
\hat{\Gamma}_{d,ij}=\frac{1}{\psi K} \sum^{m-K+1}_{k=1}\left\{\bar{Y}_i\left(t_{d,k}\right)\bar{Y}_j\left(t_{d,k}\right)-\frac{1}{2}\hat{Y}_{i,j}\left(t_{d,k}\right) \right\}\mathbf{1}\left\{ \left| \bar{Y}_{i}\(t_{d,k}\) \right| \leq u_{i,m}\right\}\mathbf{1}\left\{ \left| \bar{Y}_{j}\(t_{d,k}\) \right| \leq u_{j,m}\right\},
\end{equation}
where
\begin{eqnarray*}
	&&\bar{Y}_i\left(t_{d,k}\right)=\sum^{K-1}_{l=1}g\left(\frac{l}{K}\right)\left(Y_i(t_{d,k+l})-Y_i(t_{d,k+l-1})\right), \cr
	&&\hat{Y}_{i,j}\left(t_{d,k}\right) =\sum^{K}_{l=1}\Bigg[\left\{g\left(\frac{l}{K}\right) - g\left(\frac{l-1}{K}\right) \right\}^2 \cr
	&& \qquad \qquad \qquad  \qquad \times  \left(Y_i(t_{d,k+l-1})-Y_i(t_{d,k+l-2})\right)\left(Y_j(t_{d,k+l-1})-Y_j(t_{d,k+l-2})\right)\Bigg],
\end{eqnarray*}
$\psi=\int_0^1 g\left(t\right)^2 dt$, $\mathbf{1}\left\{\cdot \right\}$ is an indicator function, and $u_{i,m}$ is a truncation parameter.
We chose the bandwidth parameter $K = \lfloor m^{1/2} \rfloor$ and weight function $g\left(x\right) = x \wedge \left(1 - x\right)$.
For the choice of the truncation parameter, we followed the approach of \citet{ait2016increased}.
Specifically, we set 
\begin{equation*}
u_{i,m}= c_{0} \sqrt{\hat{T}_{i}}\(\dfrac{K}{m}\)^{\alpha_{u}},
\end{equation*} 
where $\alpha_{u}$ and $c_0$ are tuning parameters, and $\hat{T}_{i}$ is an estimator for
\begin{equation*}
T_i=\psi \int _{d-1}^d  \gamma _{ii}(t)  dt + \dfrac{m}{K^2}\E \left\{ e_i(t_{d,k})\right\}^2 \int_0^1 g'\left(t\right)^2 dt.
 \end{equation*}
 \citet{ait2016increased} chose $\alpha_{u}=0.47$ and recommended setting $c_0$ between 2 and 4, which is also supported by \citet{ait2014high}. 
In the numerical study, we chose $\alpha_{u}=0.47$ and $c_0$=4.
This choice is the same as in \citet{oh2024robust}.
To estimate ${T}_{i}$, we employed the following estimator \citep{christensen2014fact}:
\begin{equation*}
 \hat{T}_{i} = \dfrac{m}{m-2K+1} \cdot \dfrac{\pi}{2 K} \sum^{m-2K+1}_{k=1} \left| \bar{Y}_i\left(t_{d,k}\right) \right| \left| \bar{Y}_i\left(t_{d,k+K}\right) \right|,
 \end{equation*}
which is shown to converge to $T_i$.

\subsection{Empirical study for the constant eigenvector hypothesis}\label{empirical-study1}
\begin{figure}[!ht]
	\centering
	\includegraphics[width = 0.7\textwidth]{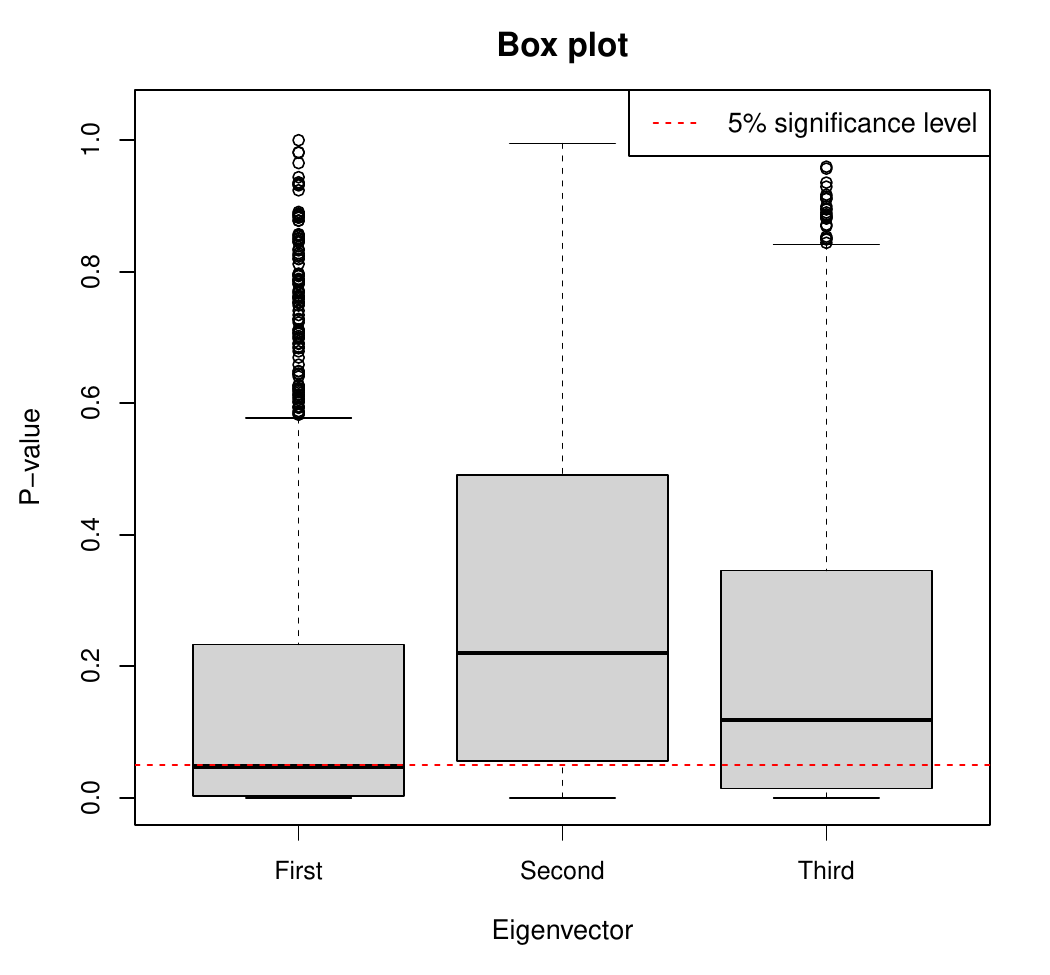}
	\caption{The boxplots of the p-values for the constant eigenvector hypothesis for three principal eigenvectors over 1498 days from 2015 to 2020.
We used 5-min log-returns of the top 200 large trading volume assets in the S\&P 500 index.
The red dash represents the 5\% significance level.}\label{Fig-6}
\end{figure}

In this section, we conducted a hypothesis test for the constant eigenvector assumption based on the procedure in \citet{fan2024tests}.
For each day, we first splited the return data into two groups as follows:
\begin{equation*}
\Delta \bY^{(1)}_{1}, \ldots, \Delta \bY^{(1)}_{T_1}, \quad \text{and} \quad \Delta \bY^{(2)}_{1}, \ldots, \Delta \bY^{(2)}_{T_2},
\end{equation*}
where $\Delta \bY^{(j)}_{i} = \(\Delta Y^{(j)}_{1,i}, \ldots, \Delta Y^{(j)}_{p,i}\)^{\top}$, $\Delta Y^{(j)}_{k,i}$ is the $i$-th return for the $k$-th asset and $j$-th group, and  $T_1$ and $T_2$ are the numbers of returns for each group.  
With these return data, we obtained the sample covariance matrices as follows:
\begin{equation*}
\hat{\bGamma}^{(j)}=\dfrac{1}{T_j}\sum_{i=1}^{T_j} \Delta \bY^{(j)}_{i} (\Delta \bY^{(j)}_{i})^{\top}  \quad \text{for } j=1,2.
\end{equation*}
Then, we decomposed the above sample covariance matrices and obtained eigenvalue and eigenvector estimates for the two groups.
With these estimates, we conducted the hypothesis test for the constancy of the eigenvector process.
The null hypothesis is that the eigenvectors of the two groups are the same. 
Detailed test procedure is presented in Section 4.3 in \citet{fan2024tests}.
The tests were conducted for three principal eigenvectors over 1498 days from 2015 to 2020. 
In this paper, to mitigate the effect of noises, we used 5-min log-returns and chose $T_1 = T_2 =38$.
As shown in Figure \ref{Fig-6}, we found that the constant eigenvector hypothesis is often rejected at 5\% significance level. 
%This implies that it is also important to handle the intraday time-varying eigenvectors.

{
\subsection{Empirical study for the structural break test}\label{empirical-study2}
In this section, we conducted the structural break test for the eigenvalue process based on the procedure in \citet{bai2003computation}.
Specifically, we considered the following linear regression model with $L$ breaks:
\begin{equation*}
1= \bb_k \hat{\bxi}_d^{\top}  + e_d, \quad d=d_{k-1}+1, \ldots, d_{k}, \quad k=1, \ldots, L+1,
\end{equation*}
where $k$ denotes the segment index, $\bb_k$ is the regression coefficient for the $k$-th segment, $\hat{\bxi}_{d}=\left(\hat{\xi}_{d,1}, \ldots, \hat{\xi}_{d,p+r}\right)^{\top}$ is a non-parametric integrated eigenvalue estimator for ${\bxi}_{d}$ whose estimation procedure is presented in Section \ref{SEC-4}, $e_d$ is a error term for the $d$-th day, $d_{0}=0$, and $d_{L+1}$ represents the end date of the data. 
The dependent variable was set as $1$ to investigate the break in the mean of the eigenvalue process.
For each $L=1, \ldots, 5$, we estimated the breakpoints, $d_{k}, k=1, \ldots, L$, by minimizing the residual sum of squares (RSS). 
Then, we chose $L \in \{0, \ldots, 5\}$ that minimizes the corresponding Bayesian information criterion (BIC).
Details can be found in \citet{bai2003computation}.
The result is $L=1$ with the breakpoint $d_1$=1044 that corresponds to March 23, 2020.
This may be due to the covid sell-off in 2020.
Thus, in the empirical study, we excluded the year $2020$  to avoid the non-stationarity.
We note that the proposed FIVAR model is based on the stationary condition and it is hard to apply the parametric model to the non-stationary period.
%We also note that the data after 2020 was excluded due to the data availability.
}

\subsection{Proof of Theorem \ref{Theorem1}}
We note that the model \eqref{eq-2.3} can be written in a VAR(1) form as follows:
\begin{equation*}
	\tilde{\bxi}_{d}=\tilde{\bnu} + \tilde{\bA}\tilde{\bxi}_{d-1}+ \tilde{\bepsilon}_{d},
\end{equation*}
where
\begin{eqnarray}\label{eq-A.3}
&&\tilde{\bxi}_{d}=\begin{pmatrix} {\bxi}_{d} \\ {\bxi}_{d-1} \\ \vdots \\ {\bxi}_{d-h+1} \end{pmatrix}_{h\(p+r\) \times 1}, \quad \tilde{\bnu}=\begin{pmatrix} \bnu \\ 0 \\ \vdots \\ 0 \end{pmatrix}_{h\(p+r\) \times 1}, \cr
&&\tilde{\bA}=\begin{pmatrix} \bA_{1} & \bA_{2} & \cdots & \bA_{h-1} & \bA_{h} \\ \bI_{p+r} & 0 & 0 & 0 & 0 \\ 0 & \bI_{p+r} & & 0 & 0 \\ \vdots & & \ddots & \vdots & \vdots \\ 0 & 0 & \cdots & \bI_{p+r} & 0 \end{pmatrix}_{h\(p+r\) \times h\(p+r\)}, \quad \tilde{\bepsilon}_{d}=\begin{pmatrix} \bepsilon_{d} \\ 0 \\ \vdots \\ 0 \end{pmatrix}_{h\(p+r\) \times 1}.
\end{eqnarray}

\begin{lemma}\label{Lemma1}
Under the model \eqref{eq-2.3} and Assumption \ref{Assumption1}(a)--(b), suppose that $s_{\beta}$ is bounded by some positive constant and $\max_{1\leq i \leq p+r}   \E (|\epsilon_{d,i}|^{c_{\epsilon}})<\infty$. 
Then, we have $\max_{1\leq i \leq p+r}\E (|\xi_{d,i}|^{c_{\epsilon}})<\infty$.
\end{lemma}

\textbf{Proof of Lemma \ref{Lemma1}.}
Since $\rho(\tilde{\bA})<1$, by Gelfand's formula,  we have
\begin{equation*}
\lim_{l\rightarrow \infty }\left\| \tilde{\bA}^{l}\right\|_{\infty }^{1/l}=\rho(\tilde{\bA}) <1.
\end{equation*}
Thus, there exists a positive integer $k$ such that $\left\| {\tilde{\bA}}^{k}\right\|_{\infty }<1$.
 Note that for any fixed matrix $\bM \in \mathbb{R}^{p_1 \times p_2}$ and multivariate random variable $\bx \in \mathbb{R}^{p_2}$, we have
\begin{equation*}
\sup_{j\leq p_1}\left\|(\bM\bx)_j \right\|_{L_{c_{\epsilon}}} \leq \left\|\bM\right\|_{\infty}\sup_{j\leq p_1}\left\|(\bx)_j \right\|_{L_{c_{\epsilon}}},
\end{equation*}
where for any vector $\bx$, $(\bx)_j$ is the $j$-th element of $\bx$.
Hence, by the fact that
\begin{equation*}
\tilde{\bxi}_{d}=\tilde{\bA}^{k}\tilde{\bxi}_{d-k}+\sum^{k}_{i=1}\tilde{\bA}^{k-i}\left(\tilde{\bnu}+\tilde{\bepsilon}_{d-k+i} \right),
\end{equation*}
we have
\begin{eqnarray*}
&& \sup_{j\leq h(p+r)}\left\|\left(\tilde{\bxi}_{d}\right)_j \right\|_{L_{c_{\epsilon}}} \cr
&& \leq \left\| {\tilde{\bA}}^{k}\right\|_{\infty }\sup_{j\leq h(p+r)}\left\|\left(\tilde{\bxi}_{d-k}\right)_j  \right\|_{L_{c_{\epsilon}}}
+\sum^{k}_{i=1}\left\| {\tilde{\bA}}^{k-i}\right\|_{\infty }\sup_{j\leq h(p+r)}\left\|\left(\tilde{\bnu}+\tilde{\bepsilon}_{d-k+i}\right)_j \right\|_{L_{c_{\epsilon}}}.
\end{eqnarray*}
Then, we have
\begin{equation}\label{eq-A.4}
\sup_{j\leq h(p+r)}\left\|\left(\tilde{\bxi}_{d}\right)_j  \right\|_{L_{c_{\epsilon}}} \leq \frac{\sup_{j\leq h(p+r)}\left\|\left(\tilde{\bnu}+\tilde{\bepsilon}_{d}\right)_j \right\|_{L_{c_{\epsilon}}}}{1-\left\| {\tilde{\bA}}^{k}\right\|_{\infty }}\left(\sum^{k}_{i=1}\left\| {\tilde{\bA}}^{k-i}\right\|_{\infty }\right) \leq C,
\end{equation}
where the first inequality is from the stationarity and the last inequality is due to the boundedness of $s_{\beta}$.
\endpf

\begin{proposition}\label{Proposition3}
Under the assumptions in Theorem \ref{Theorem1}, we have for $i \in \left\{1, \ldots, r \right\}$, with probability at least $1-3(hr+1)e^{-\delta}$,
\begin{equation}\label{eq-A.5}
\left\|\nabla {\mathcal{L}}^{F,i}_{\tau,\varpi}(\bbeta_{i0})\right\|_{\infty} \leq \eta_{F}/2.
\end{equation}
Also, we have for  $i \in \left\{r+1, \ldots, p+r \right\}$, with probability at least $1-3\(h\(p+r\)+1\)e^{-\delta}$,
\begin{equation}\label{eq-A.6}
\left\|\nabla {\mathcal{L}}^{I,i}_{\tau,\varpi}(\bbeta_{i0})\right\|_{\infty} \leq \eta_{I}/2.
\end{equation}
\end{proposition}

\textbf{Proof of Proposition \ref{Proposition3}.}
For the simplicity, we assume that $h=1$ and omit the intercept term $\bnu$.
Due to the similarity, we only provide the arguments for $\left\|\nabla {\mathcal{L}}^{I,i}_{\tau,\varpi}(\bbeta_{i0})\right\|_{\infty}$.
Note that $\hat{\bxi}_{d}^{I}=(\hat{\xi}_{d,1}, \ldots, \hat{\xi}_{d,p+r})^{\top}$. For each $1 \leq j \leq p+r$, we have
\begin{equation}\label{eq-A.7}
\left|\nabla_j {\mathcal{L}}^{I,i}_{\tau,\varpi}(\bbeta_{i0})\right|=\left|\dfrac{\partial {\mathcal{L}}^{I,i}_{\tau,\varpi}(\bbeta_{i0})}{\partial \bbeta _{j}}\right| \leq (\uppercase\expandafter{\romannumeral1})_{j} +(\uppercase\expandafter{\romannumeral2})_{j},
\end{equation}
where
\begin{equation*}
(\uppercase\expandafter{\romannumeral1})_{j} = \frac{1}{n-1}\left|\sum ^{n}_{d=2}\psi_{\tau_{I}}\left( \xi_{d,i}-\langle  \psi_{\varpi_{I}}(\bxi_{d-1}), \bbeta_{i0} \rangle \right)\psi_{\varpi_{I}}\left( \xi_{d-1,j}\right)\right|,
\end{equation*}
\begin{eqnarray*}
(\uppercase\expandafter{\romannumeral2})_{j} &=& \frac{1}{n-1}\Biggl|\sum ^{n}_{d=2}\Biggl[\psi_{\tau_{I}}\left( \hat{\xi}_{d,i}-\langle  \psi_{\varpi_{I}}(\hat{\bxi}^{I}_{d-1}), \bbeta_{i0}\rangle \right)\psi_{\varpi_{I}}\left( \hat{\xi}_{d-1,j}\right) \cr
&&\qquad \qquad \quad -\psi_{\tau_{I}}\left(\xi_{d,i}-\langle  \psi_{\varpi_{I}}(\bxi_{d-1}), \bbeta_{i0} \rangle \right)\psi_{\varpi_{I}}\left( \xi_{d-1,j}\right)\Biggr]\Biggr|.
\end{eqnarray*}
We first consider $(\uppercase\expandafter{\romannumeral1})_{j}$.
Let $\by_{d}=(y_{d,1}, \ldots, y_{d,p+r})^{\top}$, $y_{d,k}$ be the $k$-th element of   $\bxi_{d}-\psi_{\varpi_{I}}(\bxi_{d})$ for $k \in S_i$, and  $y_{d,k}=0$ for $k \in S_{i}^{c}$, where $S_i$ is defined in Assumption \ref{Assumption1}(g).
Also, let $\epsilon'_{d,i}=\epsilon_{d,i}+\langle \by_{d-1}, \bbeta_{i0}\rangle$.
Then, we have
\begin{equation*}
(\uppercase\expandafter{\romannumeral1})_{j} = \frac{1}{n-1}\left|\sum ^{n}_{d=2}\psi_{\tau_{I}}\left( \epsilon'_{d,i} \right)\psi_{\varpi_{I}}\left( \xi_{d-1,j}\right)\right| \leq (\uppercase\expandafter{\romannumeral1})_{j}^{(1)}+(\uppercase\expandafter{\romannumeral1})_{j}^{(2)},
\end{equation*}
where
\begin{equation*}
(\uppercase\expandafter{\romannumeral1})_{j}^{(1)} = \frac{1}{n-1}\sum ^{n}_{d=2}\left|\E \left\{\psi_{\tau_{I}}\left( \epsilon'_{d,i} \right)\psi_{\varpi_{I}}\left( \xi_{d-1,j}\right)\right\}\right|,
\end{equation*}
\begin{equation*}
(\uppercase\expandafter{\romannumeral1})_{j}^{(2)} = \frac{1}{n-1}\left|\sum ^{n}_{d=2}\left[\psi_{\tau_{I}}\left( \epsilon'_{d,i} \right)\psi_{\varpi_{I}}\left( \xi_{d-1,j}\right)-\E \left\{\psi_{\tau_{I}}\left( \epsilon'_{d,i} \right)\psi_{\varpi_{I}}\left( \xi_{d-1,j}\right)\right\}\right]\right|.
\end{equation*}
For $(\uppercase\expandafter{\romannumeral1})_{j}^{(1)}$, we have
\begin{equation*}
\E \left\{\psi_{\tau_{I}}\left( \epsilon'_{d,i} \right)\psi_{\varpi_{I}}\left( \xi_{d-1,j}\right)\right\}=\E \left\{\langle \by_{d-1},  \bbeta_{i0}\rangle \psi_{\varpi_{I}}\left( \xi_{d-1,j}\right) \right\}-\E\left\{\left[\epsilon'_{d,i}- \psi_{\tau_{I}}\left(\epsilon'_{d,i} \right)\right]\psi_{\varpi_{I}}\left( \xi_{d-1,j}\right) \right\}.
\end{equation*}
Let $ v_{3}= \max_{1\leq k \leq p+r}   \E (|\epsilon_{d,k}|^{3})$,   $K_{2}= \max_{1\leq k \leq p+r}   \E (|\xi_{d,k} |^{2}) $, and $K_{4}= \max_{1\leq k \leq p+r}   \E (|\xi_{d,k} |^{4})$.
Since
\begin{eqnarray*}
 \left|y_{d-1,k}  \right| \leq \left|\xi_{d-1,k}  \right| 1\left(\left|\xi_{d-1,k}  \right|> \varpi_{I} \right)\leq \varpi_{I}^{-2}\left|\xi_{d-1,k}  \right|^3 \text{ a.s.}
\end{eqnarray*}
for $k \in \left\{1, \ldots, p+r \right\}$ and
\begin{eqnarray*}
 \left|\epsilon'_{d,i}- \psi_{\tau_{I}}\left(\epsilon'_{d,i} \right) \right| \leq \left|\epsilon'_{d,i}  \right| 1\left(\left|\epsilon'_{d,i}  \right|> \tau_{I} \right)\leq \tau_{I}^{-2}\left|\epsilon'_{d,i}  \right|^3  \text{ a.s.},
\end{eqnarray*}
we have
\begin{eqnarray}\label{eq-A.8}
&& \left|\E \left\{\psi_{\tau_{I}}\left( \epsilon'_{d,i} \right)\psi_{\varpi_{I}}\left( \xi_{d-1,j}\right)\right\} \right| \cr
&&\leq  \left|\E \left\{  \langle \by_{d-1}, \bbeta_{i0}\rangle  \psi_{\varpi_{I}}\left( \xi_{d-1,j}\right)\right\} \right| + \tau_{I}^{-2} \E \left\{\left|\epsilon'_{d,i}  \right|^3 \left|\psi_{\varpi_{I}}\left( \xi_{d-1,j}\right)\right|  \right\} \cr
&&\leq \left\| \bbeta_{i0}\right\| _{1}K_4\varpi_{I}^{-2}+\tau_{I}^{-2} \E \left\{\left|\epsilon'_{d,i}  \right|^3 \left|\psi_{\varpi_{I}}\left( \xi_{d-1,j}\right)\right|  \right\} \cr
&&\leq \left\|  \bbeta_{i0}\right\| _{2}K_4 s_{\beta}^{1/2} \varpi_{I}^{-2} +4\tau_{I}^{-2}\left[ \E \left\{\left|\epsilon_{d,i}  \right|^3 \left|\psi_{\varpi_{I}}\left( \xi_{d-1,j}\right)\right|  \right\}+\left\|  \bbeta_{i0}\right\|_{2}^{3}\E \left\{\left\|  \by_{d-1}\right\|_{2}^{3} \left|\psi_{\varpi_{I}}\left( \xi_{d-1,j}\right)\right|  \right\}\right] \cr
&&\leq \left\|  \bbeta_{i0}\right\| _{2}K_4 s_{\beta}^{1/2} \varpi_{I}^{-2} +4\tau_{I}^{-2}\left\{v_{3}K_2^{1/2}+\left\|  \bbeta_{i0}\right\|_{2}^{3}K_4 s_{\beta}^{3/2} \right\}.
\end{eqnarray}
Thus, we have
\begin{equation}\label{eq-A.9}
(\uppercase\expandafter{\romannumeral1})_{j}^{(1)} \leq C\left( s_{\beta} \varpi_{I}^{-2} +s_{\beta}^3\tau_{I}^{-2}\right).
\end{equation}
For $(\uppercase\expandafter{\romannumeral1})_{j}^{(2)}$, note that the process $(\tilde{\bxi}_{d})_{d=1,2, \ldots}$ is geometrically $\alpha$-mixing and $\tilde{\bepsilon}_{d}= \tilde{\bxi}_{d} - \tilde{\bA}\tilde{\bxi}_{d-1}$.
Since each $\psi_{\tau_{I}}\left( \epsilon'_{d,i} \right)\psi_{\varpi_{I}}\left( \xi_{d-1,j}\right)-\E \left[\psi_{\tau_{I}}\left( \epsilon'_{d,i} \right)\psi_{\varpi_{I}}\left( \xi_{d-1,j}\right)\right]$ is a measurable function of $\tilde{\bxi}_{d}$ and $\tilde{\bxi}_{d-1}$, $\left\{\psi_{\tau_{I}}\left( \epsilon'_{d,i} \right)\psi_{\varpi_{I}}\left( \xi_{d-1,j}\right)-\E \left[\psi_{\tau_{I}}\left( \epsilon'_{d,i} \right)\psi_{\varpi_{I}}\left( \xi_{d-1,j}\right)\right]\right\}$ is also geometrically $\alpha$-mixing with the coefficients satisfying Assumption \ref{Assumption1}(d).
Therefore, by applying Theorem 2 in \citet{merlevede2009bernstein}, we have, for $t \geq 0$,
\begin{equation}\label{eq-A.10}
\Pr \left\{ (\uppercase\expandafter{\romannumeral1})_{j}^{(2)}  \geq t \right\} \leq  \exp\left\{-\frac{Cn^{2}t^{2}}{V^{2}n+\tau_{I}^{2}\varpi_{I}^{2}+t\tau_{I}\varpi_{I}n\left(\log n \right)^2}\right\},
\end{equation}
where
\begin{equation*}
V^{2}=\Var\left[\psi_{\tau_{I}}\left( \epsilon'_{2,i} \right)\psi_{\varpi_{I}}\left( \xi_{1,j}\right) \right]+2\sum_{d=3}^{\infty}\left| \Cov\left[\psi_{\tau_{I}}\left( \epsilon'_{2,i} \right)\psi_{\varpi_{I}}\left( \xi_{1,j}\right),\psi_{\tau_{I}}\left( \epsilon'_{d,i} \right)\psi_{\varpi_{I}}\left( \xi_{d-1,j}\right)\right] \right|.
\end{equation*}
Since the $\frac{c_{\epsilon}}{2}$-th moment of $\psi_{\tau_{I}}\left( \epsilon'_{d,i} \right)\psi_{\varpi_{I}}\left( \xi_{d-1,j}\right)$ is bounded by $Cs_{\beta}^{c_{\epsilon}/2}$, by the inequality (2.2) in \citet{davydov1968convergence}, we have
\begin{equation}\label{eq-A.11}
V^{2} \leq Cs_{\beta}^2\sum_{k=1}^{\infty}\varphi^{\left[1-(4/c_{\epsilon}) \right]k} \leq Cs_{\beta}^2,
\end{equation}
which implies
\begin{equation}\label{eq-A.12}
\Pr \left\{ (\uppercase\expandafter{\romannumeral1})_{j}^{(2)} \leq C\left(\frac{\tau_{I}\varpi_{I}\left(\log n \right)^2 \delta + s_{\beta}\sqrt{n\delta}}{n} \right) \right\} \geq 1-e^{-\delta}.
\end{equation}
Combining \eqref{eq-A.9} and \eqref{eq-A.12}, we obtain that with probability at least $1-\(p+r\)e^{-\delta}$,
\begin{equation}\label{eq-A.13}
\max _{1\leq j\leq p+r}(\uppercase\expandafter{\romannumeral1})_{j} \leq C\left(s_{\beta}^3\tau_{I}^{-2} + s_{\beta} \varpi_{I}^{-2} + \frac{\tau_{I}\varpi_{I}\left(\log n \right)^2 \delta  +s_{\beta}\sqrt{n\delta}}{n} \right).
\end{equation}

Now, consider $(\uppercase\expandafter{\romannumeral2})_{j}$.
Note that for any $x,y \in \mathbb{R}$,
\begin{equation*}
\left|\psi_{\varpi_{I}}\left(x\right) - \psi_{\varpi_{I}}\left(y\right) \right| \leq \left| x-y \right|.
\end{equation*}
Hence, by Assumption \ref{Assumption1}(e), we have
\begin{eqnarray*}
&&\left|\psi_{\tau_{I}}\left( \hat{\xi}_{d,i}-\langle  \psi_{\varpi_{I}}(\hat{\bxi}^{I}_{d-1}), \bbeta_{i0} \rangle \right)-\psi_{\tau_{I}}\left(\xi_{d,i}-\langle  \psi_{\varpi_{I}}(\bxi_{d-1}), \bbeta_{i0} \rangle \right) \right|\cr
&&\leq \left| \hat{\xi}_{d,i}-{\xi}_{d,i}- \langle  \psi_{\varpi_{I}}(\hat{\bxi}^{I}_{d-1})-\psi_{\varpi_{I}}({\bxi}_{d-1}), \bbeta_{i0} \rangle\right|\cr
&&\leq C\(b_{m,n,p}^{F}+s_{\beta}b_{m,n,p}^{I}\) \text{ a.s.}
\end{eqnarray*}
Thus, by using the fact that
\begin{equation*}
\left|x_{1}y_{1}-x_{2}y_{2}\right| \leq \left|\(x_{1}-x_{2}\)\(y_{1}-y_{2}\)\right|+ \left|\(x_{1}-x_{2}\)y_{2}\right| +  \left|x_{2}\(y_{1}-y_{2}\)\right|
\end{equation*}
for any $x_{1}, x_{2}, y_{1}, y_{2} \in \mathbb{R}$, we have
\begin{eqnarray}\label{eq-A.14}
(\uppercase\expandafter{\romannumeral2})_{j}  &\leq& C\Big[\(b_{m,n,p}^{F}+s_{\beta}b_{m,n,p}^{I}\)\max\(b_{m,n,p}^{F}, b_{m,n,p}^{I}\)\cr
&&\qquad +  \frac{b_{m,n,p}^{F}+s_{\beta}b_{m,n,p}^{I}}{n-1}\sum^{n}_{d=2} \left\{ \left|\psi_{\varpi_{I}}\left( \xi_{d-1,j}\right) \right|     \right\} \cr
&&  \qquad + \frac{\max\(b_{m,n,p}^{F}, b_{m,n,p}^{I}\)}{n-1}\sum ^{n}_{d=2}\left\{  \left|\psi_{\tau_{I}}\left(\xi_{d,i}-\langle  \psi_{\varpi_{I}}(\bxi_{d-1}), \bbeta_{i0} \rangle \right) \right|  \right\}\Big]\cr
&\leq& C\Big[\(b_{m,n,p}^{F}+s_{\beta}b_{m,n,p}^{I}\)\max\(b_{m,n,p}^{F}, b_{m,n,p}^{I}\) \cr
&& \qquad + \(b_{m,n,p}^{F}+s_{\beta}b_{m,n,p}^{I}\) \left\{(\uppercase\expandafter{\romannumeral2})_{j}^{(1)} + (\uppercase\expandafter{\romannumeral2})_{j}^{(2)} \right\} \cr
&& \qquad + \max\(b_{m,n,p}^{F}, b_{m,n,p}^{I}\)\left\{(\uppercase\expandafter{\romannumeral2})_{j}^{(3)} + (\uppercase\expandafter{\romannumeral2})_{j}^{(4)} \right\}\Big],
\end{eqnarray}
where
\begin{eqnarray*}
&&(\uppercase\expandafter{\romannumeral2})_{j}^{(1)} = \frac{1}{n-1}\left|\sum ^{n}_{d=2}\Big[ \left|\psi_{\varpi_{I}}\left( \xi_{d-1,j}\right) \right| - \E \left|\psi_{\varpi_{I}}\left( \xi_{d-1,j}\right) \right|   \Big]\right|, \cr
&&(\uppercase\expandafter{\romannumeral2})_{j}^{(2)} = \frac{1}{n-1}\sum ^{n}_{d=2} \E \left|\psi_{\varpi_{I}}\left( \xi_{d-1,j}\right) \right|, \cr
&&(\uppercase\expandafter{\romannumeral2})_{j}^{(3)} = \frac{1}{n-1}\Bigg|\sum ^{n}_{d=2}\Big[ \left|\psi_{\tau_{I}}\left(\xi_{d,i}-\langle  \psi_{\varpi_{I}}(\bxi_{d-1}), \bbeta_{i0}\rangle \right) \right| -\E \left|\psi_{\tau_{I}}\left(\xi_{d,i}-\langle  \psi_{\varpi_{I}}(\bxi_{d-1}), \bbeta_{i0} \rangle \right) \right|   \Big]\Bigg|, \cr
&&(\uppercase\expandafter{\romannumeral2})_{j}^{(4)} = \frac{1}{n-1}\sum ^{n}_{d=2} \E \left|\psi_{\tau_{I}}\left(\xi_{d,i}-\langle  \psi_{\varpi_{I}}(\bxi_{d-1}), \bbeta_{i0} \rangle \right) \right|.
\end{eqnarray*}
Consider $(\uppercase\expandafter{\romannumeral2})_{j}^{(1)}$ and $(\uppercase\expandafter{\romannumeral2})_{j}^{(3)}$.
Similar to the proofs of $(\uppercase\expandafter{\romannumeral1})_{j}^{(2)}$, we can show
\begin{equation}\label{eq-A.15}
\Pr \left\{ (\uppercase\expandafter{\romannumeral2})_{j}^{(1)} \leq C\left(\frac{\varpi_{I}\left(\log n \right)^2 \delta + s_{\beta}\sqrt{n\delta}}{n} \right) \right\} \geq 1-e^{-\delta}
\end{equation}
and
\begin{equation}\label{eq-A.16}
\Pr \left\{ (\uppercase\expandafter{\romannumeral2})_{j}^{(3)} \leq C\left(\frac{\tau_{I}\left(\log n \right)^2 \delta+  s_{\beta} \sqrt{n\delta}}{n} \right) \right\} \geq 1-e^{-\delta}.
\end{equation}
Also, we have
\begin{equation}\label{eq-A.17}
(\uppercase\expandafter{\romannumeral2})_{j}^{(2)} \leq C \text{ and }
(\uppercase\expandafter{\romannumeral2})_{j}^{(4)} \leq \frac{1}{n-1} \sum ^{n}_{d=2} \E \left|\xi_{d,i}-\langle  \psi_{\varpi_{I}}(\bxi_{d-1}), \bbeta_{i0} \rangle  \right|   \leq Cs_{\beta}.
\end{equation}
By \eqref{eq-A.14}--\eqref{eq-A.17}, we have
\begin{equation}\label{eq-A.18}
\Pr \left\{\max _{1\leq j\leq p+r}(\uppercase\expandafter{\romannumeral2})_{j} \leq Cs_{\beta}\(b_{m,n,p}^{F}+ b_{m,n,p}^{I}\)\right\} \geq 1-2\(p+r\)e^{-\delta}.
\end{equation}
Combining \eqref{eq-A.7}, \eqref{eq-A.13}, and \eqref{eq-A.18}, we obtain that with probability at least $1-3\(p+r\)e^{-\delta}$,
\begin{eqnarray}\label{eq-A.19}
&&\left\|\nabla {\mathcal{L}}^{I,i}_{\tau,\varpi}(\bbeta_{i0})\right\|_{\infty} \leq C\Bigg[s_{\beta} \(b_{m,n,p}^{F}+ b_{m,n,p}^{I}\)+ s_{\beta}^3\tau_{I}^{-2} \cr
&&\qquad \qquad   \qquad \qquad  \qquad + s_{\beta} \varpi_{I}^{-2} +\frac{\tau_{I}\varpi_{I}\left(\log n \right)^2 \delta +s_{\beta}\sqrt{n\delta}}{n} \Bigg].
\end{eqnarray}
\endpf

\textbf{Proof of Theorem \ref{Theorem1}.}
By Proposition \ref{Proposition3}, we prove the statements under \eqref{eq-A.5} and \eqref{eq-A.6}.
First, we consider $\left\| \hat{\bbeta}_{i} -\bbeta_{i0}\right\|_{2}$ for $i \in \left\{1, \ldots, r \right\}$.
Suppose  that
\begin{equation}\label{eq-A.20}
\left\| \hat{\bbeta}_{i} -\bbeta_{i0}\right\|_{2}  > \frac{\(hr+1\)^{1/2}\eta_F}{\kappa}.
\end{equation}
By the optimality of $\hat{\bbeta}_{i}$ and the integral form of the Taylor expansion, we have
\begin{eqnarray}\label{eq-A.21}
0 &\geq& {\mathcal{L}}^{F,i}_{\tau,\varpi}(\hat{\bbeta}_{i})-{\mathcal{L}}^{F,i}_{\tau,\varpi}(\bbeta_{i0}) \cr
&=& \langle \nabla {\mathcal{L}}^{F,i}_{\tau,\varpi}(\bbeta_{i0}), \hat{\bbeta}_{i}- \bbeta_{i0}\rangle \cr
&& + \int^{1}_{0} \left(1-t\right)(\hat{\bbeta}_{i}- \bbeta_{i0})^{\top}\nabla^{2}{\mathcal{L}}^{F,i}_{\tau,\varpi}(\bbeta_{i0}+t(\hat{\bbeta}_{i}- \bbeta_{i0}))(\hat{\bbeta}_{i}- \bbeta_{i0}) dt.
\end{eqnarray}
Since $\left\| \nabla{\mathcal{L}}^{F,i}_{\tau,\varpi}(\bbeta_{i0})  \right\|_{\infty} \leq \eta_{F}/2$, we have
\begin{eqnarray}\label{eq-A.22}
\left| \langle \nabla {\mathcal{L}}^{F,i}_{\tau,\varpi}(\bbeta_{i0}), \hat{\bbeta}_{i}- \bbeta_{i0}\rangle \right| &\leq& \left\| \nabla{\mathcal{L}}^{F,i}_{\tau,\varpi}(\bbeta_{i0})  \right\|_{\infty} \left\| \hat{\bbeta}_{i}- \bbeta_{i0} \right\|_{1}   \cr
&\leq& \frac{(hr+1)^{1/2}\eta_{F}}{2}\left\| \hat{\bbeta}_{i}- \bbeta_{i0} \right\|_{2}.
\end{eqnarray}
By \eqref{eq-A.20}, we have
\begin{equation*}
z=\frac{(hr+1)^{1/2}\eta_{F}}{\kappa\left\| \hat{\bbeta}_{i}- \bbeta_{i0}\right\|_{2}}<1.
\end{equation*}
Then, for any $0 \leq t \leq z$, we have
\begin{equation*}
\left\|[\bbeta_{i0} +t(\hat{\bbeta}_{i}-\bbeta_{i0} )] -\bbeta_{i0}   \right\|_{1}
\leq t(hr+1)^{1/2}\left\| \hat{\bbeta}_{i}- \bbeta_{i0}\right\|_{2} \leq \frac{(hr+1)\eta_F}{\kappa}.
\end{equation*}
Hence, we have
\begin{eqnarray}\label{eq-A.23}
&& \int^{1}_{0} \left(1-t\right)(\hat{\bbeta}_{i}- \bbeta_{i0})^{\top}\nabla^{2}{\mathcal{L}}^{F,i}_{\tau,\varpi}(\bbeta_{i0}+t(\hat{\bbeta}_{i}- \bbeta_{i0}))(\hat{\bbeta}_{i}- \bbeta_{i0}) dt \cr
&& \geq \int^{z}_{0}(1-t)\kappa \left\| \hat{\bbeta}_{i}-\bbeta_{i0} \right\|_{2}^{2}dt \cr
&&=(hr+1)^{1/2}\eta_F \left\| \hat{\bbeta}_{i}-\bbeta_{i0} \right\|_{2} -\frac{(hr+1)\eta_{F}^{2}}{2\kappa},
\end{eqnarray}
where the first inequality is due to Assumption \ref{Assumption1}(g).
Combining \eqref{eq-A.21}--\eqref{eq-A.23}, we have
\begin{equation*}
0 \geq \frac{(hr+1)^{1/2}\eta_F \left\| \hat{\bbeta}_{i}- \bbeta_{i0} \right\|_{2}}{2}-\frac{(hr+1)\eta_{F}^{2}}{2\kappa},
\end{equation*}
which contradicts to \eqref{eq-A.20}.
Thus, \eqref{eq-3.5} is showed.

Now, consider $\left\| \hat{\bbeta}_{i} -\bbeta_{i0}\right\|_{2}$ for $i \in \left\{r+1, \ldots, p+r \right\}$.
By Proposition 1 in \citet{fan2019adaptive} and Proposition \ref{Proposition3}, we can show \eqref{eq-3.6}.
\endpf

\subsection{Proof of Theorem \ref{Theorem3}.}
\textbf{Proof of Proposition \ref{Proposition1}.}
Similar to the proofs of Theorem 3 in \citet{fan2018robust}, we can show \eqref{eq-4.6} and \eqref{eq-4.7} under the event
\begin{equation*}
E=\left\{\max_{1 \leq d \leq n} \max_{1\leq i, j \leq p} \left| \hat{\Gamma}_{d,ij} - \Gamma_{d,ij} \right| \leq  C  \sqrt{\log \left( pn \vee m\right) /m^{1/2}} \right\}.
\end{equation*}
By Weyl's theorem, \eqref{eq-4.6} implies \eqref{eq-4.5}. 
Thus, it is enough to show \eqref{eq-4.4} under the event $E$.
Without loss of generality, we assume that $\text{sign}\left(\langle \hat{\bq}_i^F, \bq_i^F \rangle\right)=1$ for $i = 1, \ldots, r$.
We have for each $d \in \left\{1, \ldots, n \right\}$ and $i \in \left\{1, \ldots, r \right\}$,
\begin{eqnarray}\label{eq-A.24}
&& \left| \hat{\xi}_{d,i}-\xi_{d,i} \right| \cr 
&& \leq p^{-1}\Big|\(\bq_{i}^F\)^{\top}\left(\hat{\bGamma}_d-\bGamma_d\right)\bq_{i}^F\Big|   + p^{-1}\Big|\(\hat{\bq}_{i}^F\)^{\top}\hat{\bGamma}_d\hat{\bq}_{i}^F-\(\bq_{i}^F\)^{\top}\hat{\bGamma}_d\bq_{i}^F\Big| + p^{-1}\Big|\(\bq_{i}^F\)^{\top}\bGamma_d\bq_{i}^F-\xi_{d,i}\Big|  \cr
&& = (\uppercase\expandafter{\romannumeral1})+(\uppercase\expandafter{\romannumeral2})+(\uppercase\expandafter{\romannumeral3}).
\end{eqnarray}
For $(\uppercase\expandafter{\romannumeral1})$, we have
\begin{equation}\label{eq-A.25}
(\uppercase\expandafter{\romannumeral1}) \leq p^{-1}\left\| \hat{\bGamma}_d-\bGamma_d\right\| _{F}  \leq C  \sqrt{\log \left( pn \vee m\right) /m^{1/2}}.
\end{equation}
For $(\uppercase\expandafter{\romannumeral2})$, we  have
\begin{eqnarray*}
\left\|\bq_{i}^F-\hat{\bq}_{i}^F \right\|_{2} &\leq& Cp^{-1}\left\| \sum^{n}_{k=n-\ell+1}\left( \hat{\bGamma}_k - \mathbf{\Psi}_{k}\right)/\ell \right\|_{2} \cr
&\leq& Cp^{-1} \ell^{-1} \sum^{n}_{k=n-\ell+1} \left( \left\| \hat{\bGamma}_k - \bGamma_k \right\|_{F} +\left\| \bSigma_k \right\|_{1} \right) \cr
&\leq& C\left(   \sqrt{\log \left( pn \vee m\right)/ m^{1/2}} +p^{-1} \max_{ 1 \leq i \leq p} \sum_{j=1}^p | \Sigma_{d,ij}| ^{\Upsilon}   (\Sigma_{d,ii} \Sigma_{d,jj} ) ^{(1-\Upsilon)/2} \right) \cr
&\leq& C\left(   \sqrt{\log \left( pn \vee m\right)/ m^{1/2}} + s_I/ p \right),
\end{eqnarray*}
where the first inequality is by Theorem 2 in \citet{yu2015useful}.
Hence, we have
\begin{eqnarray}\label{eq-A.26}
&& (\uppercase\expandafter{\romannumeral2})  \cr
&& \leq   p^{-1}\Big|\left(\bq_{i}^F-\hat{\bq}_{i}^F\right)^{\top}\hat{\bGamma}_d\left(\bq_{i}^F-\hat{\bq}_{i}^F\right)\Big|  +p^{-1}\Big|\left(\bq_{i}^F-\hat{\bq}_{i}^F\right)^{\top}\hat{\bGamma}_d \bq_{i}^F\Big| +p^{-1}\Big|\(\bq_{i}^F\)^{\top}\hat{\bGamma}_d\left(\bq_{i}^F-\hat{\bq}_{i}^F\right)\Big| \cr
&& \leq p^{-1}\big\|\hat{\bGamma}_d \big\| _{F} \big\|\bq_{i}^F-\hat{\bq}_{i}^F \big\|_{2}^{2}+2p^{-1}\big\|\hat{\bGamma}_d \big\| _{F} \big\|\bq_{i}^F-\hat{\bq}_{i}^F \big\|_{2} \cr
&& \leq C  \sqrt{\log \left( pn \vee m\right)/ m^{1/2}}.
\end{eqnarray}
For $(\uppercase\expandafter{\romannumeral3})$, we have
\begin{eqnarray}\label{eq-A.27}
(\uppercase\expandafter{\romannumeral3})&=& p^{-1}\(\bq_{i}^F\)^{\top}\bSigma_d \bq_{i}^F \leq p^{-1}\big\|\bSigma_d \big\|_{2} \leq p^{-1}\big\|\bSigma_d \big\|_{\infty} \leq p^{-1}\max_{ 1 \leq i \leq p} \sum_{j=1}^p | \Sigma_{d,ij}| ^{\Upsilon}   (\Sigma_{d,ii} \Sigma_{d,jj} ) ^{(1-\Upsilon)/2} \cr
&\leq& C s_I/p.
\end{eqnarray}
Combining \eqref{eq-A.24}--\eqref{eq-A.27}, we have
\begin{equation*}
\left| \hat{\xi}_{d,i}-\xi_{d,i} \right|\leq C \(\sqrt{\log \left( pn \vee m\right)/ m^{1/2}} +s_I/ p  \),
\end{equation*}
which completes the proof.
\endpf

\textbf{Proof of Theorem \ref{Theorem3}.}
We show the statements \eqref{eq-4.10}--\eqref{eq-4.12} under \eqref{eq-A.5}--\eqref{eq-A.6} and \eqref{eq-4.4}--\eqref{eq-4.9}.
For simplicity, we assume that $h=1$ and omit the intercept term $\bnu$.
Note that $\hat{\bbeta}_i = \left(\hat{A}_{1,i1}, \ldots, \hat{A}_{1,ir} \right)^{\top}$ for $1 \leq i \leq r$ and $\hat{\bbeta}_i = \left(\hat{A}_{1,i1}, \ldots, \hat{A}_{1,i\(p+r\)} \right)^{\top}$ for $r+1 \leq i \leq p+r$.
First, we consider \eqref{eq-4.10}.
By \eqref{eq-4.4} and \eqref{eq-4.8}, we have for any $1 \leq i \leq r$,
\begin{eqnarray*}
\left| \hat{\xi}_{n+1,i}-\E \left(\xi_{n+1,i} | \FF_{n} \right) \right|&=& \left|\sum^{r}_{j=1}\left(\hat{A}_{1,ij}\hat{\xi}_{n,j}-{A}_{1,ij}{\xi}_{n,j}\right) \right| \cr
&\leq& \sum^{r}_{j=1}\Big[\left|\left(\hat{A}_{1,ij}-{A}_{1,ij}\right)\left(\hat{\xi}_{n,j}- {\xi}_{n,j}\right)\right| \cr
&& \qquad \quad + \left|{A}_{1,ij}\left(\hat{\xi}_{n,j}- {\xi}_{n,j}\right)\right| + \left|\left(\hat{A}_{1,ij}-{A}_{1,ij}\right){\xi}_{n,j}\right|\Big] \cr
&\leq& C \Big[H_m + \left(\log n\right)^2\sqrt{ \log p / n}\Big].
\end{eqnarray*}

Consider \eqref{eq-4.11}.
Similar to the proofs of Proposition 1 in \citet{fan2019adaptive}, we can show $\hat{\bbeta}_i-\bbeta_{i0} \in \mathcal W_{i}$ for any $r+1 \leq i \leq p+r$, where $\mathcal W_{i}$ is defined in Assumption \ref{Assumption1}(g).
Thus, we have
\begin{eqnarray*}
\max_{r+1 \leq i \leq p+r}\sum^{p+r}_{j=1}\left|\hat{A}_{1,ij}-{A}_{1,ij}\right| &\leq& \max_{r+1 \leq i \leq p+r} 4 \sum_{j \in S_i}\left|\hat{A}_{1,ij}-{A}_{1,ij}\right| \cr
&\leq &  C\left\{s_{\beta}^{2}  s_{I}H_m^{1-\Upsilon} + s_{\beta}^{2}\left(\log n\right)^2 \sqrt{\log p/ n}\right\},
\end{eqnarray*}
where the last inequality is due to the Cauchy--Schwarz inequality and \eqref{eq-4.9}.
Then, by \eqref{eq-4.5}, we have, for any $r+1 \leq i \leq p+r$,
\begin{eqnarray*}
\left| \hat{\xi}_{n+1,i}-\E \left(\xi_{n+1,i} | \FF_{n} \right) \right|&=& \left|\sum^{p+r}_{j=1}\left(\hat{A}_{1,ij}\hat{\xi}_{n,j}-{A}_{1,ij}{\xi}_{n,j}\right) \right| \cr
&\leq& \sum^{p+r}_{j=1}\Big[\left|\left(\hat{A}_{1,ij}-{A}_{1,ij}\right)\left(\hat{\xi}_{n,j}- {\xi}_{n,j}\right)\right| \cr
&& \qquad \quad + \left|{A}_{1,ij}\left(\hat{\xi}_{n,j}- {\xi}_{n,j}\right)\right| + \left|\left(\hat{A}_{1,ij}-{A}_{1,ij}\right){\xi}_{n,j}\right|\Big] \cr
&\leq& C\Big[s_{\beta}^{2}  s_{I}H_m^{1-\Upsilon} + s_{\beta}^{2}\left(\log n\right)^2 \sqrt{\log p/ n}\Big].
\end{eqnarray*}

For \eqref{eq-4.12}, we have
\begin{eqnarray}\label{eq-A.28}
\|\tilde{\bGamma}_{n+1} - \E \left(\bGamma_{n+1} | \FF_{n} \right) \|_{\bGamma^{\ast}} &\leq&  \| \hat{\mathbf{\Psi}}_{n+1}-\E \left(\mathbf{\Psi}_{n+1} | \FF_{n} \right)  \|_{\bGamma^{\ast}} \cr
&&+ \| \hat{\mathbf{\Sigma}}_{n+1}-\E \left(\bSigma_{n+1} | \FF_{n} \right)  \|_{\bGamma^{\ast}} \cr
&=& \(\uppercase\expandafter{\romannumeral1}\) +\(\uppercase\expandafter{\romannumeral2}\).
\end{eqnarray}
Consider $\(\uppercase\expandafter{\romannumeral1}\)$. 
We have
\begin{eqnarray*}
 \left\| \hat{\mathbf{\Psi}}_{n+1}-\E \left(\mathbf{\Psi}_{n+1} | \FF_{n} \right)  \right\|_{F} &=& p\left\| \sum_{i=1}^{r}\left\{\hat{\xi}_{n+1,i}\hat{\bq}^F_{i} \(\hat{\bq}_{i}^F\)^{\top}  - \E \left(\xi_{n+1,i} | \FF_{n} \right)\bq_{i}^F\(\bq_{i}^F\)^{\top} \right\}  \right\|_{F} \cr
&\leq& p\Bigg[\left\| \sum_{i=1}^{r}\left\{\hat{\xi}_{n+1,i}-\E \left(\xi_{n+1,i} | \FF_{n} \right)\right\} \hat{\bq}_{i}^F \(\hat{\bq}_{i}^F\)^{\top}   \right\|_{F} \cr
&&  \qquad + \left\| \sum_{i=1}^{r}\E \left(\xi_{n+1,i} | \FF_{n} \right)\left\{\hat{\bq}_{i}^F \(\hat{\bq}_{i}^F\)^{\top}  - \bq_{i}^F\(\bq_{i}^F\)^{\top} \right\} \right\|_{F} \Bigg] \cr
&\leq& p\Bigg[ \sum_{i=1}^{r}\left| \hat{\xi}_{n+1,i}-\E \left(\xi_{n+1,i} | \FF_{n} \right)\right| \left\|\hat{\bq}_{i}^F \(\hat{\bq}_{i}^F\)^{\top}   \right\|_{F} \cr
&&  \qquad +  \sum_{i=1}^{r}\E \left(\xi_{n+1,i} | \FF_{n} \right) \left\|\left\{\hat{\bq}_{i}^F \(\hat{\bq}_{i}^F\)^{\top}  - \bq_{i}^F\(\bq_{i}^F\)^{\top} \right\} \right\|_{F} \Bigg] \cr
&\leq& Cp\Bigg[H_m + \left(\log n\right)^2\sqrt{ \log p / n}+  \sum_{i=1}^{r}\left\| \hat{\bq}_{i}^F \(\hat{\bq}_{i}^F\)^{\top}  - \bq_{i}^F\(\bq_{i}^F\)^{\top} \right\|_{F} \Bigg],
\end{eqnarray*}
where the last inequality is due to \eqref{eq-4.10}.
For the last term, by Theorem 2 in \citet{yu2015useful}, we have
\begin{eqnarray*}
\left\| \hat{\bq}_{i}^F \(\hat{\bq}_{i}^F\)^{\top}  - \bq_{i}^F\(\bq_{i}^F\)^{\top} \right\|_{F} &\leq& Cp^{-1}\left\| \sum^{n}_{d=n-\ell+1}\left( \hat{\bGamma}_d - \mathbf{\Psi}_{d}\right)/\ell \right\|_{2} \cr
&\leq& Cp^{-1} \ell^{-1} \sum_{d=n-\ell+1}^n \left( \left\| \hat{\bGamma}_d - \bGamma_d \right\|_{F} +\left\| \bSigma_d \right\|_{1} \right) \cr
&\leq& C\left(   \sqrt{\log \left( pn \vee m\right)/ m^{1/2}} + s_I/ p \right).
\end{eqnarray*}
Thus, we have
\begin{equation*}
\left\| \hat{\mathbf{\Psi}}_{n+1}-\E \left(\mathbf{\Psi}_{n+1} | \FF_{n} \right)  \right\|_{F} \leq Cp\Big[H_m + \left(\log n\right)^2\sqrt{ \log p / n}\Big].
\end{equation*}
Then, similar to the proofs of Theorem 4.1 in \citet{fan2018robust}, we can show
\begin{eqnarray}\label{eq-A.29}
\(\uppercase\expandafter{\romannumeral1}\) &\leq& C\Bigg[\left\{p^{-3/2}+p^{-1}  \hat{\xi}_{n+1,1}\right\} \left\| \hat{\mathbf{\Psi}}_{n+1}-\E \left(\mathbf{\Psi}_{n+1} | \FF_{n} \right)  \right\|_{F} \cr
&& \qquad \qquad \quad+p^{-3/2}  \hat{\xi}_{n+1,1} \left\| \hat{\mathbf{\Psi}}_{n+1}-\E \left(\mathbf{\Psi}_{n+1} | \FF_{n} \right)  \right\|_{F}^2 \Bigg] \cr
&\leq& C\Big[H_m + p^{1/2}H_m^{2}+ \left(\log n\right)^2\sqrt{ \log p / n} +p^{1/2} \log p \left(\log n\right)^4/n \Big].
\end{eqnarray}
Consider $\(\uppercase\expandafter{\romannumeral2}\)$. 
We have
\begin{eqnarray}\label{eq-A.30}
\(\uppercase\expandafter{\romannumeral2}\)^2 &\leq& p^{-1} \left\| \left\{\hat{\mathbf{\Sigma}}_{n+1}-\E \left(\bSigma_{n+1} | \FF_{n} \right) \right\}\bGamma^{-1}\right\|_{F}^{2}\cr
&\leq&  p^{-1} \left\| \hat{\mathbf{\Sigma}}_{n+1}-\E \left(\bSigma_{n+1} | \FF_{n} \right) \right\|_{F}^{2}  \left\| \bGamma^{-1} \right\|_{2}^{2} \cr
&\leq&  p^{-1} \left\| \hat{\mathbf{\Sigma}}_{n+1}-\E \left(\bSigma_{n+1} | \FF_{n} \right) \right\|_{F}^{2}\cr
&=&  p^{-1} \left\| \sum_{i=1}^{p}\left\{\hat{\xi}_{n+1,i+r}\hat{\bq}_{i}^I \(\hat{\bq}_{i}^I\)^{\top}  - \E \left(\xi_{n+1,i+r} | \FF_{n} \right)\bq_{i}^I\(\bq_{i}^I\)^{\top} \right\} \right\|_{F}^{2} \cr
&\leq& Cp^{-1}\Bigg[\left\| \sum_{i=1}^{p}\left\{\hat{\xi}_{n+1,i+r}-\E \left(\xi_{n+1,i+r} | \FF_{n} \right)\right\} \hat{\bq}_{i}^I \(\hat{\bq}_{i}^I\)^{\top}   \right\|_{F}^2 \cr
&& \qquad \qquad + \left\| \sum_{i=1}^{p}\E \left(\xi_{n+1,i+r} | \FF_{n} \right)\left\{\hat{\bq}_{i}^I \(\hat{\bq}_{i}^I\)^{\top}  - \bq_{i}^I\(\bq_{i}^I\)^{\top} \right\} \right\|_{F}^2 \Bigg] \cr
&=& \(\uppercase\expandafter{\romannumeral3}\)+\(\uppercase\expandafter{\romannumeral4}\).
\end{eqnarray}
For $\(\uppercase\expandafter{\romannumeral3}\)$, we have
\begin{eqnarray}\label{eq-A.31}
&& \(\uppercase\expandafter{\romannumeral3}\) \cr
&& = Cp^{-1}\sum_{i=1}^{p}\sum_{j=1}^{p} \mathrm{tr} \left(\{\hat{\xi}_{n+1,i+r}-\E \left(\xi_{n+1,i+r} | \FF_{n} \right)\} \{\hat{\xi}_{n+1,j+r}-\E \left(\xi_{n+1,j+r} | \FF_{n} \right)\}\hat{\bq}_{i}^I \(\hat{\bq}_{i}^I\)^{\top} \hat{\bq}_{j}^I \(\hat{\bq}_{j}^I\)^{\top}\right) \cr
&& = Cp^{-1}\sum_{i=1}^{p} \mathrm{tr} \left( \{\hat{\xi}_{n+1,i+r}-\E \left(\xi_{n+1,i+r} | \FF_{n} \right)\}^2 \hat{\bq}_{i}^I \(\hat{\bq}_{i}^I\)^{\top} \right)\cr
&& = Cp^{-1}\sum_{i=1}^{p} \{\hat{\xi}_{n+1,i+r}-\E \left(\xi_{n+1,i+r} | \FF_{n} \right)\}^2 \cr
&& \leq C \left(s_{\beta}^{2}  s_{I}H_m^{1-\Upsilon} + s_{\beta}^{2}\left(\log n\right)^2 \sqrt{\log p/ n}\right)^2,
\end{eqnarray}
where the last inequality is due to \eqref{eq-4.11}.
For $\(\uppercase\expandafter{\romannumeral4}\)$, we have
\begin{eqnarray*}
&& \(\uppercase\expandafter{\romannumeral4}\) \cr
&& = Cp^{-1} \left\|\sum_{i=1}^{p}\left\{\E \left(\xi_{n+1,i+r} | \FF_{n} \right) - \E\left(\xi_{n+1,p+r} | \FF_{n} \right)\right\}\left\{\hat{\bq}_{i}^I \(\hat{\bq}_{i}^I\)^{\top}  - \bq_{i}^I\(\bq_{i}^I\)^{\top} \right\}\right\|_{F}^2  \cr
&& \leq Cp^{-1}\sum_{i=1}^{p} \mathrm{tr}\Bigg[ \left\{\E \left(\xi_{n+1,i+r} | \FF_{n} \right) - \E\left(\xi_{n+1,p+r} | \FF_{n} \right)\right\}^2 \left\{\hat{\bq}_{i}^I \(\hat{\bq}_{i}^I\)^{\top}  - \bq_{i}^I\(\bq_{i}^I\)^{\top} \right\}^2\Bigg] \cr
&& \quad +Cp^{-1}\sum_{i=1}^{p-1}\sum_{j=i+1}^{p}\mathrm{tr}\Bigg[  \left\{\E \left(\xi_{n+1,i+r} | \FF_{n} \right) - \E\left(\xi_{n+1,p+r} | \FF_{n} \right)\right\} \left\{\E \left(\xi_{n+1,j+r} | \FF_{n} \right) - \E\left(\xi_{n+1,p+r} | \FF_{n} \right)\right\}  \cr
&&  \qquad \qquad \qquad \qquad \quad \times \left\{\hat{\bq}_{i}^I \(\hat{\bq}_{i}^I\)^{\top}  - \bq_{i}^I\(\bq_{i}^I\)^{\top} \right\}\left\{\hat{\bq}_{j}^I \(\hat{\bq}_{j}^I\)^{\top}  - \bq_{j}^I\(\bq_{j}^I\)^{\top} \right\}\Bigg]\cr
&& \leq Cp^{-1}\sum_{i=1}^{p} \left\{\E \left(\xi_{n+1,i+r} | \FF_{n} \right) - \E\left(\xi_{n+1,p+r} | \FF_{n} \right)\right\}^2   \left\|\hat{\bq}_{i}^I \(\hat{\bq}_{i}^I\)^{\top}  - \bq_{i}^I\(\bq_{i}^I\)^{\top} \right\|_{F}^2,
\end{eqnarray*}
where the first equality is due to the fact that $\sum_{i=1}^{p}\hat{\bq}_{i}^I \(\hat{\bq}_{i}^I\)^{\top} =\sum_{i=1}^{p}\bq_{i}^I\(\bq_{i}^I\)^{\top}$ and the last inequality is from the positiveness of $\E \left(\xi_{n+1,i+r} | \FF_{n} \right)-\E\left(\xi_{n+1,p+r} | \FF_{n} \right)$ for $1 \leq i \leq p-1$ and \eqref{eq-A.32} below.
For $i \neq j$, we have
\begin{eqnarray}\label{eq-A.32}
&&\mathrm{tr}\Bigg[\left\{\hat{\bq}_{i}^I \(\hat{\bq}_{i}^I\)^{\top}  - \bq_{i}^I\(\bq_{i}^I\)^{\top} \right\}\left\{\hat{\bq}_{j}^I \(\hat{\bq}_{j}^I\)^{\top}  - \bq_{j}^I\(\bq_{j}^I\)^{\top} \right\} \Bigg] \cr
&& =-\mathrm{tr}\Bigg[ \hat{\bq}^I_{i} \(\hat{\bq}^I_{i}\)^{\top}\bq_{j}^I\(\bq_{j}^I\)^{\top} + \bq_{i}^I\(\bq_{i}^I\)^{\top}\hat{\bq}^I_{j} \(\hat{\bq}_{j}^I\)^{\top}  \Bigg] \cr
&& = -\mathrm{tr}\Bigg[\(\hat{\bq}_{i}^I\)^{\top}\bq_{j}^I\(\bq_{j}^I\)^{\top}\hat{\bq}_{i}^I + \(\bq_{i}^I\)^{\top}\hat{\bq}_{j}^I \(\hat{\bq}_{j}^I\)^{\top}\bq_{i}^I  \Bigg] \cr
&& = -\mathrm{tr}\Bigg[\left\{\(\hat{\bq}_{i}^I\)^{\top}\bq_{j}^I\right\}^2+ \left\{\(\bq_{i}^I\)^{\top}\hat{\bq}^I_{j} \right\}^2\Bigg] \leq 0.
\end{eqnarray}
Thus,  we have
\begin{eqnarray}\label{eq-A.33}
\(\uppercase\expandafter{\romannumeral4}\) &\leq& Cp^{-1} \sum_{i=1}^{p} \chi^{2\(i+r\)} \left\|\hat{\bq}_{i}^I \(\hat{\bq}_{i}^I\)^{\top}  - \bq_{i}^I\(\bq_{i}^I\)^{\top}\right\|_{F}^2 \cr
&\leq& Cp^{-1} \sum_{i=1}^{p} \chi^{2\(i+r\)} \left\|\sum_{d=n-\ell+1}^{n}\(\hat{\bSigma}_d  - \bSigma_d  \)/\ell \right\|_{2}^{2}/\chi^{2\(i+r\)} \cr
&\leq&  C  s_{I}^2   H_m^{2-2\Upsilon},
\end{eqnarray}
where the first inequality is due to Assumption \ref{Assumption3},  the second inequality is from Theorem 2 in \citet{yu2015useful}, and the last inequality is due to \eqref{eq-4.6}.
By \eqref{eq-A.30}, \eqref{eq-A.31}, and \eqref{eq-A.33}, we have
\begin{equation}\label{eq-A.34}
\(\uppercase\expandafter{\romannumeral2}\) \leq C\(s_{\beta}^{2}  s_{I}H_m^{1-\Upsilon} + s_{\beta}^{2}\left(\log n\right)^2 \sqrt{\log p/ n}\).
\end{equation}
Combining \eqref{eq-A.28}, \eqref{eq-A.29}, and \eqref{eq-A.34}, we have
\begin{eqnarray}\label{eq-A.35}
\|\tilde{\bGamma}_{n+1} - \E \left(\bGamma_{n+1} | \FF_{n} \right) \|_{\bGamma^{\ast}} &\leq& C\Big[p^{1/2}H_m^{2}+ p^{1/2} \log p \left(\log n\right)^4/n \cr
&& \quad + s_{\beta}^{2}  s_{I}H_m^{1-\Upsilon} + s_{\beta}^{2}\left(\log n\right)^2\sqrt{ \log p / n} \Big].
\end{eqnarray}
\endpf

\newpage

\subsection{Miscellaneous materials}
\begin{algorithm}\caption{Parameter estimation  procedure}   \label{algorithm1}
\begin{algorithmic}
\State \textbf{Step 1 } Decompose the input volatility matrix:
$$\hat{\bGamma}_d= \sum_{k=1}^p \bar{\xi}_{d,k} \bar{ \bq}  _{d,k}  \bar{\bq}_{d,k} ^\top,$$  where $\bar{\xi}_{d,k}$ is the $k$-th largest eigenvalue of $\hat{\bGamma}_d$ and $\bar{\bq}_{d,k}$ is its corresponding eigenvector.

\State \textbf{Step 2 } (factor components)
Calculate $r$ eigenvectors, $\hat{\bq}_1^F, \ldots, \hat{\bq}_{r}^F$, of $\frac{1}{\ell}\sum_{d=n-l+1}^{n}\hat{\bGamma}_d$ and obtain the eigenvalues $\hat{\xi}_{d,i}=\(\hat{\bq}_{i}^F\)^{\top}\hat{\bGamma}_d \hat{\bq}^F_i/p$ for $d=1, \ldots, n$ and $i=1, \ldots, r$.

\State  \textbf{Step 3 } Obtain the input idiosyncratic volatility matrix estimator:
$$
\bar{\bSigma}_d =(\bar{\Sigma}_{d,ij})_{1\leq i,j\leq p} = \hat{\bGamma}_d - \sum_{k=1}^r \bar{\xi}_{d,k} \bar{\bq} _{d,k} \bar{\bq}_{d,k} ^\top.
$$

\State  \textbf{Step 4 } Threshold the input idiosyncratic volatility matrix estimator:
\begin{equation*}
 \hat{\Sigma}_{d,ij} =
\begin{cases}
 \bar{\Sigma}_{d,ij} \vee 0   & \text{ if } i= j\\
 g_{ij} ( \bar{\Sigma}_{d,ij}) \1 ( |\bar{\Sigma}_{d,ij}| \geq \upsilon_{ij} )  & \text{ if } i \neq j
\end{cases}
\quad \text{ and } \quad \hat{\bSigma}_d =(\hat{\Sigma}_{d,ij})_{1\leq i,j\leq p},
\end{equation*}
where $g_{ij} (\cdot) $ satisfies $|g_{ij} (x) -x |\leq \upsilon_{ij}$,  and  $\upsilon_{ij} = \upsilon_{m}\, \sqrt{ ( \bar{\Sigma}_{d,ii} \vee 0 ) ( \bar{\Sigma}_{d,jj} \vee 0 )}$.

\State  \textbf{Step 5 } (idiosyncratic components)
Calculate $p$ eigenvectors, $\hat{\bq}_{1}^I, \ldots, \hat{\bq}_{p}^I$, of $\frac{1}{\ell}\sum_{d=n-l+1}^{n}\hat{\bSigma}_d$ and obtain $\hat{\xi}_{d,i+r}=\(\hat{\bq}_{i}^I\)^{\top}\hat{\bSigma}_d \hat{\bq}_i^I$ for $d=1, \ldots, n$ and $i=1, \ldots, p$.

\State  \textbf{Step 6 } Estimate the factor coefficient:
\begin{equation*}
\hat{\bbeta}_{i} = \arg \min_{\bbeta_i \in \mathbb{R}^{hr+1}}{\mathcal{L}}^{F,i}_{\tau,\varpi}(\bbeta_i) \quad  \text{ for } i=1, \ldots, r,
\end{equation*}
where ${\mathcal{L}}^{F,i}_{\tau,\varpi}(\bbeta_i)$ is  defined in \eqref{eq-3.4}.

\State  \textbf{Step 7 } Estimate the idiosyncratic coefficient:
\begin{equation*}
 \hat{\bbeta}_{i} = \arg \min_{\bbeta_i \in \mathbb{R}^{h(p+r)+1}}{\mathcal{L}}^{I,i}_{\tau,\varpi}(\bbeta_i) + \eta_I \left\| \bbeta_i \right\| _{1} \quad \text{ for } i=r+1, \ldots, p+r,
\end{equation*}
where $\eta_I>0$ is the regularization parameter, and ${\mathcal{L}}^{I,i}_{\tau,\varpi}(\bbeta_i)$ is defined in \eqref{eq-3.2}.

\State  \textbf{Step 8 }
Calculate the future eigenvalue  and conditional expected volatility matrix:
\begin{eqnarray*}
&&\hat{\bxi}_{n+1}=\left(\hat{\xi}_{n+1,1},\ldots, \hat{\xi}_{n+1,p+r}\right)^{\top}=\hat{\bnu}+\sum_{k=1}^{h}\hat{\bA}_{k}\hat{\bxi}_{n+1-k} \quad \text{ and}\cr
&&\tilde{\bGamma}_{n+1}=\hat{\mathbf{\Psi}}_{n+1} + \hat{\mathbf{\Sigma}}_{n+1} = p\sum^{r}_{i=1}\hat{\xi}_{n+1,i}\hat{\bq}_{i}^F \(\hat{\bq}^F_{i}\)^{\top}+\sum^{p}_{i=1}\hat{\xi}_{n+1,i+r}\hat{\bq}_{i}^I \(\hat{\bq}_{i}^I\)^{\top}.
\end{eqnarray*}
\end{algorithmic}
\end{algorithm}

\begin{figure}[!ht]
\centering
\includegraphics[width = 0.93\textwidth]{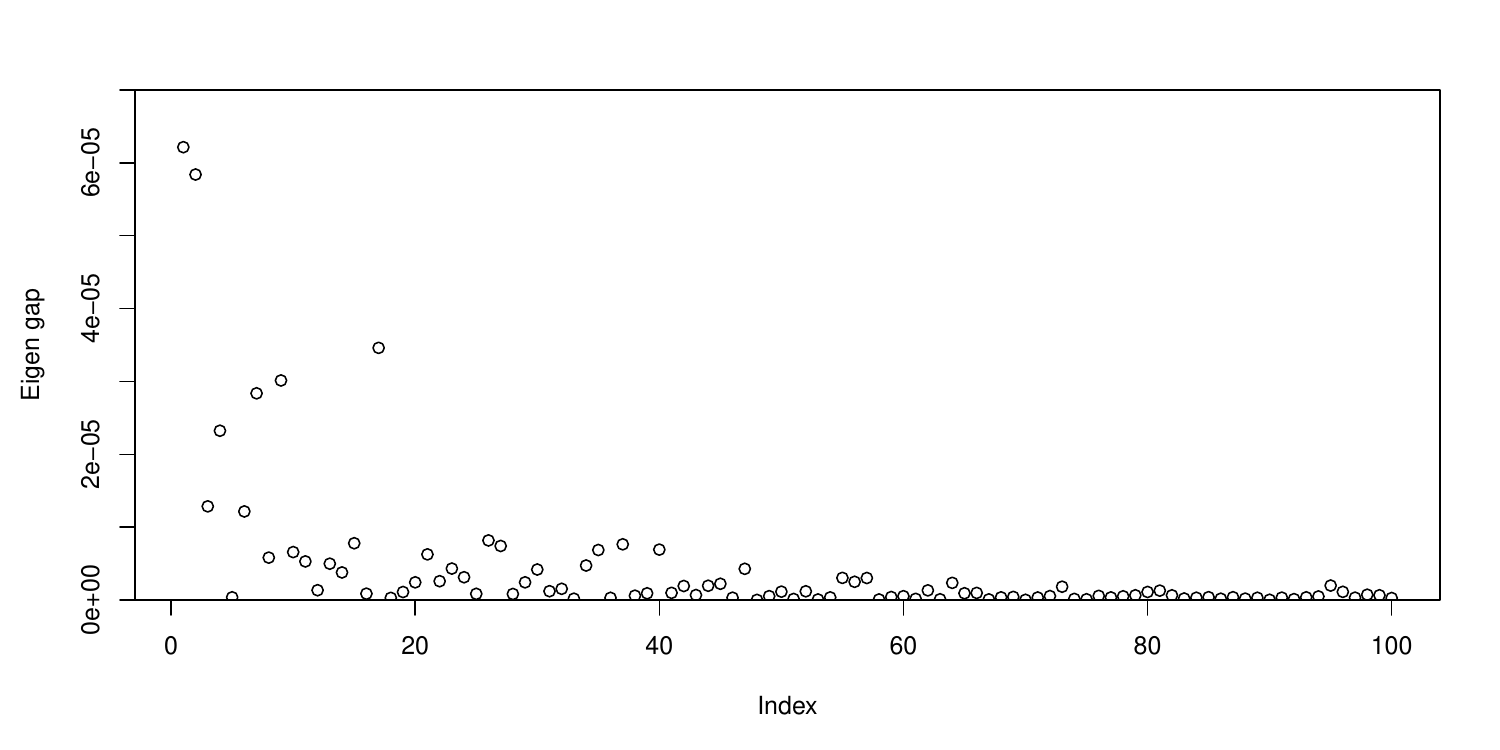}
\caption{The plot of the first 100 differences between the consecutive eigenvalues of the average of 997 idiosyncratic volatility matrix estimators. 
{We estimated the rank $r$ based on the rank estimation procedure in \eqref{eq-5.1} with $n=997$.
The result is $\hat{r}=3$.}
We used 1-min log-returns of the top 200 large trading volume stocks among the S\&P 500 from January 2016 to December 2019.}\label{Fig-7}
\end{figure}

\end{spacing}
\end{document}